# HIERARCHICAL MODELING

# OF MULTIDIMENSIONAL DATA

# IN REGULARLY DECOMPOSED SPACES

## TOME 2 : IMPLEMENTATION ON COMPUTER

## (1988 – 1992)

- 2016 -

Olivier Guye

# Table of Contents

# Table of Figures

# Introduction

This document presents how the software designed for hierarchical modeling of multidimensional data has been developed and implemented on different computers, especially on parallel architecture computers.

It describes how have been implemented the operations described in the previous tome and what are the complementary operators that have been designed for having at one's disposal a coherent functions set enabling to pre-process, to analyze and to solve problems using geometrical reasoning.

In data pre-processing, a specific effort has been provided on topological analysis in the aim of generalizing the filtering functions proposed in mathematical morphology.

The first version of the software has been written in Fortran IV programming language so as to be used on the scientific computers available at that time: it has been necessary to simulate in this language the mechanisms enabling to implement recursive calls of functions and procedures as well as data structures with simple or double links in order to manipulate lists, stacks, queues and binary trees. These mechanisms are partly described underneath because they are essential for the implementation of algorithms describes in annex of these documents.

Finally, it is shown how the software has been implemented on parallel computers so as to take in account the existing needs as far as the processing of huge numerical data bases is concerned. The works for parallelizing the software have been done with the support of the Direction de la Recherche et des Etudes Technologiques (DRET-ADERSA contract n°90/34/106) and has been led on two different computing systems:

– a synchronous computer with distributed memory, the CONNECTION MACHINE 2 from the firm THINKING MACHINES;

– an asynchronous computer with distributed memory, the T.NODE from the manufacturer TELMAT INFORMATIQUE.

Indeed, starting from its first beginnings ADERSA has created and developed competences in the field of parallel computing; especially with Pr Claude Timsit who was a pioneer in this domain with the development of the first synchronous computer in France, named PROPAL, and to whom was awarded the CNRS bronze medal for these works in 1978.

It must be kept in mind that if the software has been named KDTREE, it is different from the one developed by J. L. Bentley according to the fact that the modeling space is regularly decomposed and that if the word of pyramid is used for pointing out a valued tree, it does not correspond to the classical meaning for a pyramid where all the branches are fully developed down to its building precision.

In spite of the fact that this software has migrated into different programming languages starting from Fortran IV up to ANSI C, it is no more existing today: only the algorithms written in pseudo-code are still remaining and they can be found in annex of the two first tomes. For writing them, it has been used a tridimensional terminology in the aim to lighten the result ; but it must be kept in mind that these algorithms apply on data sets belonging to spaces of any dimension (but being integer and finite). The reader will find in the first part a glossary which gather and explain all the terms that are used in the three tomes of the present work set.

# 1. Overall presentation

The KDTREE software is a multidimensional modeling software for volume (or hyper-volume) data. Its design and its development have taken benefit of the support of the Délégation Générale à l'Armement.

It is focusing on a wide range of application domains and is relying on a specific principle for solving problems : the « divide and conquer » paradigm from which it is known that it is providing an optimal bound for time processing and memory space consumption for a given class of problems. The paradigm is the following one: it consists in dividing a problem that cannot be directly solved into sub-problems and in iterating this approach until that all the problems are solved. Well-known authors have linked their names to theses works, especially in computer geometry and the results of some of their works are now daily used in display and printing systems.

In fact, the KDTREE software has existed in different releases, from which several of them were working on parallel architecture computers and have enabled to perform experiments about massive parallelism.

In this document, it will be talked about pyramids for naming trees modeling the evolution of a functional relatively to its support: this term is not truly appropriate but it will allow to quickly distinguish between:

- trees enabling to model volume data sets (the $2^k$-trees at their own meaning) ;

- trees that are used for modeling surface manifolds (the inappropriately named pyramids, because it is describing complete trees with fully developed branches and on which a specific focus will be applied during the chapter dealing with parallelism).

# 2. Representation model

## *2.1. Nature of the modeled information*

The KDTREE software enables to represent pieces of information described in a space of any dimension:

- along a straight line (mono-dimensional information) ;
- inside a plane (bi-dimensional information) ;
- in the space (tri-dimensional information) ;
- in a hyper-space of higher dimension.

It enables to model numerical pieces of information of two different kinds:

- volume pieces of information, by representing the indicator function of a data set ;
- surface pieces of information, by representing the simple function (constant by pieces) associated to the data set.

Let us assume that we have at our disposal a data set belonging to a space of dimension k:

$$V = \{v_j, j = 1, \cdots, N\},$$

where $v_j$ is a coordinate vector $\left(v_1^j, \cdots, v_i^j, \cdots, v_k^j\right)$

then the indicator function of this set is the membership function :

$$\delta : \{v\} \rightarrow \{0,1\}, \text{ such as } S = \{v / \delta(v) = 1\},$$

that is to say, the function which is true for all the points of the set $V$ and false for its complementary set (once more named background of the space) :

$$\overline{S} = \{v / \delta(v) = 0\}$$

This describing model suits very well to data of volume kind.

Let us try now to define a straight line in a plane, a surface in the space, a hypersurface in a hyper-space.

In a first step, it can be represented by the functional:

$$f : \{v\} \rightarrow \{f(v)\},$$

where $v$ is describing the support of $f$ and $f(v)$ is finding its values in a scalar value set.

In a second step, it can once more represented by its indicator function :

$$\delta : \{v, f(v)\} \to \{0,1\}$$

describing a $k+1$ dimensional set made up from the functional support to which $v$ is belonging and from the set of scalar values on which the functional $f(v)$ can be expressed.

It is then possible to come back to a volume representation for data of surface kind.

For representing such data sets, KDTREE is based on a hierarchical description data structure, in which each branch is encoding a particular path enabling to reach a set of coordinates in the representation space.

This path enables to reach a particular point or a sub-set in the space.

When this one is belonging to the modeled object or the functional support, the corresponding terminal node is colored in black, else in white for describing the space background.

When the indicator function is modeled, tree nodes are not valued.

At the opposite, when a functional is modeled, tree nodes are valued with the values locally taken by the functional.

Concerning data of surface kind, it is possible to move from one representation kind to another one, then to come back to the previous one.

Let us illustrate theses concepts using a real example. Let us assume that we have got a digital elevation model, this one is usually represented by a matrix of altimetry data regularly sampled over a planar coordinate lattice.

For instance, a digital terrestrial model usually represented by a matrix of altimetry data computed by starting from the sea level and sampled along latitude and longitude, that is to say into a planispheric reference frame (Earth is spherical, but coordinates are handled in a planar manner). It is a surface representation which is describing the position the earth-atmosphere interface comparatively to the sea elevation. The functional is here the altimetry.

If we are only concerned by this single interface where the functional remains positive, the functional support represents the lands above the sea and the planisphere background the submerged ground. They are data of surface kind, because at any point of the support, there is one piece of data and only one. This altimetry matrix can be represented by a bi-dimensional multi-valued tree whose nodes will be colored in black for definite positive values, this is the land over the sea, and the background in white for describing the lack of information above the seas and oceans Rendered on a display, this matrix would appear using the form of a map whose local coloring will be tuned by the elevation of the present point on the lattice of planar coordinates (hypsometry).

By adding the functional axis to the first two axes of the support, it will be got a volume representation of the terrestrial surface above the land above the sea. Rendered on a display, the terrestrial surface is then shown in the form of a uniformly colored cover in sustenance at a given elevation, varying in accordance to the position of the observed point. It corresponds to the indicator function of the emerged land surface: an unvalued tri-dimensional tree enables to model these data.

In this kind of representation, it will be seen in a next future that it will be possible to rebuild the relief laying under this surface, in order to handle the terrestrial relief as a solid shape, then how to perform the inverse operation in order to come back to the initial representation. This relief may itself be colored with auxiliary information, as for instance the nature of objects on the ground with the help of planetary data linked to altimetry data

Therefore, the representation model that will be described is dealing with the modeling of numerical data sets that can be found in a wide range of scientific and technical problems:

- planar image analysis (monospectral, multispectral images) ;
- tomography image analysis ;
- reconstruction of objects from tridimensional scans ;
- tridimensional solid object reconstruction ;
- environment modeling ;
- observation of evolving systems (by inserting time as one dimension of the representation space) ;
- cartography ;
- earth resource observation ;
- statistical data analysis ;
- numerical database management ;
- geometrical reasoning;
- decision making.

Most of these application fields will be quoted during this presentation.

| *Name* | *Function* |
|---|---|

| KDCLAL | Coloring an altimetry tree by a planimetry tree |
| KDDETZ | Retrieve the elevation of a point in a tree |

## *2.2. Data representation model*

The representation model comes from the decomposition principle induced by the paradigm "divide and conquer". Given a set of numerical data, it can be found a box of a given shape that is including data population, then it can be applied a regular dividing procedure enabling to divide boxes into sub-boxes and once more these ones iteratively until a uniform result can be obtained in each of these boxes :

- either the box is empty or full for an indicator function ;

- or all the functional values are defined and equal, or not at all for a simple function.

Compared to a cellular representation model which is exhaustively describing the space holding the data, in such a model, to be able and keep in packets the homogenous data, enables to include a compression operator inside the representation model.

Moreover, the ability to regularly divide a data collection into sub-collections and to iterate this process leads to provide a tree-like structure for organizing the data: it is consequently a hierarchical representation model.

To keep in an aggregated manner the sub-sets of homogenous data results in the fact that the different branches of this structure will be irregularly developed: it is then about incomplete trees, but information compression needs it.

Let us take in account a set of scalar data spread along an axis and bounded inside an interval of values. Let us have a look to the content of this interval and if this one is not homogenous, then let us divide it by the middle in two parts. Let us once more examine each of these two half-intervals and let us divide them once more if they do not seem homogenous. Let us carry on such a way, until reaching elementary pieces of data if necessary. At the end of this processing, the scalar data collection will be modeled by a binary tree, where all the non terminal nodes will have two sons.

Let us take in account now a planar binary image, it can be regularly divided half by half along the two dimensions of its support, that is to say into four quadrants:

- north-west ;

- north -east ;

- south- east;

- south - west.

If the image data is not homogenous in each quadrant, it can be applied once more this division operation in four equal parts for each implied quadrant. By carrying on until reaching the digital resolution of the image, it will be provided a quaternary tree, where all the non terminal nodes will have four sons for representing this image.

If the distance between the bounding values according each axis is the same, then the initial bounding box and all of its sub-divided boxes will be squares. That is the result when the coordinates of the data set have been normalized.

The same building process can apply on a tri-dimensional object. The quadrants become octants in order to take in account the complementary dimension and the resulting data structure is an octernary tree.

This dividing process generalizes itself on data sets of any dimension k. If the coordinates are normalized, then the initial cube becomes an unitary hypercube, this one is divided into $2^k$-ants and the corresponding tree is a $2^k$-tree.

The number of division levels $r$ sets the representation precision, each of the k dimensions will have been discretized:

- if the data are integers belonging to the following sub-set of N $\left\{ 0, 1, ..., 2^r - 1 \right\}$ ;

- if the data is normalized inside the following sub-set of Q or R $\left\{ 0, \dfrac{1}{2^r}, ..., \dfrac{2^r - 1}{2^r} \right\}$.

The data is then referring to :

- the integer hypercube $\left\{ 0, 1, ..., 2^r - 1 \right\}^k$ ;

- the unitary hypercube $\left\{ 0, \dfrac{1}{2^r}, ..., \dfrac{2^r - 1}{2^r} \right\}^k$ .

In practice, in order to use only a single structure whatever is the dimension of the modeling space, the dividing process has been modified so as only to provide binary trees. It is managed by sequentially dividing half by half intervals according to each space dimension.

So, the previous binary image is divided into two rectangular sub-images according to the first axis, then these two rectangles recover the shape of quadrants by dividing them according to the second axis.

In three dimensions, cubes are temporally divided into rectangular parallelepipeds when the sequential diving process is performed.

This operation is equivalent to map the unitary hypercube $\left\{0, \frac{1}{2^r}, ..., \frac{2^r-1}{2^r}\right\}^k$ into the unitary line segment $\left\{0, \frac{1}{2^{kr}}, ..., \frac{2^{kr}-1}{2^{kr}}\right\}$.

It can be then noticed that the size of the handled data will be straightly proportional to the product of the space dimension $k$ by the analysis precision $r$ along each axis.

The processing time of processing operators will depends on the resulting value of this product; so it will be only got a sustainable response time when it is handling data sets in spaces of low dimension at middle resolution and dealing with spaces of mid resolution at low resolution. It will be not so easy to model spaces of higher dimension according to this approach.

So according to the space dimension, a data set modeled at precision 8 will hold at most:

- $2^8$, that are 256 elements in dimension 1 ;
- $2^{16}$, that are 65 536 elements in dimension 2 ;
- $2^{24}$, that are 16 777 216 elements in dimension 3 ;
- $2^{32}$, that are 4 294 967 296 elements in dimension 4.

At the opposite $2^{18}$, that are 262 144 elements that represent a space analyzed at precision:

- 18 in dimension 1 ;
- 9 in dimension 2, but also a planar mesh of 512 x 512 cells;
- 6 in dimension 3, but also a solid mesh of 64 x 64 x 64 cells.

## *2.3. Parameterization of processing operators*

The performance of processing operators depends on these two parameters: space dimension and computing precision

It can be distinguished two classes of operators:

- those that enable to build trees modeling data sets ;
- those that are transforming these trees.

It is not necessary to use the latter ones at the same precision than the one used for generating their initial representation models.

When an operator is applied with a precision coarser than the building precision, the result is provided according to its upper hull: if a non terminal node is met during the tree analysis it will be processed as a terminal black node at this precision. So in case of uncertainty, its bounding box will be taken in account at the requested precision.

It will be done in the same way for a valued tree: at each non terminal node, the maximum of the son functional values will be registered in the corresponding node and it will be this given value that will be taken in account when a tree will be examined at an intermediate precision. In such a manner, the functional is approximated with this precision bi its upper bounding half-continuous function.

With a variable precision, volume data is modeled by the nesting of volume structures that are holding the initial set and surface data by the set of simple functions that are bounding up the initial data set. This approach enables to preserve set compactness

Among the operators that are transforming trees, it will be found a particular application case with geometric transformations where the initial tree is examined in the aim to produce a new tree; in this case it will be necessary to specify together the precision to which the original tree must be analyzed and the precision to which the resulting tree must be built.

Varying precision computing allows to develop an application by starting to experiment it at a coarse precision and then to test it at a finer precision and to exploit finally it at full precision with changing the applied data set.

It can also be noticed that if the application must be used in constrained time, it will be ever possible to find the right precision value respecting such an exploitation condition.

## *2.4. Representation model registration*

Two different techniques are used under the KDTREE software for recording valued or not binary trees.

The first technique is relying on a data sequential allocation system the linear lists. It is a tree coding using an alphabet made from three basic colors enabling to characterize nodes:

- 0 and 1, for white and black terminal nodes ;

- 2 for non terminal nodes (grey).

If a tree is visited according to an exhaustive traversal, it will be provided a tree code made from a sequence of 2, 1 and 0 that will sequentially follow according to the order with which each nodes belonging to the tree will be met.

For a single tree, different traversals may be envisioned according to the priority puts favorably on:

- the filiation order over the sibling order of nodes (in depth or in breadth first) ;

- upwards over downwards moves (by climbing up or down in the tree).

Whatever is the chosen traversal, the number of visited nodes remains the same. Nodes are coded using only two bits in order to provide compact tree codes. In addition, the functional values are recorded apart concerning valued trees, according to the same traversal but only for black terminal nodes.

This data recording technique is only used for data archiving within the KDTREE software: although the resulting codes are very compacts, it is a data structure enabling only a sequential access for which it is necessary to look over the whole information in order to reach an single piece of data.

During processing, the KDTREE software is using a data structure with a fastest data access based on indexed allocation: the linked lists. This one is relying on a dynamic memory allocation system and including linking information enabling to link data one to another one according to an order defined in advance.

It might be envisioned to implement a double indexing system enabling to handle in parallel the two main orders of displacement into a tree. Only the filiation order is explicitly managed with the KDTREE software with the help of simple linkage lists, favoring then depth-first tree traversals.

So for coding a non terminal node using an indexed way, it will be necessary to get:

- two double words, containing applicative data and pointers towards left and right sons, for an unvalued binary tree ;

- three double words for handling the functional associated to the present node for a valued binary tree.

With the help of self-referring addresses for pointing white and black nodes, it is only necessary for coding a terminal node:

- one double word for an unvalued binary tree;

- two double words for a valued binary tree..

| *Name* | *Function* |
|---|---|
| KDWRBT | Write on disk a tree |
| KDRDBT | Read on disk a tree |
| KDWRPY | Write on disk a pyramid |
| KDRDPY | Read on disk a pyramid |

# 3. Data management

## *3.1. Memory management*

The KDTREE software includes its own management functions of its memory space. This one relies on two elementary mechanisms :

- a virtual addressing system ;

- a dynamic memory allocation system.

The virtual addressing system is built over a mapped memory system. The unit of memory allocation is the double-word, couple consisting in a value position and a link position enabling to manage all the free memory according to a linked list.

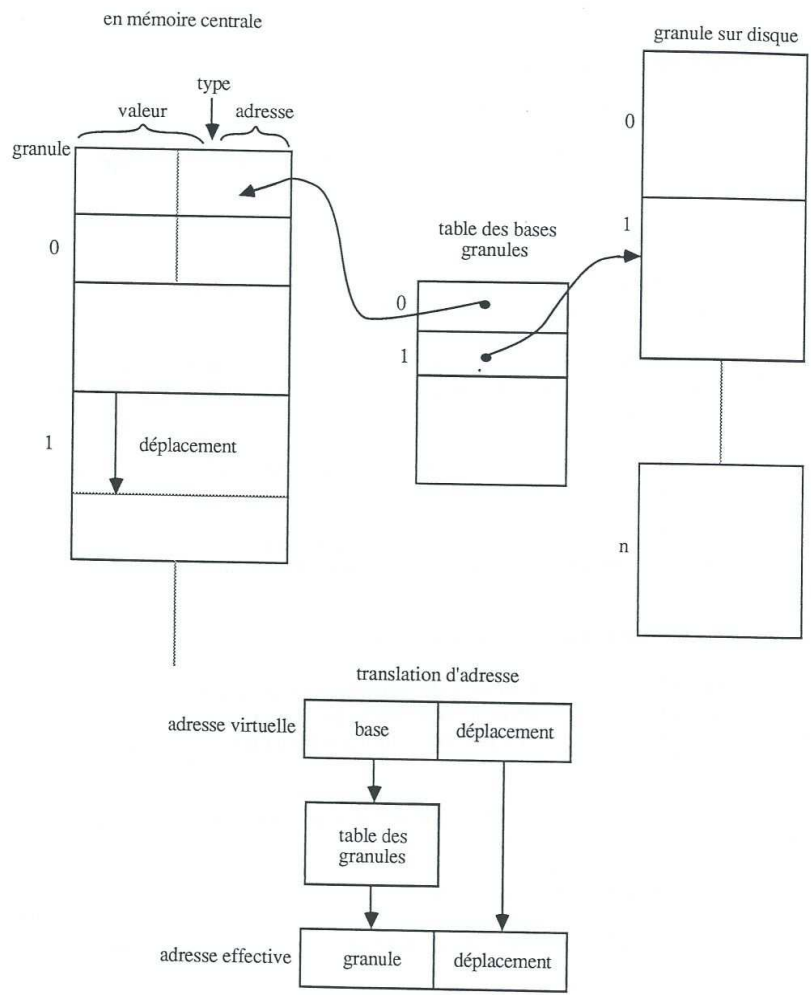

**Figure 1 : Doublet management in a virtual memory**

It is a fixed-length memory allocation system, working without any mechanism for data reorganization («garbage collector»). Double-words are gathered in memory granules or fixed-length memory pages, representing the memory extending unit enabling to satisfy allocation requests that cannot be straightly satisfied and the memory exchange unit – on external memory when the virtual memory is activated.

This one enables to manage data sets that cannot directly stand in the processor physical memory, with the help of the computer mass storage. In this case, the data stored on the disk non-volatile memory is representing the memory space addressable by software and only a subset of this data is standing at a given time in the processor volatile memory.

Volatile and non-volatile memories are exchanging pieces of information according to the processing needs.

The localization of data in computer is handled with the help of' a memory granule table informing where the granules are standing. A translating address mechanism allows to substitute in the granule addressing field a virtual piece information by a real one in order to point out the actual granule localization.

Rewriting in non-volatile memory is postponed until the performance end of a process, when this one is unnecessary, in order to keep the coherence of data between volatile and non-volatile memories.

Concerning critical applications needing to be performed in high reliability conditions, the virtual memory management mechanism is complemented with a mirror recording system enabling to double the copy of data in non-volatile memory.

The activation of memory storage is set up at the end of the processing after having checked that no processing error occurred. In this situation, the saving is applied on all the granules in non-volatile memory that would have been modified during the processing. If its exploitation must restart, then the data in mirror storage are automatically copied back before restarting data processing.

In this way, the software always restarts on the last validated memory state :

- before error detection ;
- before computer failure.

Three processing modes are then available within the KDTREE software:

- direct handling of data in volatile memory ;
- data management in virtual memory above the addressing capabilities of the computer volatile memory;
- virtual memory data management with a safety mirror recording system.

It is also possible to tell to the software the maximum number of granules that are expecting to be used during a work session.

| *Name* | *Function* |
|--------|------------|
| KDMDWK | Define the management mode of the work area |
| KDINWK | Initialize memory space for the work area |
| KDALLO | Allocate a memory double-word |
| KDFREE | Free a memory double-word |
| KDWTYP | Modify the type of a double-word |
| KDRTYP | Read the type of a double-word |
| KDWVAL | Write a value in a double-word |
| KDWLNK | Modify the link of a double-word |
| KDRVAL | Read the value in a double-word |
| KDRLNK | Read the link of a double-word |

## *3.2. Work session*

When data is only staying in the processor memory, its availability does not last more than the processing duration of a program. It is then necessary to archive trees that have been created if it is expected to use them once more in the future.

At the opposite, if it is handled with the help of virtual memory, with or without mirror copy, its life duration will be longer than the program processing time. It is corresponding to the duration of the work session:

- it starts with a session creation ;

- it stops for a moment with the session suspension ;

- it restarts with the session recovery ;

- it ends with a session deletion.

Each session requires to provide a name that will be used as a prefix for defining the names of files that will store its associated data. The session duration is corresponding to the life duration of these files.

| *Name* | *Function* |
|--------|------------|
| KDCRWK | Create a work session |
| KDRSWK | Restart a work session |
| KDSUWK | Suspend a work session |
| KDDEWK | Delete a work session |

## *3.3. Data structures*

Three addresses have got a specific status in the doublet memory: the addresses NIL, WHITE and BLACK. They are self-referring addresses and are used as end addresses by data structures of list or tree types.

Within the KDTREE software, the elementary structure is the simple linkage list: at the initialization of a session, the doublet memory is only made up from the three self-referring addresses and the list of free words of the memory allocation system.

Each list has got a head, except trees whose root is a particular node of the structure, its top. A head is a doublet dedicated for storing the address of the first list element, and for some ones for storing also the address of the last list element (in a queue).

There are four kinds of lists:

- the lists that are managed element by element ;

- the queues where elements are extracted from the head and where elements are inserted by the end ;

- the stacks where elements are inserted and extracted from the head ;

- the circular lists, where the end element is referring the head of the list.

Except in the last ones, the last element of a list is always referring the end of a list with NIL.

The circular lists, handled as queues, are used for modeling coordinate vectors and matrices under the KDTREE software.

As already quoted, trees may have two distinguished types: unvalued and valued binary trees.

Each word of the doublet memory is tagged with the structure type to which it belongs at a given while, this mechanism enables to build structures of structures:

- lists of lists for handling a matrix for instance ;

- trees for which lists are hanged to their nodes ;

- lists of trees for handling a forest of connected components.

These functions enable to globally manage these structures without taking notice of their own architecture.

| *Name* | *Function* |
|---|---|
| KDNIL | Comparison of an address with the value NIL |
| KDWHIT | Comparison of an address with the value WHITE |
| KDBLAC | Comparison of an address with the value BLACK |

| *Name* | *Function* |
|---|---|
| KDCRLS | Create a list |
| KDEMLS | Check if a list is empty |
| KDDHLS | Delete the head of a list |
| KDINLS | Insert an element in a list |
| KDSULS | Suppression of an element in a list |
| KDDELS | Delete a list |

## 3.4. Elementary operators

If the lists have at their disposal a head enabling to point at an element, they enable to handle the elementary status which is empty list. It is characterized by the fact that its head is pointing at nothing, that is to say at the address NIL.

On the contrary concerning the trees, a tree root is not handled in a distinctive manner from the other tree nodes: a tree has no head. It allows so to choose a recursive definition for a tree where no node can basically be distinguished from another one.

An empty tree has no reality, but:

- a white node will ever represent an empty set of any dimension and at any precision ;

- a black node, the entire space (full set) of any dimension and at any precision.

It is then the terminal or non terminal status that will prevail in a tree. This recursive definition enables to transform a tree during a traversal allowing to visit all its nodes.

At the traversal time, two transformations can be applied on a node without questioning the whole data structure:

- the merge of its filial nodes, if it is non terminal ;

- its own fission into two new nodes, if it is terminal.

The fission generates by decomposition two new elements in the tree-like structure. The merging deletes by aggregation the same elements.

It can then be set up a straight correspondence between the elementary operators that can apply on a list and those used on a binary tree:

- create a list/create a terminal node of a given color (white/black) ;

- check if a list is empty/check if a node is terminal ;

- next list element/address of a son of a given side (left/right) ;

- insert a list element/node fission or sub-trees union ;

- suppression of a list element/merging of a non terminal node ;

- delete a list/delete a tree.

Delete operations are summarized to the deletion of a list head or a terminal node. By combining deletion and merging operators, complex structures can be deleted.

Concerning the insertion of a new element in a tree, it has been distinguished two different ways to perform it: fission of a preexisting node or union of two sub-trees.

The fission of a node into two new nodes is performed during a processing aiming in modifying a preexisting tree.

At the opposite, the union of two sub-trees occurs when the processing is building a new tree deduced from a first one which is currently analyze: it is usually a transformed tree of the first one.

The fission of a node is an operator used during the descent phase in the analysis of a tree, while the union of two sub-trees occurs more especially on the ascending return of the same traversal, when a tree is synthesized.

| Name | Function |
|---|---|
| KDCRBT | Create a tree node |
| KDTERM | Check if a node is terminal |
| KDSON | Read the address of a node son |
| KDFIBT | Fission of a terminal node |
| KDMERG | Merge of two terminal nodes |
| KDUNBT | Union of two sub-trees |
| KDDEBT | Delete an unvalued tree |

| Name | Function |
|---|---|
| KDCRPY | Create a pyramid node |
| KDISOC | Check if two nodes are iso-colored |
| KDWFCT | Write a functional value in a node |
| KDRFCT | Read a functional value in a node |
| KDWCOL | Modify the color of a node |
| KDRCOL | Get the color of a node |

| Name | Function |
|---|---|
| KDBTPY | Conversion of a tree into a pyramid |
| KDPYBT | Conversion a pyramid into a tree |

| Name | Function |
|---|---|
| KDDEST | Delete a structure of any kind |
| KDCPST | Copy a structure of any kind into another one |
| KDLGST | Length a structure of any kind |
| KDLSST | List a structure of any kind |

# 4. Constructive geometry

## *4.1. Programming model*

Within the KDTREE software, the processing functions are designed according to the following meta-algorithm:

    PROCEDURE process (tree, level, depth)

    BEGIN

        IF (*terminal* (tree) OR (level = depth))

        THEN terminal processing

        ELSE DO

            descending pre-processing

            CALL process (*left son* (tree), level +1, depth)

            CALL process (*right son* (tree), level +1, depth)

            ascending post-processing

        END

    END

This meta-algorithm produces a depth-first traversal of the visited data structure and chains up the following phases on the met nodes:

- descending pre-processing of non terminal nodes ;

- terminal processing of a node terminal or not (if the traversal stops at an intermediate depth)

- ascending pre-processing of non terminal nodes.

If the processing is performed at an intermediate depth, then this one is applied according to its upper hull precision.

Nodes are created at the terminal processing step.

Node fission or division is performed during the descending pre-processing step. Union of sub-trees and nodes merge are done during the ascending return step.

The processing of a tree starts by executing the instruction:

        result <— CALL process (tree root, 0, dimension * precision).

Some processes require the parallel examination of two trees, and then the previous procedure becomes:

```
    PROCEDURE compare (tree 1, tree 2, level, depth)
BEGIN
        IF (iso-colored (tree 1, tree 2) OR (level = depth))
        THEN terminal processing
        ELSE DO
                descending pre-processing
                CALL compare (left son (tree 1), left son (tree 2), level +1, depth)
                CALL compare (right son (tree 1), right son (tree 2), level +1, depth)
                ascending post-processing
        END
END
```

In this procedure, the traversal control is no more performed by checking the terminal status of a tree branch, but by testing the iso-coloring of two nodes.

For valued trees, this test is extended to the control of the equality of functional values when to black nodes are compared.

The recursive processing of a tree-like data structure enables to remember the node that has been visited and not to have to explicitly handle an upward link in order to be able to come back to the tree root. So the recursion allows to use a single linkage structure for implementing a tree.

## *4.2. Generation of the tree of a data set*

Several methods can be envisioned for building the tree of a data set.

The most general method proposed by the software proceeds by enriching a preexisting set.

This one is created empty (white terminal node). Then, element by element, it is added to the set the point represented by a coordinate vector:

- for an unvalued binary tree, either by a vector whose coordinates are integer values included between 0 and $2^r-1$, or by a vector whose coordinates are floating point values normalized between 0 and 1 ;

- for a valued tree, the functional value is given as a parameter at the same time as the floating point coordinate vector.

This operation is named addition of a vector to a tree.

When all the points of a set have been introduced in the initially empty set, the tree of this set has been built.

This procedure is enough general for taking in account:

- discrete sets, irregularly sampled and not ordered ;

- overcrowded data sets where several occurrences may appear at a same point in the space.

| *Name* | *Function* |
|--------|------------|
| KDICVC | Generation of an integer coordinate vector |
| KDIRVC | Generation of a real coordinate vector |
| KDAIVT | Addition of an integer vector to a tree |
| KDARVT | Addition of a real vector to a tree |
| KDARVP | Addition of a real vector to a pyramid |

## 4.3. Boolean operations

Boolean operations enable to carry out the set operations belonging to the Boole algebra on data sets modeled by trees:

- assertion of a set ;

- negation of a set ;

- union of two sets ;

- intersection of two sets ;

- exclusion of two sets ;

- difference of two sets.

The assertion of a set enables to provide a copy of the set but at a precision different from its building precision.

The negation of a set is the tree of its complementary set.

The union of two sets is the set of their common parts and their different parts.

The intersection is restricted to the set of the common parts of these sets.

The exclusion is the set of the non common parts of these two sets.

The difference is a copy of the first set where are lacking its common parts with the second set.

When it is dealing with a valued tree:

- its complementary set is an unvalued tree ;

- the maximum of functional values is stored when common parts of the union and the intersection are processed.

| *Name* | *Function* |
|--------|-----------|
| KDASS  | Assertion of a tree |
| KDNOT  | Negation of a tree |
| KDUNIO | Union of two trees |
| KDINTR | Intersection of two trees |
| KDEXCL | Exclusion of two trees |
| KDDIFF | Difference of two trees |

## 4.4. Handling of slices orthogonal to space reference axes

Two functions enable to handle slices orthogonally to the space reference system in which data is represented: there are slice inserting and extracting.

The slice insertion enables to rebuild objects in a space of a given dimension from parallel observations defined in a lower dimension space.

The slice extraction enables to extract such a sub-set at a given set of coordinates without applying a projection on this sub-space that would mix all the entities belonging to the data cloud.

The first situation occurs in tomography when it is tried to sum up parallel planar slices orthogonally to a third axis so as to build the tri-dimensional model of a physical object in order to have at one's disposal a representation for looking at it using different viewpoints and for measuring its size.

The second situation rather occurs in statistical data analysis, in multi-criteria selection or in decision making when data is analyzed in a given plane at a given localization.

When it will be dealt with geometric transforms, it will be told how to move, to turn and to perform sub-space projections without altering any data in the data set.

Let it be $E^k$, the space of dimension k in which are inserted or extracted slices $C^m$ of dimension m, it is necessary to give the coordinates of a point P belonging to this slice for performing such an operation : only the k-m point coordinates belonging to the axes orthogonal to the slice are needed for defining P.

It is the reason why for inserting or extracting a slice, it is needed to specify or to deduce:

- the data set and the dimension of the space in which it will take place;

- the slice and the dimension of the slice that will be inserted or extracted ;

- the point and the number of its significant coordinates (slice codimension).

The point can be described as:

- a list of coordinates (a vector) ;

- a tree (the point tree developed in the space complementary to the slice one).

| *Name* | *Function* |
|--------|------------|
| KDEXSP | Extraction of a slice parallel to the axes |
| KDINSP | Insertion of a slice parallel to the axes |

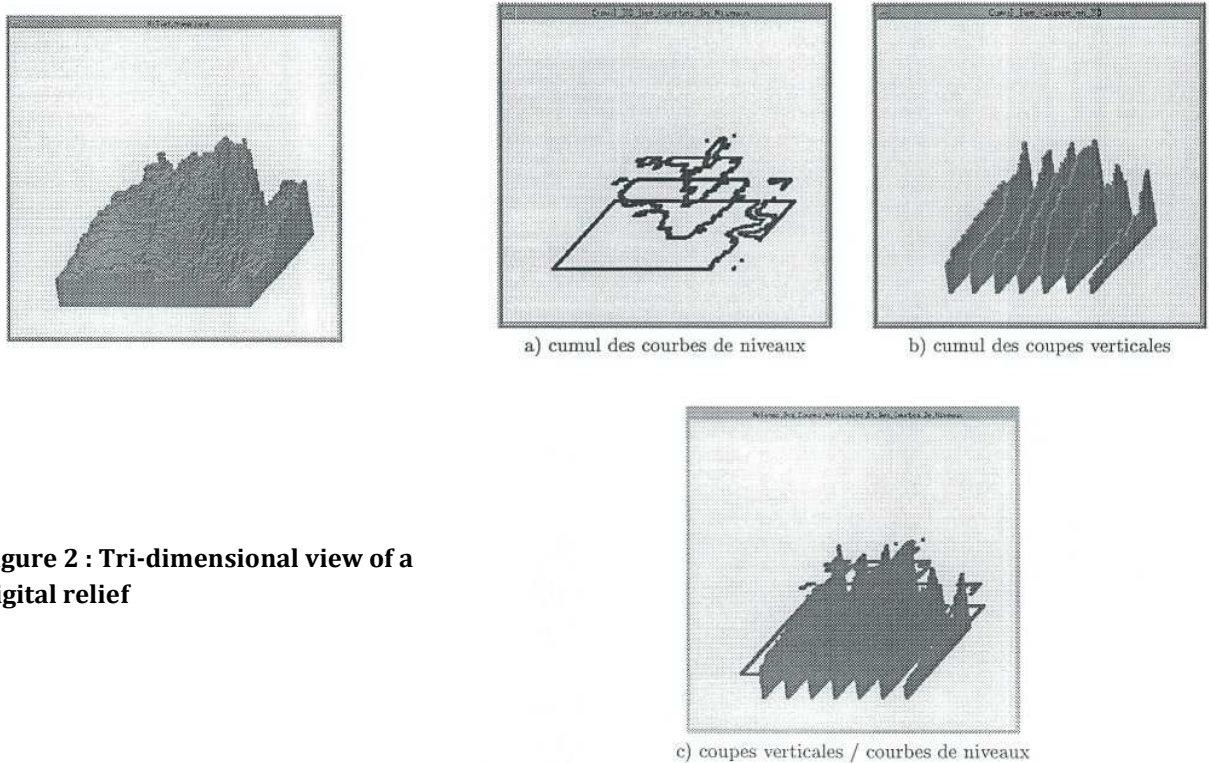

**Figure 2 : Tri-dimensional view of a digital relief**

**Figure 3 : Parallel slices put together using Boolean operations**

## *4.5. Reference hypercube and inductive limit calculation*

Let us come back for a few instants to the initial description of data sets modeled by a $2^k$-tree, in order to complete the information about this representation model.

In the paragraph that has been dedicated to it, it has been mainly paid attention on multidimensional data sets discretized over $N$ or $Q$ et $R$ when they have been normalized.

In a more general way, modeling using $2^k$-trees can be applied on bounded multidimensional data sets for which is existing a partial order relation according to each space axis.

If all these separate order relations can be quantized, the space becomes measurable. In this precise case, it is not necessary to know explicitly the lower and upper bounds of the modeling space in order to be able to build the tree representing a data set.

The multidimensional spaces regularly sampled have got a structuring of Borel algebra that can be extended to the infinite inductively by nesting the bounded spaces in each other like "Russian dolls".

In this structuring, each hypercube fits itself into another one and is holding as many other ones as it can be defined by recursive dividing. Each of these hypercubes that are belonging to such spaces, can be distinguished from its neighbors only by its position or its diameter, while keeping parallel their (hyper-) faces. That is to say, that they are similar each other more or less a translation and a homothety.

By using $N$, $Q$ or $R$ as quantification fields, starting from the unitary hypercube of dimension k it is possible to induce a topological structuring all over the measured space, that enables to make compliant two measurements made over two sets of measures or over multiple sets of measures captured at different moments.

The hypercube including the data set and belonging to this structuring of the measured space will then provide the necessary information that will allow in the future to compare this set of realizations with every new collection of data: it is then named a referring hypercube.

Then for comparing two distinct sets of realizations, it is only needed to reseat each of them in the hypercube of the structuring including the two reference hypercubes linked to the data sets.

So by assigning an implicit structuring of the modeling space compliant with the tree-like organization of data, it will make comparable two data sets captured in distinct conditions.

It will not be achieved in bounding the two data sets using their exscribed hull, that is to say using the parallelotope (hyper-parallelogram) defined by the lower and upper values of a data set according each quantization axis of the space.

It would be then necessary to resample the data sets at each comparison trial, because the hierarchical organization of the structures is directly depending on set constitutive data that they must describe. In other words, according to the interval of lower and upper bounds selected for each space axis, it may provide unlike trees for representing a same data set.

At the opposite, if it can be restricted to use only reference hypercubes each other homothetic by power of 2 according to more or less a translation, all the produced trees will be comparable because they are only describing a given branch of a wider tree in which every data set can take place.

Having at one's disposal the enough information enabling to describe the reference hypercubes linked to two distinct data sets, the hypercube linked to the space area enabling the comparison of two trees is deduced from the two first ones by an inductive limit calculation.

Boolean operations have been modified so as to work in inductive limit: the operand reference frames are given as parameters and the processing result is delivered with the hypercube which is linked to.

| Name | Function |
|---|---|
| KDCTIL | Creation of a tree in inductive limit |
| KDAVIL | Addition of a vector to a tree in inductive limit |
| KDCPIL | Creation of a pyramid in inductive limit |
| KDVPIL | Addition of a vector to a pyramid in inductive |
| KDUNIL | Union of two trees in inductive limit |
| KDINIL | Intersection of two trees in inductive limit |
| KDEXIL | Exclusion of two trees in inductive limit |
| KDDFIL | Difference of two trees in inductive limit |

# 5. Management of particular structures

## 5.1. Vectors and generation of primitive shapes

It has been previously sawn that the simple linkage lists enable to specify data using the form of vectors before performing the building of trees by enrichment.

These same vectors can be used for other usages. In planar spaces, they can be gathered under the form of a vector list – that is to say a list of lists – in order to specify:

- a broken line for which each vertex is described by a coordinate vector ;
- a closed polygonal line describing a polygon.

These two representations enable to specify shapes of line or surface nature in order to build the corresponding trees.

In a space of any dimension, vectors are used for pointing out the beginning and the ending of a line segment: the tree representing this segment can be built on the basis of this single information. Combined with the computation of the epigraph and the hypograph, these trees enable to build randomly located slices in the observed space.

Two primitive shapes can be directly built in a space of two or three dimensions: a sphere or a cone (a disk or a disk sector).

All these operations are relying on an algorithm of testing the intersection between a straight line and a parallelotope.

Knowing a point $A$ belonging to the straight line and its direction vector $u$, a straight line can be parametrically designated as the set of points :

$$M(t) = A + t \cdot u, \qquad \text{where } t \in [-\infty, +\infty]$$

For each parallelotope stemming from the regular decomposition of the modeling space, it is easy to define the normal to each face and to find out the values for which the straight line goes across the hyper-planes of its hyper-faces.

Concerning a line segment for which the parameter value spectrum is finite, it is only needed to check if the found values are belonging to the value interval of the parameter.

The general algorithm enabling to check if a line segment intersects a tree-like modeled set, is then the following one:

> PROCEDURE intersection of a line segment with a tree
>
> (tree, lower faces, upper faces, origin, end, level, depth, dimension)

```
BEGIN
    intersection <- intersection test(origin, end, lower faces, upper faces, dimension)
    IF (intersection) THEN DO
        /* the line segment is included in the $2^k$-ant */
        IF ((level <> depth) AND (NOT terminal (tree)))
        THEN DO
            /* dividing the $2^k$-ant by halves */
            CALL block division (lower faces, upper faces,
                left lower faces, left upper faces,
                right lower faces, right upper faces, level, dimension)
            /* tree descending */
            left intersection <- CALL intersection of a line segment with a tree
                (left son (tree), left lower faces, left upper faces,
                origin, end, level + 1, depth, dimension)
            right intersection <- CALL intersection of a line segment with a tree
                (right son (tree), right lower faces, right upper faces,
                origin, end, level + 1, depth, dimension)
            RETURN (left intersection OR right intersection)
        END
        ELSE DO
            /* terminal node processing */
            IF (NOT white (tree))
            THEN RETURN (true)
            ELSE RETURN (false)
        END
    END
END
```

The procedure is processed by using the following calling sequence :

> lower faces <- minimal coordinate vector of the unitary hypercube
>
> upper faces <- maximal coordinate vector of the unitary hypercube
>
> intersection <- CALL intersection of a line segment with a tree (tree, lower faces, upper faces, vector origin, vector end, 0, computation precision * dimension, space dimension)

This algorithm shows how to build a primitive geometrical shape with the KDTREE software.

| *Name* | *Function* |
|---|---|
| KDITST | Intersection test of a line segment with a tree |
| KDITSP | Intersection test of a line segment with a pyramid |
| KDEXTS | Extraction of a slice along a line segment |
| KDBRLI | Generation of the tree of a broken line |
| KDPOLY | Generation of the tree of a polygon |
| KDSPBT | Building of the tree of a sphere |
| KDCOBT | Building of the tree of a cone |

## *5.2. Geometrical transformation matrices*

Transformation matrices enable to define the geometrical transformations that can be applied on a data set represented by a tree.

They are built as lists of line vectors and they are expressed using homogenous coordinates:

- on one hand, in order to take in account transformations in projective geometry as well as in affine geometry ;
- on the other one, in order to specify generalized transformations built from elementary transformations.

So it will be possible to build by matrix concatenation, a direct transformation matrix composed of the following transformations:

- homothety ;

- translation ;

- rotation ;

- and perspective.

As the matrix multiplication is not commutative, it can be got the transformation inverse matrix by concatenating, in the inverse occurring order when the direct transformation matrix has been built, the inverse matrices of the elementary transformations.

The elementary transformations can be expressed for:

- a homothety by specifying the homothety rates according to each space axis ;

- a translation by specifying the moves according to every space dimensions;

- a rotation by specifying the angle and the plane, defined by its two axes, according to the operation to be processed ;

- a perspective by specifying the perspective center.

Except for the rotation, all these transformations are described by a real coordinate vector

Each of these elementary transformations has got for inverse:

- a homothety, the matrix of its inverse values ;

- a translation, the matrix of its contrary values;

- a rotation, the matrix of contrary angle ;

- a perspective, the matrix of its contrary values.

So for each elementary transformation, it can be directly computed its inverse transformation.

The inverse of a generalized transformation will be the multiplication of the inverse matrices of the elementary transformations whose are composing it, applied in the inverse occurring order of the direct transformation.

| *Name* | *Function* |
|---|---|
| KDMTAN | Generation of a homothety matrix |
| KDMTTR | Generation of a translation matrix |
| KDMTPR | Generation of a perspective matrix |
| KDMTRT | Generation of a rotation matrix |

| KDMTOP | Computation of the contrary of a matrix |
| KDMTIV | Computation of the inverse of a matrix |
| KDMTTP | Computation of a transposed matrix |
| KDCMTH | Concatenation of two matrices |

## *5.3. Parallelotopes and polytopes*

Applied on a unitary hypercube, these transformation matrices enable to generate:

- parallelotopes in affine geometry ;

- polytopes in projective geometry.

Within the KDTREE software, parallelotopes and polytopes being the homographic transformed images of the unitary hypercube, are based on its description and have at their disposal two complementary modes of representation:

- a direct representation where the polytope is described by its $2^k$ vertices (mixed up or not) ;

- a dual representation where the polytope is designated by its 2k hyper-faces (parallel for affine transformed images).

So they can be represented by:

- the list of coordinate vectors of its vertices ;

- the list of its lower faces and the list of its upper faces.

In this last representation, the polytope can be interpreted as the intersection of the right half-spaces of its lower faces with the left half-spaces of its upper faces.

The geometrical transformations that has been presented can directly apply on the two representations of a same polytope by following the following rule: if the direct transformation matrix is applied on the vertices coordinates, the inverse transformation matrix must be applied on the coefficients of the equations of the hyper-planes defining the polytope faces for getting the same result after transformation (dual space).

These transforming rules being known, it is now possible to show how can be performed the building of the tree of a polytope. This one is relying on the test of the intersection between a $2^k$-ant stemming from the regular dividing of the modeling space and the polytope whose tree must be built.

A polytope is a convex set owning the following properties:

- every point internal to the polytope is a positive linear combination of its vertices (barycentric coordinates) ;

- every polytope is entirely located on a same side of each hyper-plane defining its faces.

So, for two polytopes to be compared:

- if all the vertices of a polytope are located in one of the external half-spaces of the other polytope, it will be the same for all convex combination of these points and the two polytopes will not intersect ;

- if all the vertices of a polytope are located in all the half-spaces internal to the other one, it will be then included in this one ;

- finally, if the vertices of a polytope are distributed on both sides of one of its faces, there will be only an intersection.

| *Name* | *Function* |
| --- | --- |
| KDPESP | Generation of the unitary hypercube defined by its vertices |
| KDTRHP | Homographic transformation of a polytope defined by its vertices |
| KDLOHP | Generation of the lower hyper-planes of the unitary hypercube |
| KDUPHP | Generation of the upper hyper-planes of the unitary hypercube |
| KDTRHH | Homographic transformation of a hyper-plane list |
| KDPOLT | Building of the tree of a polytope |

# 6. Geometric transformations

## *6.1. Homographic transformation of a $2^k$-tree*

The principles followed for building the tree of a polytope can be generalized so as to lead to the computation of the homographic transformation (homothety, translation, rotation, perspective) of a $2^k$-tree.

This transformation is based on the inverse image of the unitary hypercube including the modeled tree. In fact, for getting the direct image of a tree using such a transformation, it easier to decompose the original tree in the inverse image of the reference frame than to try and compute the direct image of each $2^k$-ant belonging to the initial tree.

This operation is enabled due to the regularity of the decomposition: vertices and faces of the homographic image of the unitary hypercube remain in harmonic range after decomposition and insure the existence of the bijection between the homographic images of the hypercube decomposed blocks and the recursively decomposed blocks of the homographic image of the same hypercube.

So, the transformation applied to a $2^k$-tree can be defined by specifying:

- either the list of the inverse images of the unitary hypercube vertices ;
- or the list of the direct images of the lower and upper hyper-planes of the unitary hypercube.

If a perspective is introduced in the generalized transform matrix, it can be performed a space closing bounded by front and back planes around the perspective center, by moving lower and upper faces of the unitary hypercube according to the first space dimension.

The perspective center materializes the viewpoint of an observer present in the modeling space. The closing then shortens its observation field.

It can be delivered the lists of the vertices or faces direct or inverse images of the unitary hypercube by building the direct and inverse generalized matrices and by applying them straightly to the describing lists of the unitary hypercube created in its own reference system.

Two complementary transformations enable to perform simplified transforms without going across all these steps:

- symmetry (hyper-planar) ;
- and translation.

Complementary to the transformations belonging to the positive linear group (homothety, translation, rotation), symmetries enable to provide the full set of linear transformations in the modeling space.

| *Name* | *Function* |
|--------|-----------|
| KDTHOM | Homographic transformation of a tree |
| KDSYMT | Symmetric image of a tree |
| KDTRAN | Translated image of a tree |

## 6.2. Projective transformations

The perspective is only implying a deformation on the objects lying in the initial space.

The perspective views are including a complementary transformation: a projection that enables to gather the information lying in the initial space into a hyper-plane normal to a perspective axis.

Before performing this projection, it must be done a removal of hidden parts along this axis, in order to select the information that must be preserved by the transformation.

To reach it, the objects lying in the modeling space must be redirected so as to be aligned along the observer viewing axis. According to its localization in the space, it is performed the perspective whose center is this position. The viewing axis is aligned on one of the space dimensions : the hidden part removal and the projection are applied in taking in account of this one. It will then be got the planar view that an observer captures about space from the point where he is staying and according to the direction in which he is looking at.

If the observed scene is lightened by one or several light sources, it can be rendered the resulting shading by doing:

- for one omnidirectional source located at a finite distance, a move in order to be located in this place, a perspective and a hidden part removal for extracting les shaded parts ;

- for a source located at the infinite, the same operations by moving towards the lighting direction, but without applying any perspective (orthographic transformation).

A shaded view is produced using the double-visibility method by confronting the views obtained from the two viewpoints: the observer one and the lighting source one.

So, let us assume that it is willing to know what could be seen by an observer placed in $P_o$ space position and looking in the direction $\vec{o}$ at a scene lighted by a luminous source located in $P_e$ and irradiating in direction $\vec{e}$, the following steps will enable to know it :

- computation of the direct and inverse matrices in order to move oneself in the lighting reference frame ;

- computation of the inverse image of the unitary cube according to this transformation and application to the modeling tree of the scene in its initial reference frame;

- hidden part removal for getting the tree of the lighted parts ;

- computation of the direct and inverse matrices enabling to move from the lighting source to the observation point ;

- visible space closing (using front and back planes) ;

- computation of the inverse image of the unitary cube and application to the tree of the lighted parts ;

- not visible part removal ;

- projection in the viewing plane of the resulting tree.

The projection operation can be performed in a space of any dimension and provides a hyper-plane.

This one can be repeated several times and provides other usages in different fields. For instance in statistical data analysis, it can be found the principal axes of a data cloud (cf. attribute calculation and Eigen spaces afterwards), then the cloud is turned according to these axes and the necessary number of projections is performed in order to bring back the cloud in the sub-space of its principal axes. Restricted to two or three dimensions, it can be then envisioned to display the cloud on a flat screen while knowing that the most significant part of the information will appear in views restricted to its principal sub-space.

| *Name* | *Function* |
|---|---|
| KDHIPR | Hidden part removal |
| KDPRVA | Projection according to a viewing axis |

## *6.3. Integral transformations*

### 6.3.1. Epigraph and hypograph

When the tree of a set is modeling a hypersurface in a space of k dimensions, attention may be focused on what is lying under or above this surface.

The hypersurface represents a functional $f$ that is dividing the space into two sub-spaces:

- the hypograph of the functional linked to the hypersurface :

$$\{(x,r) \in E^k \: / \: r \leq f(x), x \in [0,1]^{k-1}\}$$

- the epigraph of the same functional :

$$\{(x,r) \in E^k \: / \: r \geq f(x), x \in [0,1]^{k-1}\}$$

For instance, a digital terrain model is usually represented in digital geography by a matrix of altimetry regularly sampled in a geographical coordinate reference frame.

It can be provided a digital relief by computing the hypograph of the altimetry matrix, on which it is then easy to perform the following operations using geometrical reasoning :

- intersection of mobiles objects with the terrain;
- slices extraction along some given profiles ;
- looking for optimal paths.

Computing epigraph and hypograph can be also seen as the inverse operations of the hidden part removal.

| *Name* | *Function* |
| --- | --- |
| KDHYPG | Compute the hypograph of a tree |
| KDEPIG | Compute the epigraph of a tree |

### 6.3.2. Convex hull

It has been seen that testing the intersection of two polytopes is greatly eased if it is relying on convexity properties. It can be computed the convex hull of any set, which is the smaller convex set including it.

This operator may be very useful, when only a small number of points is available for describing an entire set:

- for describing simple shapes from a minimum number of information pieces ;
- for building back the shape of a set from a discrete set of points randomly sampled while using a minimal number of distribution hypotheses about the available data.

This operator may show a significant response time when it is required to get a result of a good precision.

The implementation method is the following one : based on the tree-like structure of the representation, it is performed a depth-first traversal of the initial data tree :

- terminal nodes are convex by definition ;

- on the traversal ascending return, it is computed the convex hull of two convex sub-trees represented by the son nodes of the current node.

The computation of the convex hull of two convexes sub-trees consists in three steps:

- relatively to the separation hyper-plane associated to the current node, it is computed the lower and upper hulls of the future convex set (minimum and maximum according to the separation) ;

- compute a covering of sub-hulls by convex segments ;

- sample the segments of the convex covering.

| *Name* | *Function* |
|---|---|
| KDCVXH | Compute the convex hull of a tree |

## 6.4. Inter-visibility graphs

One of the applications of geometrical reasoning is the inter-visibility test: it is consisting in determining if an observer can see an object lying on a digital terrain without that any other object interposes itself.

The computation primitives of a sphere or a cone enable to model an observation unit omni-directional or directional but including a restricted range.

Their intersection with the terrain enables to know the areas covered by these units, but without insuring that an obstacle may or does not damage their observation capabilities.

By testing the intersection of segments stemmed from each observation center with every point inside the covered area, it can be known the points visible by the observer.

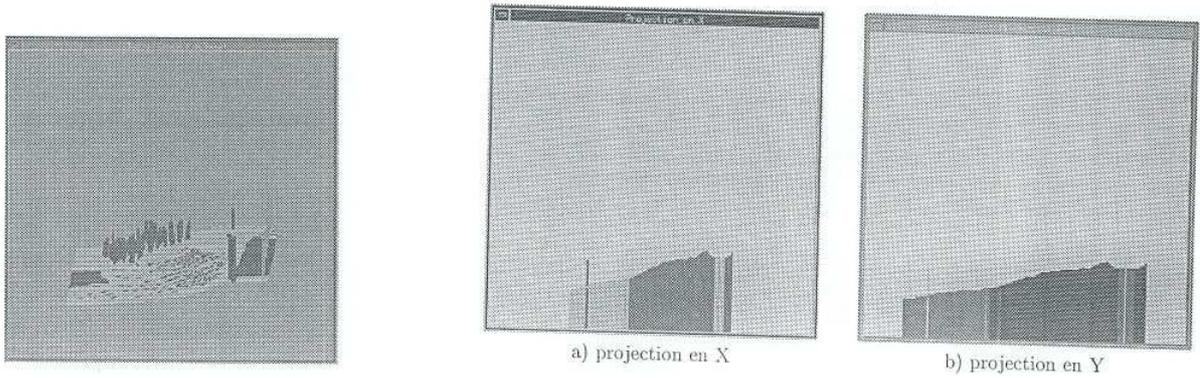

**Figure 4 : Projections extracted from a relief enriched with planimetric information**

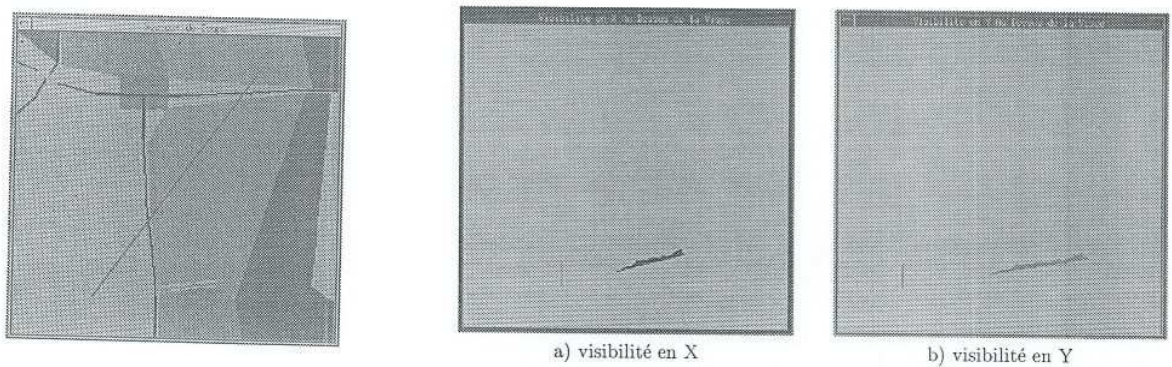

**Figure 5 : Slicing vector and visibility above a sight line**

This procedure can be iterated:

- on the set of points belonging to a path ;
- on the set of points lying inside a bounded area.

These operators enable so:

- to determine the coverage of a multiple observation system ;
- to determine the coverage of a mobile communication system along a path or inside an area.

Linked to an optimization procedure, it is then possible to compute:

- the best localization for a surveillance network relatively to a terrain ;

- the optimal frequency allocation of a communication network enabling to better cope with the relief irregularities in order to reuse emitting frequencies.

| *Name* | *Function* |
|---|---|
| KDPTTM | Compute the propagation area provided by an emitter |
| KDCRTM | Compute the propagation area provided by emitters lying in a maneuvering area |

# 7. Topological analysis

## 7.1. Neighborhoods

In a meshed space, it can be defined a notion of neighborhood by using the distances induced by the mesh structure. In the case of regularly sampled spaces, they are metric distances. The most common metric distances are :

- $d_\infty$, the distance of the maximum of absolute values of the coordinate differences of two points ;

- $d_1$, the one of the sum of absolute values of the same differences ;

- $d_2$, the Euclidean distance, represented by the square root of the sum of the square values of the same differences.

Two points will be adjacent or neighbors if they are distant from one measure unity according to the chosen distance. The nearest neighbors of a point will make a neighborhood of this point and the smallest one for the topology induced by this distance.

On meshes regularly sampled according to each dimension of the space (square, cubic or hyper-cubic meshes), it can be easily found $d_1$- or $d_\infty$- neighborhoods. On the other hand, the neighborhood relations are more uneasy to highlight using the Euclidean distance in such spaces.

The number of neighbors, which can be found for a given point, is varying according the space dimension and the used distance:

- in two dimensions, a point may have 4 $d_1$- neighbors and 8 $d_\infty$- neighbors;

- in three dimensions, 6 $d_1$- neighbors and 26 $d_\infty$- neighbors;

- in k dimensions, 2k $d_1$- neighbors and $3^k-1$ $d_\infty$- neighbors.

## 7.2. Looking for adjacencies

Let us now look for and retrieve the adjacency relations existing inside a set of points belonging to a space of any dimension. Modeled using the structure of $2^k$-tree, these points are arranged in the tree according to their positions relatively to the separation planes (hyper-planes) that have enabled to regularly divide the space in which is standing the set to be analyzed

Two points will be $d_1$-adjacent if it exists a separation plane for which they are symmetrical, that is meaning locally located on left and right sides of this same plane.

It is the case in particular, when:

- the two points are sibling tree nodes, the symmetry plane is then the separation plane that has enabled to divide the paternal node into two nodes ;

- the two points are sharing a same ancestor, the symmetry plane is the one that has enabled to divide this ancestor in two half-spaces in which each of them are lying.

In this last case, the traversals that enable to reach these two neighbors from this common ancestor are made from symmetrical moves, consequently opposed, along the space dimension that supports the symmetry plane and parallel, consequently identical, along the other space dimensions so as to reach points with identical coordinates according these axes. It can be still spoken about "mirror effect" for it

So, from an algorithmic viewpoint, when it is looked for $d_1$-adjacencies in a space of any dimension, the tree must be first visited for retrieving non terminal nodes bearing all the possible symmetry planes. For each found node, it is stored in an indicator vector, the dimension on which symmetry plane leans (orthogonally), by marking each vector coordinate according to the following way:

- N : neutral, for a dimension parallel to the hyper-plane;

- S : symmetrical, for the dimension supporting the hyper-plane.

Then it can be listed all the $d_1$-adjacencies stemmed from a non terminal node by applying the following algorithm:

PROCEDURE search for $d_1$-adjacencies (node1, node2, (..., ..., ...))

BEGIN

    IF (*terminal*(node1) AND *terminal*(node2))

    THEN store adjacency (node1, node2)

    ELSE DO

        IF (*first list element* (..., ..., ...) = S)

        THEN search for $d_1$ - adjacencies (*right son (*node1), *left son* (node2),

            *rotation* (..., ..., ...))

        ELSE DO

            search for $d_1$ - adjacencies (*left son* (node1), *left son* (node2),

            *rotation* (..., ..., ...))

            search for $d_1$ - adjacencies (*right son* (node1), *right son* (node2),

```
            rotation (…, …, …))
        END
    END
END
```

performed with the help of the instantiation :

APPEL search for $d_1$-adjacencies (*left son* (tree root)), *right son* (tree root), (S, …, N))

The $d_\infty$-adjacencies represent a generalization of the $d_1$-adjacencies: they allow taking in account symmetrical effects not only around a given plane, but also around a combination of orthogonal planes.

So in a space of k dimensions, the $d_\infty$-adjacencies are provided by the symmetries:

- around each of the k possible axes ($C_k^1$ combinations) ;

- around 2 among k axes ($C_k^2$ combinations) ;

- …

- around k-2 among k axes (planar symmetries) ;

- around k-1 among k axes (axial symmetries) ;

- around k axes of the space (point symmetries).

If there are $C_k^j$ combinations of j symmetry planes in a space of dimension k, at maximum resolution, it will be found no more than $2^j$ neighbors to a given point.

By adding the contributions of all the possible combinations of symmetry planes, the number of neighbors in a $d_\infty$-neighborhood can be enumerated. This value exponentially grows according to the space dimension and leads to restrict oneself to perform operations in $d_1$-adjacency for high dimension problems, if it is willing to keep sustainable response times.

Looking for $d_\infty$-adjacencies looks like doing it for $d_1$-adjacencies, but by enabling the progressive introduction of new symmetry planes. During this introduction, it will be registered the kind of pairing made between left and right sons of symmetrical nodes:

- S, for couples made from right and left sons (symmetrical) ;

- A, for couples made from left and right sons (anti- symmetrical).

The previous algorithm then becomes:

```
PROCEDURE search for d∞-adjacencies (node1, node2, (..., ..., ...))

BEGIN

    IF (terminal (node1) AND terminal (node2))

    THEN store adjacency (node1, node2)

    ELSE DO

        IF (first list element (..., ..., ...) = N)
        THEN DO

            search for d∞ - adjacencies (left son (node1), left son (node2),

            rotation (..., ..., ...))

            search for d∞ - adjacencies (right son (node1), right son (node2),

            rotation (..., ..., ...))

            store ((..., ..., ...))

            search for d∞ - adjacencies (right son (node1), left son (node2),

            rotation (A, ..., ...))

            search for d∞ - adjacencies (left son (node1), right son (node2),

            rotation (S, ..., ...))

            restore ((..., ..., ...))
        END

        IF (first list element (..., ..., ...) = S)

        THEN search for d∞ - adjacencies (right son (node1), left son (node2),

            (..., ..., ....))

        IF (first list element (..., ..., ...) = A)

        THEN search for d∞ - adjacencies (left son (node1), right son (node2),

            (..., ..., ....))

    END

END
```

| *Name* | *Function* |
|--------|-----------|
| KD1ANR | Search for $d_1$-adjacencies in a tree |
| KD0ANR | Search for $d_\infty$- adjacencies in a tree |

## *7.3. Local topological analysis*

### 7.3.1. Boundary of a set

In local topological analysis, it can be locally acted according to the relative position of a point in comparison with a set. It can be distinguished three different situations:

- the point is inside the set when itself and all its neighbors are belonging to this set;

- the point is on the boundary of the set when itself and only one part of its neighbors are belonging to this set;

- the point is outside the set when it is not belonging to the set.

The notion of outside can be refined by distinguishing points of the outside from the points belonging to the exo-boundary of the set: they are outside points that share some neighbors with the concerned set. It is also the boundary of the complementary of the initial set which is also named space background.

The nature of boundaries is conditioned by the chosen neighborhood system which is meaning here by the distance used to define neighborhoods.

A space closure operation enables to take or not to take in account the space borders when it must be performed a local topological analysis of a set.

One must remember that in regular meshes:

- a $d_1$-boundary is $d_\infty$-connected ;

- a $d_\infty$- boundary is $d_1$- connected.

The $d_1$- and $d_\infty$-topologies are playing roles symmetrical from one to the other one in local topological analysis.

Having to its disposal the boundary of a set, a filling operator enables to come back to its initial shape by regenerating the set of its interior points

| *Name* | *Function* |
|---|---|
| KDSCLO | Space closure |
| KD1BND | Compute the $d_1$-boundary of a tree |
| KD0BND | Compute the $d_\infty$- boundary of a tree |
| KDFILL | Regenerate the interior of a set |

### 7.3.2. Homotopic transformations

Knowing how to identify points belonging to the boundary or to the exo-boundary of a set, it is now possible to envision the use of two continuous topological transformations:

- the erosion of a set, operation that gives back the boundary of this set to the space background;

- the dilation of a set, operation that gives the boundary of the space background with this set to this set.

During the first operation the set is losing its boundary, during the second one it gets back its exo-boundary.

When these two operations are alternatively used, they become operators that enable to stabilize the shape of these sets by increasing their compactness:

- the opening of a set is the dilated release of its eroded version;

- the closing of a set is the eroded release of its dilated version.

The initial set is then included between its opened set and its closed set that are themselves included between its eroded set and its dilated set.

Opening and closing slightly modify the volume or the hypervolume of the initial set.

Moreover, although it has been seen that it is not possible to perform neighborhood analysis with the help of the Euclidean distance over regularly sampled data, similar results can be reached with continuous transformations when it is alternatively used operators according to $d_1$ and $d_\infty$.

Some authors have sent out that the $d_1$ and $d_\infty$-topologies may bound the $d_2$-topology (or more exactly that the $d_p$-topologies are bounding each others): it can then approach homotopic transformations according to $d_2$ by alternating the use of $d_1$- and $d_\infty$-transforms.

| Name | Function |
|------|----------|
| KD1ERO | $d_1$-erosion of a tree |
| KD0ERO | $d_\infty$-erosion of a tree |
| KD1DIL | $d_1$-dilation of a tree |
| KD0DIL | $d_\infty$-dilation of a tree |

### 7.3.3. Median transformations

They are still homotopic transformations, but where the neighborhood analysis of a point is performed with a deeper manner.

The first transformation is the median filtering of a set. This operation consists in attributing to each point the status which has a major position among all its neighbors, without taking in account its own one, except in case of equality. It is then simultaneously performed on points belonging to the boundary as well as the exo-boundary of a set:

- a boundary point, whose neighbors are mainly belonging to the space background, is got back to the background ;

- an exo-boundary point, whose neighbors are mainly belonging to the processed set, is given back to this one.

The effect of this operation is similar to the opening or to the closing of a set and as those ones the result is bounded by the eroded and the dilated sets of the initial set.

Likewise, the compactness of the transformed set is better than the initial set one. This one can be evaluated by comparing the volume set relatively to the surface of its boundary. More precisely, a measure of the compactness of a set is got by calculating the ratio of the k-th root of its hypervolume with the (k-1)-th root of the hypersurface of its boundary: more the ratio is close to the unit more the set is compact.

From the median filtering, it can be then paid attention to median sets. The median set of a set is the locus of the set points that are located at equal distance to the boundary of this set.

The computation of a median set relies on the infinite iterated process of a new homotopic transformation: the thinning is a kind of erosion which is preserving the middle points of a set. It works as an erosion but by restricting itself in maintaining points of a given connectivity degree. When the iterated process of thinning does not deliver any more observable modifications, the median set is achieved.

In order to preserve the structures of unitary thickness towards the process is converging, thinning is performed in two steps:

- a first step where are eroded the points left connected with the outside (lower hull of the boundary) ;

- a second one where are eroded the points right connected with the outside (upper hull of the boundary).

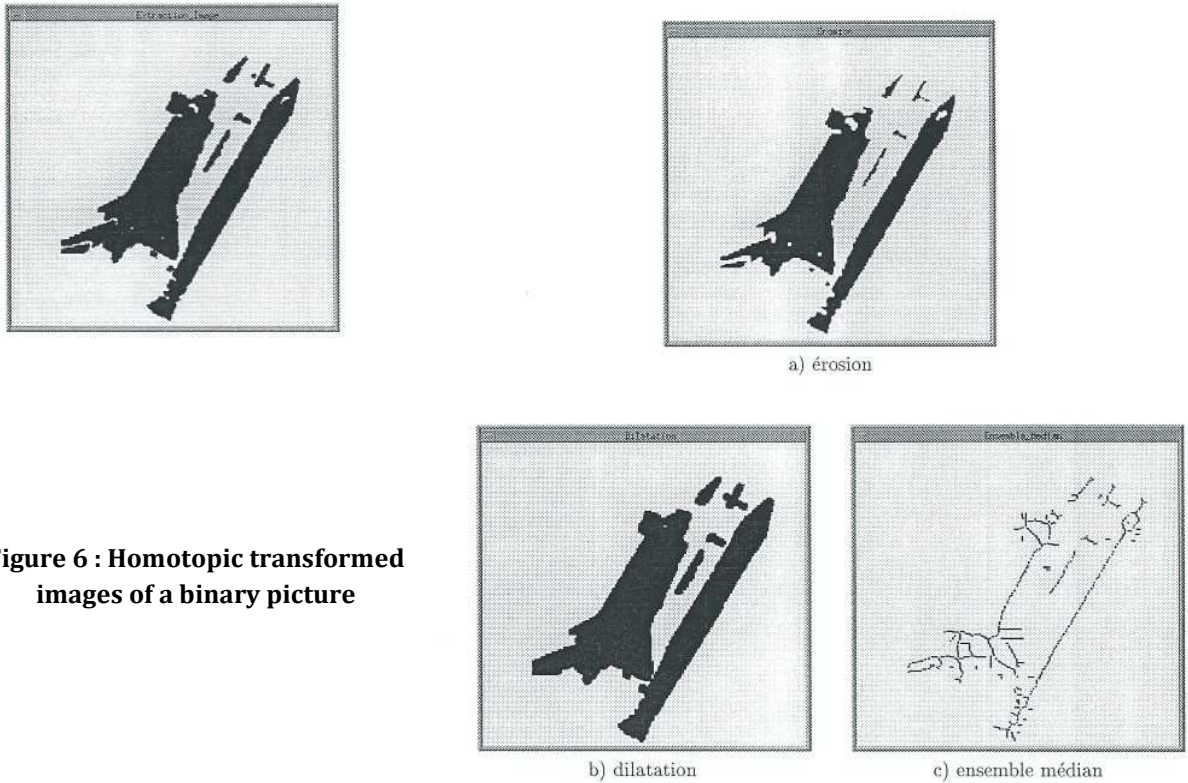

**Figure 6 : Homotopic transformed images of a binary picture**

The median sets are calculated according to the two $d_1$- and $d_\infty$-topologies, but one must be aware that in $d_\infty$-topology the median set of a connected set does not necessarily remain a connected set and that in $d_1$-topology only the $d_\infty$-connectivity of the result is insured.

Finally, by locally analyzing the connectivity degree of each point, it can be estimated the intrinsic dimension of a set. The intrinsic dimension of a set is its connectivity order, which is meaning the maximum connectivity degree of each of its points. It enables to know if the set is a point, a curve, a surface, ..., a hypersurface of a given dimension in a space of higher dimension.

| *Name* | *Function* |
|---|---|
| KD1MDF | $d_1$-median filtering of a tree |
| KD0MDF | $d_\infty$- median filtering of a tree |
| KDTHIN | Thinning of a tree |
| KDMEDS | Compute the median set of a tree |
| KDIDIM | Compute the intrinsic dimension of a tree |

## *7.4. Regional topological analysis*

### 7.4.1. Segmentation

The aim of the segmentation is to succeed in decomposing a data set into homogenous components. More precisely, the segmentation of a set is a partition of this set into sub-sets verifying a given property, named predicate:

– the predicate is checked over each partition sub-set ;

– the predicate is not checked on the union of two sub-sets ;

– the union of all the sub-sets reconstructs the initial set.

The most often used predicate is the iso-coloring predicate:

– for data of volume nature, it is corresponding to belong or not to the initial set ;

– for data of surface nature, where each space point is colored with a functional value, the predicate enables to decompose data sets where functional values are varying (pieces identification of a simple function).

This last specificity concerns multi-valued sets, for instance data sets that would be labeled by a teacher in supervised classification.

It is based on the neighborhood relations that points are sharing between them: a set partition will be provided by looking for the set connected components satisfying to the chosen predicate.

The connected components are the equivalence classes for the connectivity relation induced by the kind of selected neighborhood: they are sub-sets for which it can ever find a string of connected points enabling to link every couple of points of this same sub-set. So according to the used distance, the result of a segmentation shall vary.

A segmentation according to $d_\infty$ enables to keep into a single component manifolds whose intrinsic dimension is lower than the modeling space one, although it is not the case for a segmentation according to $d_1$ that will scatter components.

On the other hand, in order to preserve the space isotropy property, if a set is segmented according to $d_1$, its complement (space background) should be processed according to $d_\infty$ and vice-versa.

### 7.4.2. Connected components labeling

For labeling connected components, two methods are proposed:

- the first one is relying on the bucket algorithm which is implementing a simple procedure for listing connected components;

- the other one is relying on an regional hierarchical growing algorithm.

The bucket algorithm provides in practice better response times than region growing ; on the other hand, it is necessary to explicitly handle adjacencies as links between nodes in a given tree and increases greatly the size of temporary data.

The regional hierarchical growing follows the same principle as the convex hull computation:

- at beginning, all terminal tree nodes are marked with a distinct label, as they were isolated components ;

- on the tree traversal back return, it is looked at all the symmetries provided by the separation plane associated to the current non terminal node and connected components are re-labeled with one of the two labels marking the couple of adjacent nodes.

The algorithm follows the underneath script:

PROCEDURE labeling a tree (tree, symmetry axis, level, depth)

BEGIN

    IF ((NOT *terminal* (tree)) AND (level <> depth))

    THEN DO

        /* labeling left and right sub-trees */

        CALL labeling a tree (*left son* (tree), symmetry axis, level + 1)

        CALL labeling a tree (*right son* (tree), symmetry axis, level + 1)

        /* determination of minimal labels of the two sub-trees */

CALL search for adjacencies (*left son* (tree), *right son* (tree), symmetry axis)

/* spreading minimal labels in the sub-trees */

CALL labeling a tree (*left son* (tree), symmetry axis, level + 1)

CALL labeling a tree (*right son* (tree), symmetry axis, level + 1)

END

ELSE DO

/* labeling an isolated component */

IF (NOT *white* (tree))

THEN give a new label to the node

END

END

Step by step, the entire tree is so labeled according to the chosen distance for performing the adjacency search by invoking the following instruction:

CALL labeling a tree (tree, (S, …, …, …), 0, dimension * precision)

For minimizing the variation range of label values, labels are counted from the beginning and the minimum of label numbers is kept when a component merge occurs.

When the tree connected components have been labeled using the bucket method, a complementary operator enables to build the forest of segment trees: it is a list including the series of the different components modeled in the initial description space of the set to be labeled.

Then, according to the followed labeling method, components can be extracted until their exhausting:

- for the bucket method, the first component of the forest ;

- for the growing method, the component of lower label.

| Name | Function |
|---|---|
| KDLBCC | Labeling connected components in a tree |
| KDSGTF | Building the forest of segment trees |
| KDEXSG | Extraction of a segment tree from the forest |
| KD1LAB | Search and labeling of $d_1$-connected components |
| KD0LAB | Search and labeling of $d_\infty$- connected components |
| KDLLCC | Extraction of the lower label connected component |

### 7.4.3. Classification

In order to illustrate these regional analysis techniques, it is proposed to have a look to the example of the performance of a thematic classification applied on a multispectral image. It is a central question in earth resource observation: in such a context frame, it is looking after areas of vegetal coverage from shots of middle resolution.

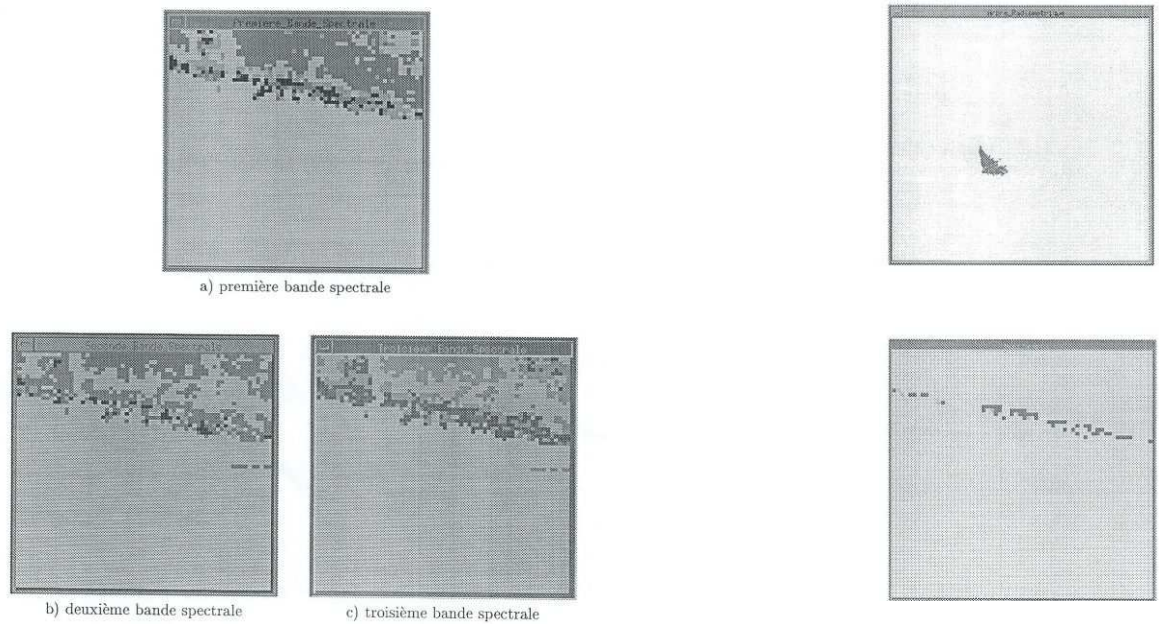

**Figure 7 : Multispectral image**

**Figure 8 : Radiometric tree and pseudo-colored image**

In the space of luminescence, water, cultivations, meadows, forests and urban areas are providing distinct response vectors, independent from their position in the image but depending on the date of data capture (seasonal variations).

For retrieving these different themes, their response vectors are gathered from every point of the image planar support.

The vector dimension is equal to the number of spectral bands according to which data have been captured. For a standard color image, they are three, corresponding to the three primary colors that are composing it.

Response vectors are then analyzed in their radiometric space and the uniform wide areas of coverage create accumulation points in this new space.

If the capture sensors are enough sensitive, these points are isolated from each others in the response space; it may be then performed a search for localizing the connected components for identifying these accumulation points and labeling them: it has been then performed a thematic classification of the multispectral image. It is then only needed to read once more the original image and to replace in each image point the response vector by its label: it will be got a image labeled according to the ground coverage classification that has been built. The areas with a homogenous coverage can be spatially localized and the coverage nature has been identified with the help of the average response vector linked ot the connected component belonging to the multispectral space.

| *Name* | *Function* |
|---|---|
| KDRADT | Building a radiometric tree |
| KD1CLA | Classification in $d_1$-connected components |
| KD0CLA | Classification in $d_\infty$- connected components |
| KDCVMT | Conversion of a multispectral image into a thematic image |

# 8. Attribute calculus

## *8.1. Generalized moments and Eigen representation*

The attribute calculus enables to associate a measurement vector with a set, a shape, a component.

This measure vector, when it consists in an quite representative expression of the modeled shape, can be used as a storing key or an identification label in order to sort and classify the shapes handled in an application.

The developed attribute calculus relies on the generalized moments. These ones offer some particular advantages:

- they are integral measures, that are applying on regions and not on their boundaries, and which are then less sensitive to digitalization noise ;

- they are statistically mutually independent ;

- they are also geometrically independent from a sub-set of affine transforms, the similarities.

In a space with k dimensions, the generalized moments are the following measures:

$$M_{(object)}\left(X_1^{n_1}, X_2^{n_2}, ..., X_k^{n_k}\right) = \int_{X_1} \int_{X_2} ... \int_{X_k (X_1, X_2, ... X_k) \in object} X_1^{n_1} X_2^{n_2} ... X_k^{n_k} dX_k ... dX_2 dX_1$$

where: $\quad n_i \geq 0, \ \forall i \in \{1, 2, ..., k\}$

and: $\quad p = \sum_{i=1}^{k} n_i \quad$ is called order of the moment.

In a digitalized space, this measure can be rewritten:

$$M_{(object)}(X_1^{n_1}, X_2^{n_2}, \cdots, X_k^{n_k}) = \sum_{X \in object} X_1^{n_1} X_2^{n_2} \cdots X_k^{n_k} dm$$

where $dm$ is the unitary element mass of the digitalized space .

In a more compact manner, moments will be invoked by: $M_{(object)}\left(\prod_{i=1,k} X_i^{n_i}\right)$

and *object* will be the support space on which the measure is computed, which is the set, the shape or the component that it is tried to measure.

It will be progressively computed the moments of an object in increasing order.

At order 0, it will be got the mass of the object: $M_{(object)}(1)$

At order 1, it will be deduced the gravity center of the object :

$$XG_i = M_{(object)}(X_i) / M_{(object)}(1), \ i \in \{1, 2, ..., k\}$$

In order to lighten the presentation, it will be referred no more to the measure support.

It will be then got the centered values of order 2 moments in the following way:

$$\text{if } x_i = X_i - XG_i, \ i \in \{1, 2,, k\} \text{ then}$$
$$M(x_i x_j) = M(X_i X_j) - XG_i M(X_j) - XG_j M(X_i) + XG_i XG_j M(1)$$

The centered order 2 moments enable to compute the rotation matrix for going into the Eigen reference frame of the object whose origin is the gravity center.

In this aim, it must be set up the inertia matrix of the object:

$$In_{k \times k}(i, j) = M(x_i x_j), \ i \in \{1, 2, ..., k\}, j \in \{1, 2, ..., k\}$$

It is a positive definite square matrix that can be rewritten after diagonalization:

$$In_{k \times k} = V^T \Lambda V$$

where $\Lambda_{k \times k}(i, j) = M(u_i, u_j) / M(u_i^2) \geq 0$ and $M(u_i u_j) = 0$ for $i \neq j$

is the matrix of inertia axes of the object represented in its Eigen reference frame, and $V$ is the matrix of Eigen vectors enabling to go after centering, from the capture reference frame of the object to is Eigen reference frame.

The object can then be assimilated to its inertia ellipsoid and its Eigen vectors are defined more or less π.

The moments of order 3 allow to solve the uncertainty about the direction of the Eigen axes. Restricted to its Eigen directions, they can be written after centering:

$$M(u_i^3) = \sum_j \sum_m \sum_n v_{ji} v_{mi} v_{ni} M(x_j x_m x_n)$$

where $v_{ji}, v_{mi}, v_{ni}$ are the components of the Eigen vector $v_i$ of the matrix $V$.

They represent the object asymmetries according each of its Eigen axes and measure the eccentricity along each axis.

The uncertainty of the axis directions is removed by orientating the axes in the direction of the strongest eccentricity, that is to say in such a meaning where:

$$M(u_i^3) \geq 0, \quad \forall i \in \{1, 2, ..., k\}$$

that is by performing a hyper-planar symmetry when $M(u_i^3) < 0$.

The transformation for going from the observation reference frame to the Eigen reference frame of the object can be expressed in homogenous coordinates as it follows:

$$\begin{bmatrix} U \\ 1 \end{bmatrix} = \begin{bmatrix} V & -XG^T \\ 0 & 1 \end{bmatrix} \begin{bmatrix} X \\ 1 \end{bmatrix}$$

By normalizing according to the principal inertia axis, it can be then provided a representation independent to the similarities:

$$\begin{bmatrix} U \\ 1 \end{bmatrix} = \begin{bmatrix} \dfrac{1}{M(u_1^2)}V & -XG^T \\ 0 & 1 \end{bmatrix} \begin{bmatrix} X \\ 1 \end{bmatrix}$$

This transformation enables to reconstruct an object represented in its Eigen reference frame : its Eigen tree.

If a cloud of multidimensional data is analyzed, it is equivalent to do a statistical factorial analysis. By standing it in its Eigen reference frame and by using a projective operator, it is possible to represent the cloud in its principal axes. It can be then display in two or three dimensions using rendering tools.

The Eigen vectors highlight the correlations between the base variables and the Eigen values measuring the explanation rates of these variable combinations.

| *Name* | *Function* |
|---|---|
| KDMOMG | Compute the moments of a tree |
| KDCTRM | Center a moment list |
| KDNRMG | Normalize a moment list |
| KDEIGT | Generate the Eigen tree of a tree |

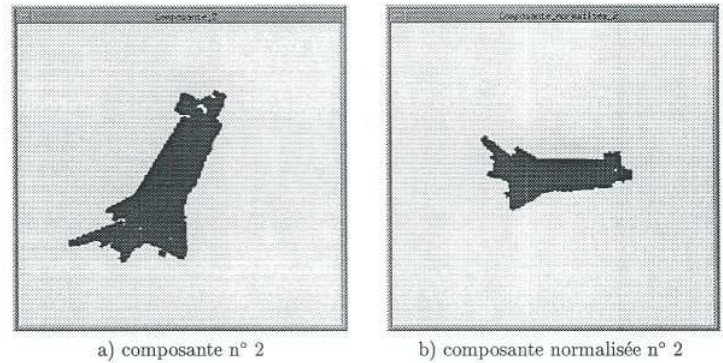

Figure 9 : Image of a connected component

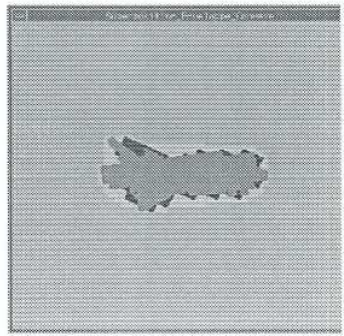

Figure 10 : Tree of Eigen components and its convex hull

## *8.2. Pattern recognition and similarity measure*

The computation of centered and normalized moments allows to propose two new models for representing a numerical object:

- its attribute vector, made from measures invariant in position, orientation and observing distance (only in orthography, not in perspective, that is the case in a distant position from the observed object) ;

- its Eigen tree where the object is reconstructed independently it these observing conditions.

Having at one's disposal these ones, it can be envisioned to perform statistical pattern recognition procedures. They consist in two steps:

- the first one is the learning phase during which the system calibrates its recognition procedure ;

- the second one is the recognition which is corresponding to the operational phase for exploiting captured data.

During the learning phase, it is gathered a learning set built from experiences that are usually made from observations performed over the studied process.

Two different approaches can be envisioned: perform a supervised or an unsupervised learning.

During a supervised learning, a teacher will put a label on each experience in the aim to make the system able to repeat by itself the teacher gesture.

During an unsupervised learning, the system will perform by itself the labeling of experiences in order to be able after it to classify by itself each new experience according to the labels computed over the learning set.

For performing an unsupervised classification, it is applied a segmentation of the space of measures or attributes belonging to the learning set.

In supervised classification, it can be built a base of Eigen trees labeled by a teacher. For each new observation, this one is moved in its Eigen reference frame and it is an exclusive or operation between this representation and the recognition base.

After this operation, the base is sorted according to increasing weight of the labels: it has then be got a data base sorted by decreasing similarity with the captured experience. In other words, it has just been performed a content-based query in the data base.

It can be provided a simplified version of this procedure by restricting it to the usage of attribute vectors: the attribute tree becomes then a shape indexing system enabling to perform content-based queries:

- a multi-criterion selection by specifying a varying value interval for each attribute;

- a nearest neighbor query by selecting the sub-tree of a given diameter around an observation.

Experiences are then sorted according to their similarity with this last one.

| Name | Function |
|---|---|
| KDCOLT | Coloring a tree |

# 9. Surface manifolds

## 9.1. Handling and transformation

The $2^k$-trees are favoring the handling of volume data, but in some circumstances it may be preferable to distinguish the evolution of one variable from the other ones:

- in image analysis, the luminosity is a variable evolving according to a planar support;

- in data statistical analysis, labels are used to identify sets of experiences observed with the help of explicative variables.

In these two cases, a given variable seems to express oneself according to its location inside a bi- or multi-variable support space:

- continuously for a mono-spectral image;

- discontinuously in statistical data classification (it is not necessary that an order relation rules its expression domain).

Distinguishing a variable from the other ones may be highly useful in the analysis of an observable phenomenon. So, if it is tried to apply an approach used in multispectral analysis for performing a thematic classification over a monochrome image, it would be equivalent to look for accumulation points in the luminescence space and to re-label the image according to the response class to which is belonging every image point, which is equal to segment a multi-threshold image.

The sets that are classified in such a way are not always regularly sampled. Learning sets are usually discrete sets made from data randomly spread in their representation sets (defining set of the variables).

The techniques of majority voting used in median filtering can be extended to discrete multi-valued sets so as to fill the missing data in learning spaces.

Attribute calculus and scaling operators enable to know the lower and higher bounds of functional values, their mean and dispersion, and to modify the value interval of this data.

| *Name* | *Function* |
|---|---|
| KDMIPY | Minimum of a pyramid |
| KDMAPY | Maximum of a pyramid |
| KDCTDP | Compute center and dispersion of a pyramid |
| KDSCAL | Scale a pyramid |

## 9.2. Median transforms

Applied on surface manifolds, median filtering enables to reallocate functional values according to their neighbor values. In a similar way to the median filtering applied on volume data, this value reallocation process is aimed to stabilize the shape of curved surfaces in increasing their continuity.

If it is looked for identifying the set of median locations of a set of volume nature by performing progressive thinning steps, concerning surface manifolds it is rather aimed to regenerate the locations where information is lacking. It is a diffusion process, done by examining the neighborhood of each point lacking information and by allocating to it the most prevalent value from its neighborhood found by a majority vote. As each the neighborhoods are not fully filled, the expansion process is iterated until no unaffected point can be found. It is an extension operation, enabling to convert randomly sampled discrete spaces into regularly sampled sets.

It is a deeply useful operator for building continuous spaces for pattern recognition from scattered spaces of learning data.

| *Name* | *Function* |
|---|---|
| KD1MFP | $d_1$-median filtering of a pyramid |
| KD0MFP | $d_\infty$-median filtering of a pyramid |
| KD1EXT | $d_1$- extension of a pyramid |
| KD0EXT | $d_\infty$- extension of a pyramid |

## 9.3. Polynomial fitting of a simple function

Coming back to the first works performed by ADERSA in the field of hierarchical multidimensional modeling, these ones were relying on the development of an innovative technique of piecewise multiple regression based on the recursive dividing of a data set orthogonally to its main inertia axis by the hyper-plane crossing the center of gravity of the point cloud until satisfying to a minimal approximating error. This method of piecewise multi-linear approximation is providing some troubles when interpolation is performed far from the gravity centers of data clouds stemmed from the recursive dividing. In order to reduce the generation of singularities coming from this modeling approach, it has been developed a technique for creating a continuous matching relying on the barycentric interpolation of data estimated on two connected pieces in the modeling tree.

The continuous matching is computed as the composition of shapes of straight higher order in the following way:

$$f = \frac{\left|\overrightarrow{C_g M} \cdot \overrightarrow{C_g C_d}\right|}{\left\|\overrightarrow{C_g C_d}\right\|^2} \cdot f_g + \frac{\left|\overrightarrow{MC_d} \cdot \overrightarrow{C_g C_d}\right|}{\left\|\overrightarrow{C_g C_d}\right\|^2} \cdot f_d$$

where $M$ is a space point about which it is expected to get an estimated value,

$C_g$ and $C_d$ are the gravity centers of the point clouds associated to the left and right sons of a node in the modeling tree,

$f_g$ and $f_d$ are the regression forms computed over the two clouds.

It is also a multidimensional fitting scheme of a B-Spline function.

If on the contrary to apply it simply locally, it is generalized to the whole set of modeled data by coming back successively up to the root of the representation tree, this interpolation method enables to recursively determine the polynomial approximation degree that better fits the data set used for building the model: it is equal to the number of levels met in the modeling tree.

If now, the same data set is modeled by a pyramid representing a simple function, when a paternal node is divided into two filial sons according to a direction defined by the level reached in the tree, the interpolation can be then written in the following way:

$$f = \frac{x_i - x_{g,i}}{x_{d,i} - x_{g,i}} \cdot f_g + \frac{x_{d,i} - x_i}{x_{d,i} - x_{g,i}} \cdot f_d$$

where $i$ is the index of this direction and $x_i$ is the corresponding coordinate of point $M$,

$x_{g,i}$ and $x_{d,i}$ are the coordinates of same index of the gravity centers of the point clouds associated to the left and right sons of a paternal node,

$f_g$ and $f_d$ are the forms recursively computed from the functional values registered in the terminal nodes with the help of this interpolation scheme.

By coming back up in the pyramid tree, if it occurs that the forms associated to the filial nodes are identical then the recursive scheme will provide the same form for the paternal node and a maximal degree of interpolation will be locally achieved. If the simple function is corresponding to the digitalization of a continuously differentiable function, it will be provided its polynomial expansion of maximal degree in its digitalization reference frame.

# 10. Implementation on parallel computer

## 10.1. Complements about the software structuring on sequential computer

### 10.1.1. Main organization

The KDTREE software is a library of functions to which have been added user interface modules.

Three kinds of modules can be distinguished:

- modules for building commands enabling a user to interactively specify an action to perform ;

- modules for analyzing commands that decode the action specified by a user or a file of pre-registered commands describing a series of actions to perform;

- modules for processing commands that constitute the library of functions accessible by the software.

Disconnected from building and analysis command modules, the function library can be used for developing particular applications.

### 10.1.2. Structure of a command

The syntax of a command is the following one :

    function (parameter 1',     ..., parameter n') [= result] ;

which is the succession of :

- a function name ;

- between parentheses, a list of parameters separated by commas that will condition the function performance ;

- the equal mark and the name of a variable that will receive the performance result, if this function is delivering a result and if it is expected to store it ;

- a semicolon as terminal mark.

Parameters are either immediate values or preexisting variables.

The variables are handled (creation, modification, duplication, suppression) by the KDTREE software and enable to refer using a mnemonic:

- an immediate value (a logical, integer or floating point value) ;

– a data structure stored in virtual memory (simple linkage list or binary tree).

This manner to invoke the processing of a function is very flexible and will enable by sending a message to distinct processors to activate its processing on an asynchronous parallel computer.

Commands can be gathered into a file closed off by the end of file command named:

    KDEND.

The software is delivered with a command file that is specifying all the tags used for defining the different kinds of structures handled in virtual memory (cvtype.cmd).

| *Name* | *Function* |
|---|---|
| KDSUVR | Suppression of a variable |
| KDINVR | Insertion of a variable |
| KDCPVR | Duplication of a variable |
| KDMDVR | Modification of a variable |
| KDLSVR | Attribute list of a variable |
| KDPRVR | Print the variable list |

| *Name* | *Function* |
|---|---|
| KDPAUS | Pause during a given while |
| KDSTOP | Stop software processing |
| KDEND | End of a command file |

### 10.1.3. Organization of the function library

The function library has got a ring structuring: it is made up from successive layers in such a way that functions from one layer can only call functions belonging to lower layers.

The complexity of performed operations increases when moving towards the highest layers.

Four successive layers can be distinguished:

– access to the lower level of the doublet memory (*allocation/freeing of a doublet, reading/writing*) ;

- access to primary data structures (*list, queue, stack* and *binary tree*) ;

- access to tree simple processing functions (Boolean operations, …) ;

- access to tree complex processing functions (search for connected components, …).

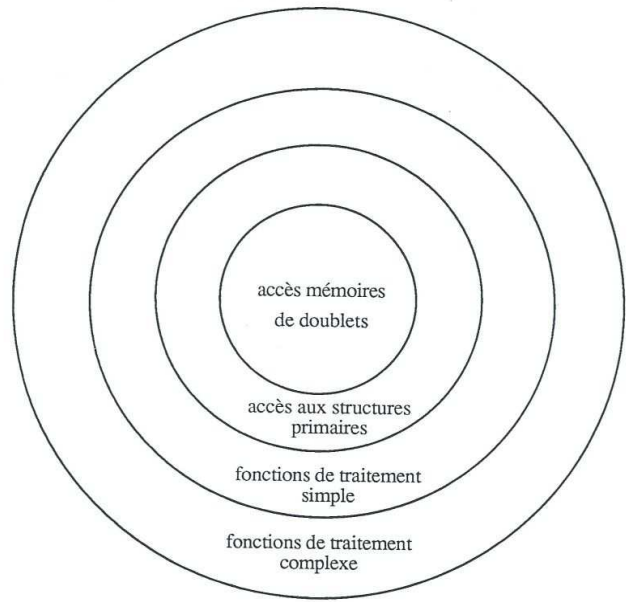

**Figure 11 : Ring organization of the processing library**

The two last layers are made from parallelizable tree processing operators:

- the first one consisting in operators providing a response time asymptotically linear with the data set size (simple recursion) ;

- the first one consisting in operators providing a response time asymptotically quadratic with the data set size (double or nested recursion).

## *10.2. Implementation methodology*

### 10.2.1. Parallel computer architectures

Parallel computers are made up from a collection of processors and memories able to work simultaneously and having at their disposal communication capabilities between them. The way that they are working together and the apparatus enabling them to communicate between them will define their own kind of architecture.

Usually a front-end computer is associated to a parallel computer, named host computer: it carries out communications with the outside and can take in account sequential computing parts of an application.

A computing processor is usually subdivided in two units:

- a decoding and control unit for the processing of instructions ;
- an arithmetic and logic unit which is processing instructions.

Two ways may then be envisioned for developing a parallel architecture computer:

- computing units are sharing a same control unit which is decoding a single stream of instructions ;
- each computing unit has got its own control unit which is decoding its own stream of instructions.

It is then told about single instruction stream computer in the first case and multiple instruction stream computer in the second one. By essence, it is a synchronous computer in the first case (only one clock runs a same instruction set) and an asynchronous computer in the other one (several clocks run different instruction sets). In the last case, the programmer must take in charge the program control by using different synchronization mechanisms :

- message passing ;
- semaphore for shared memory access control;
- processing interruption-based synchronization.

The second element which is characterizing the parallel computer architecture is the chosen access mode enabling processors to reach data in memory. From this point of view, it can be distinguished:

- distributed memory systems ;
- from shared memory systems.

In a distributed memory system, each processor has at its own disposal a local memory and communicates directly with the other processors.

In a shared memory system, all the processors share a global memory and communicate between them with the help of this single memory.

In these two cases, a communication network is included in the system:

- in a distributed memory system so as to enable processors to communicate between them ;

- in a shared memory system so as to enable several processors to access simultaneously at the same memory.

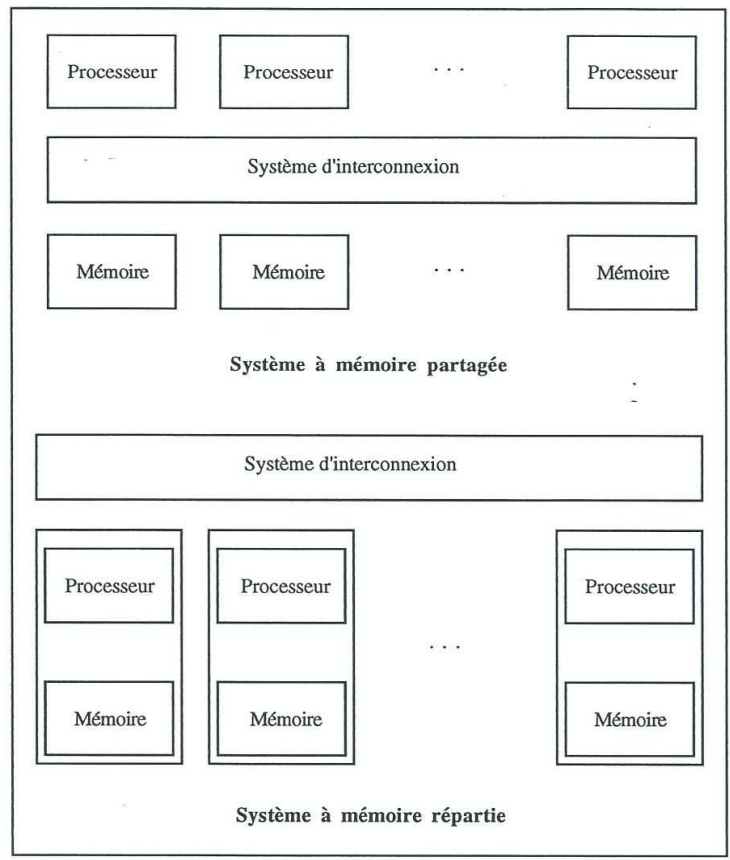

**Figure 12 : Parallel computer architectures**

It will not go further for describing parallel computer architectures, except for pointing out two particularities that are usually taken in account for developing parallel computing applications.

On one hand, it will be achieved a high degree of parallelism if the application is able to run in parallel over most of the processors, which is leading to design algorithms for which a same copy is able to run simultaneously over any number of processors for building the application. If different programs must collaborate to the building of a same solution, efforts will be quickly restricted for developing applications with a high degree of parallelism.

On the other hand, it often happens that communications show a bottleneck in this kind of architecture. The weight of this bottleneck varies according to the fact that the process to be performed requires or not:

- communications with other processors ;

- local communications only with neighbor processors ;

- global communications with all the other processors according to regular access schemes ;

- global communications with all the other processors according to irregular access schemes.

In order to set up simultaneous communications between several units, communication networks are implemented. These ones are sensitive to communication patterns that can be at the origin of collisions in information transport.

These networks may have very simplified architectures, but the message avoidance of routing collision must be handled by the programmer. Progresses in micro-electronics have progressively smoothed out these problems.

### 10.2.2. Methodology followed for software parallelization

For implementing algorithms on parallel computers, it has been carried out the following steps:

- development of parallelizable sequential algorithms ;

- adaptation of sequential algorithms to data partitioning in processor memories;

- algorithm effective parallelization;

- analysis of algorithm dynamic behavior.

Testing data sets are developed on sequential computers and are used to check the good progress along all these steps.

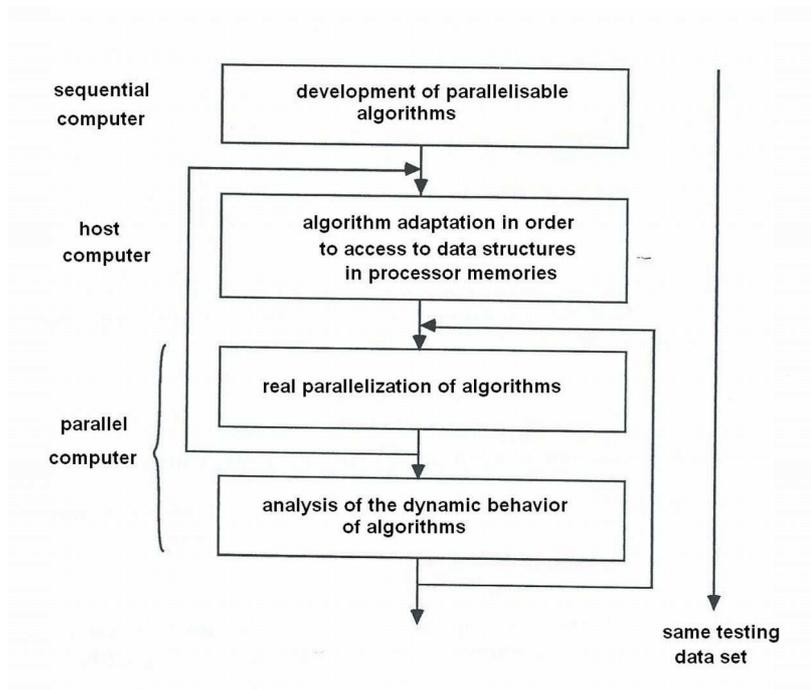

**Figure 13 : Parallelization methodology**

This methodology is deeply relying on the processing capabilities of the host computer and distinguishes the development of an implementation model of data in parallel memories with the effective implementation of parallel algorithms. All the algorithms are not systematically parallelizable, but for a given function to implement there is often a parallel algorithm that can perform it.

According to the kind of parallel architecture, it is more or less easy to carry out this exercise. If the control of a function is easy to carry out on a synchronous computer, it is restricted the nature of algorithms that can be implemented on this kind of architecture.

On asynchronous computers, the control being dependent on the algorithm, a wider range of possibilities is offered to the programmer, but this freedom create more problems at development : the result is certainly not deterministic and some authors raised the anxiety that may feel a programmer during development (a dead-lock fear).

Besides controlling algorithms, the other source of troubles comes from the communication schemes used by them. Even progresses have been achieved concerning the development of networks enabling to simultaneously interconnect electronic sub-sets in parallel, the data through-puts that can be achieved is depending on the size of communications to perform and the regularity of communication schemes to set up (that is by restricting the use of common routing paths in order to avoid collisions, during information transport). A good example can be found with the search of connected components:

- the bucket algorithm has higher performances on sequential computer than the hierarchical growing of regions ;

- but, this last one provides regular communication schemes and is effectively parallelizable.

So, the bucket algorithm cannot be implemented on a synchronous computer and has not a deterministic response time on a synchronous computer.

Giving some parallelizable algorithm, its sequential release is then on parallelized data structures:

- the adequacy of the algorithm with the model of data in parallel memories is checked ;

- it is always possible to come back to the full sequential release, in case of bad performances.

The software then runs on the host computer while accessing to the parallelized data.

The next step is the effective software parallelization: the necessary modifications for running on parallel processors are brought.

The last step is the analysis of the dynamic system behavior which is leading to optimize algorithms in order to take in account of the disparity of response times between processors, memories and communication network.

## *10.3. Distributed memory synchronous computers*

### 10.3.1. Data representation model in memory

The KDTREE software has been ported on a fine-grain synchronous computer, made from a very high number of elementary processors with a little memory.

In order to avoid irregular schemes of communication, it has been used a pyramidal model for managing trees: all branches are developed until the computing precision. The property of information compression existing in $2^k$-trees is then lost.

It can be associated to each node a virtual or physical processor in the following way:

- the processor number 0 is left free ;

- the processor number 1 is holding the tree root or the root of any other tree ;

- if a processor n is holding some node, the left and right sons will stay respectively in processors 2n and 2n+1.

In such a way, a complete tree is mapped on a linear network of processors.

### 10.3.2. Rewriting the general resolving algorithm

In a synchronous algorithm, the activity of each processor is controlled by a state latch: after selection or deselecting, it becomes active or inactive. The general resolving algorithm, adapted to the pyramidal representation model, can then be written:

```
PROCEDURE process (tree number, level, depth)
BEGIN
    IF (level = depth) THEN
    DO IN PARALLEL /* on all active processors */
        terminal processing
    ELSE DO
        DO IN PARALLEL descending processing
        DO IN PARALLEL
            activate present processor number * 2
            activate present processor number * 2 + 1
            deselect present processor
        END
        CALL process (tree number, level + 1, depth)
        DO IN PARALLEL ascending processing
    END
    DO IN PARALLEL /* ascending return */
        WHERE (even number (present processor))
        THEN activate present processor number / 2
        deselect present processor
    END
END
```

For a synchronous parallel architecture computer, it must be distinguished instructions that have to run on a sequential unit (usually the host computer), with those that have to run on a parallel unit:

- DO becomes DO IN PARALLEL for parallel computing units ;
- IF is changed into WHERE for running tests in parallel.

The recursion is controlled by the host computer and trees are traversed in breadth first.

The control of processor activity depends on the level reached by the recursion in the tree and on the node location in computer memory.

## 10.4. Distributed memory asynchronous computers

### 10.4.1. Data representation model in memory

In a coarse grain system, it can be put a full tree branch in the memory of one processor.

In an asynchronous system, it is not necessary to wholly develop tree branches. The management model based on linked lists implemented in a virtual addressing memory can be kept and extended to the data management in a distributed memory: the addressing system (page and location in the page) is complemented with the processor number in whose memory data is staying.

According to the place where the processor field is standing in a word address, it can be defined different manner for distributing data in memory:

- by contiguous addressing, when the processor field is belonging to higher order addressing bits;

- by interlaced addressing, when this one is standing in the lower order addressing bits.

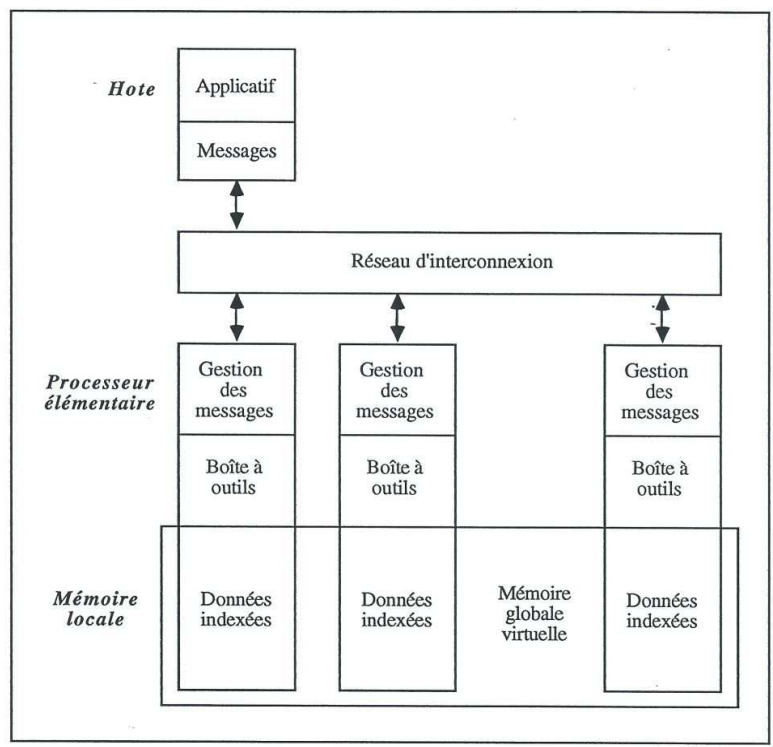

**Figure 14 : Software implementation on distributed memory asynchronous architecture**

At the contrary of the pyramidal implementation way, a current processor can be reused for processing other tree nodes, especially the left branch of its sons and its grandsons: the right branch is then out-sourced to another processor.

The right branch of its sons is sent to another processor located at a position depending on the level reached in the tree: at the $2^{p-n+1}$-th neighbor, if there are $2^p$ processors in the network and if the level n has been reached in the tree.

Starting from processor 0, step by step the tree will be distributed over the full set of memories associated to the parallel processors.

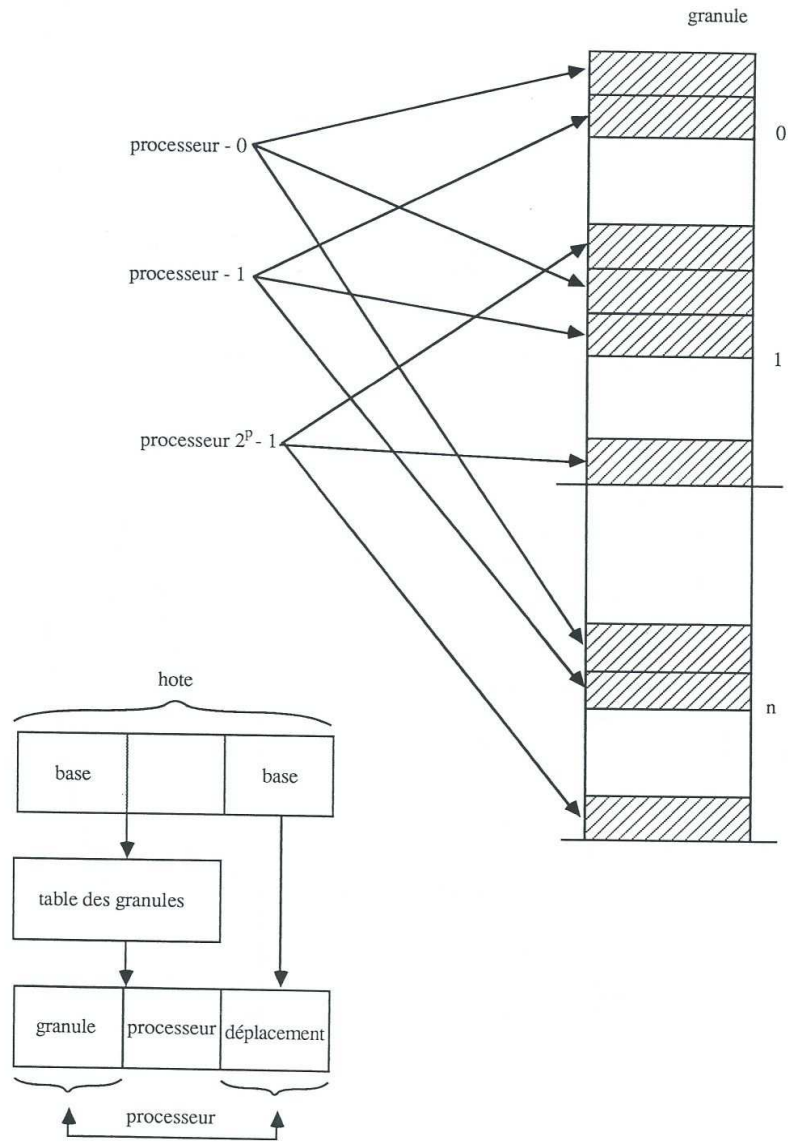

**Figure 15 : Virtual memory adaptation to a distributed memory system**

When the processor network will be completely filled, it will be carrying on to locally allocate in memory of the current processor the nodes that would remain to store.

## 10.4.2. Rewriting the general resolving algorithm

In a distributed memory asynchronous computer, the processor activity is controlled by passing messages between processors.

In order not to depend on the number of available processors, each processor will have at its disposal a same copy of the program.

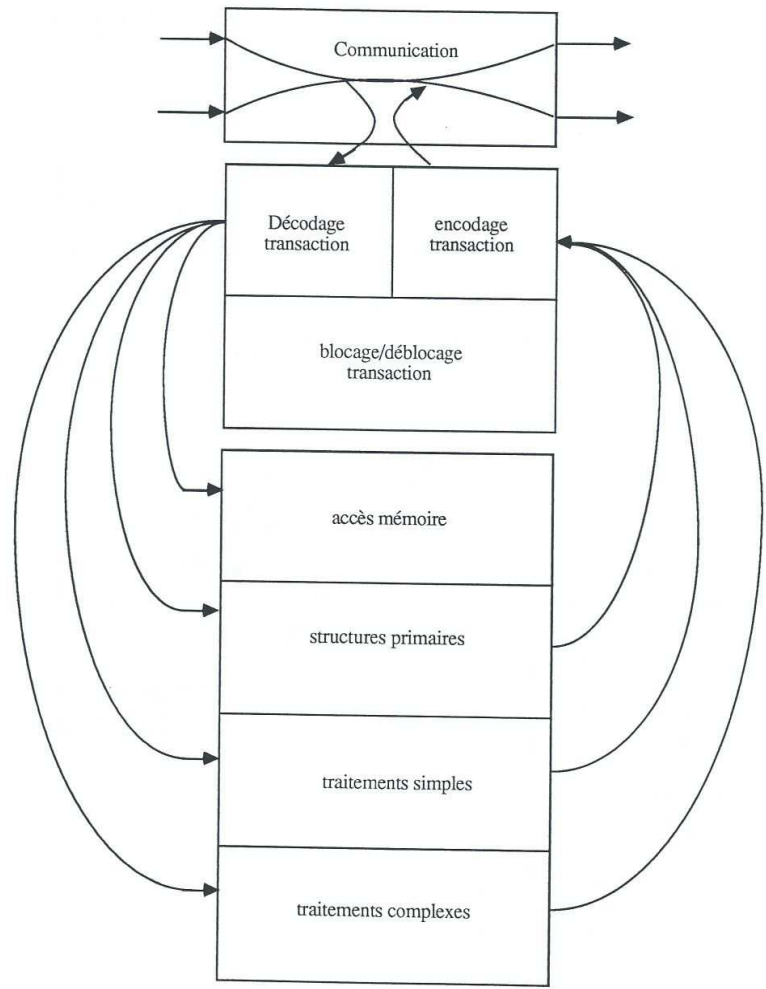

**Figure 16 : Active tasks on a processor belonging to an asynchronous architecture**

The implementation of the general resolving algorithm is then the following one :

- on the host processor :

BEGIN

    transaction number <- *send message* (process,

        processor (*localize* (root)), root, 0, dimension, precision)

*wait for acknowledgement* (transaction number)

END

- on parallel processors :

*receive message* (function, parameters)

BEGIN

    IF (function = process) THEN DO

        locally register dimension and precision

        depth <- dimension * precision

        CALL process (root, level)

    END

    *acknowledge message*

END

PROCEDURE process (root, level)

BEGIN

    IF (*terminal* (root) OR (level = depth))

    THEN terminal processing

    ELSE DO

        descending preprocessing

        IF (NOT *local* (*right son* (root)))

        THEN transaction number <- *send message* (process,

            processor (*localize* (*right son* (root))),

            *right son* (root), level + 1, dimension, precision)

        CALL process (*left son* (root), level + 1)

        IF (NOT *local* (*right son* (root)))

        THEN *wait for acknowledgement* (transaction number)

        ELSE CALL process (*right son* (root), level + 1)

        ascending post-processing

```
        END
    END
```

Trees are traversed in breadth first during the phase of computing distribution, then in depth first when the phase of local computing is reached on the processors.

If the right branch processing is distributed, it is anticipated compared to the left by sending a message in the network for requesting the processing of the right branch.

Synchronization is performed in return at the end of the left branch processing.

## *10.5. Experimentation on two distributed memory computers*

### 10.5.1. Presentation of the target systems

The KDTREE software has been ported on two target systems: the CONNECTION Machine of THINKING MACHINES and the computer T.NODE of TELMAT Informatique.

The Connection Machine is a distributed memory synchronous computing system, that can be partitioned into four sub-sets.

A partition is made from:

- a host computer including a connection electronic board ;
- a sequencer for parallel instructions ;
- parallel processors.

Parallel processors perform simultaneously the same instruction on multiple data (parallel computing architecture of SIMD kind).

They have the following features:

- a single-bit computing unit ;
- each processor is controlled by an activity indicator ;
- each processor has at its disposal a reduced local memory ;
- serial communication links connect one processor with its n nearest neighbors in a hypercube of dimension n.

Communications are realized:

- in a single step to the nearest neighbors,
- in no more than n steps to any processor, when they are no message collision in a computing network with $2^n$ processors.

The Connection Machine shows the complementary particularities:

- the possibility to simulate a processor number higher than its physical processor number (virtual processors),
- parallel inputs/outputs on disk storage and graphic screen,
- a design favoring reliability over computing fastness.

The T.NODE computers are distributed memory asynchronous computing systems.

They are made from:

- a host computer including a connection board,
- a general control system of parallel processors.

The parallel processors are TRANSPUTER processors manufactured by INMOS Ltd each executing their own instruction codes on multiple data streams (parallel computing architecture of MIMD kind).

They have the following features:

- the computing unit processes words of 16 or 32 bits ;
- each processor has at its disposal a wide local memory ;
- four communication links per processor enabling to perform point to point links (in cross-bar organization up to 64 processors and in using a tree-like organization scheme above) ;
- the communication network between processors is reconfigurable and its topology is defined directly by the user ;
- communication routing can be taken in charge by the software environment.

Three different software environments are proposed:

- the OCCAM language from which the TRANSPUTER has been designed ;
- stand-alone compilers integrating a minimal set of operating system functionalities;
- the multiprocessor operating system HELIOS which is compliant with UNIX.

Parallel inputs/outputs on disk storage or graphic screen can also be used.

The TRANSPUTER includes by manufacturing functionalities enabling to develop applications based on concurrent programming:

- inter-task communications using physical or logical channels,

- parallel handling of communications and computations with the help of a micro-coded monitor,

- internal multi-tasking management,

- access control to shared data using semaphores.

### 10.5.2. Implementation specificities

The communication network of T.NODE computers is reconfigurable.

Each processor has at its disposal four communication links that can be connected with the links of four other processors in the network.

In order to set up an optimal static communication procedure, it has been chosen to configure the network into a network of Omega kind.

It is a logarithmic communication architecture owning the following properties:

- four unidirectional communication paths are enough for its implementation whatever is the number of processors to be connected ;

- it is reentrant and it does not need to be reconfigured for insuring the data recirculation up to the recipient processors ;

- it is optimal in the meaning of logarithmic communication networks, that is in the meaning where a message for moving from one processor to another one will need to traverse no more than p processors inside a network of $2^p$ processors.

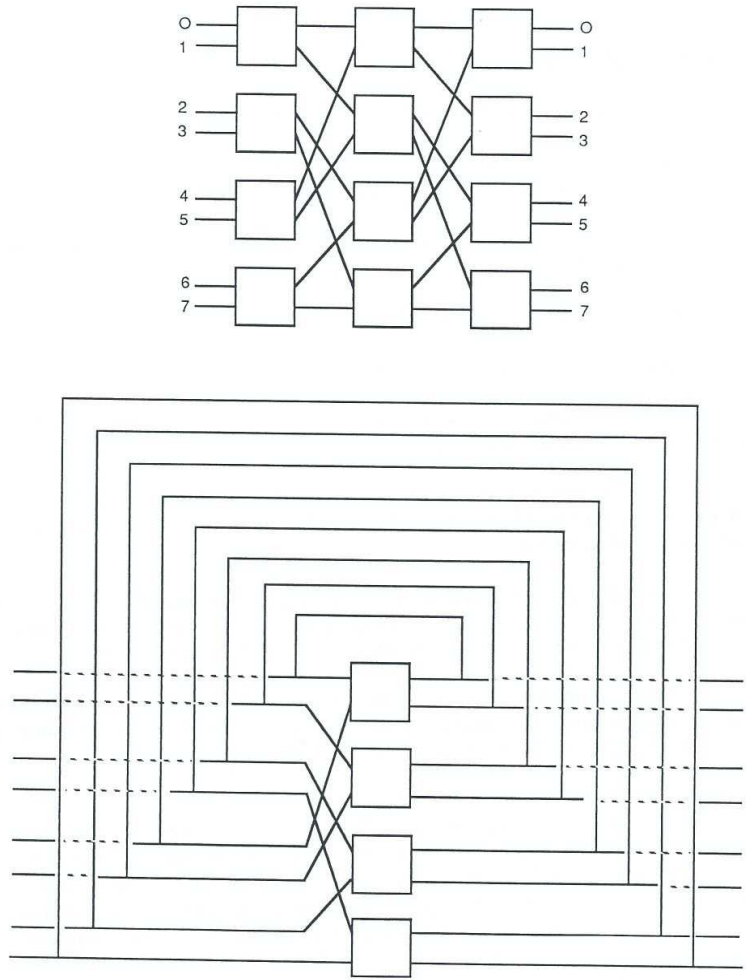

**Figure 17 : Reentrant Omega network connecting 8 processors**

The communications between processors being asynchronous, collisions are implicitly handled using a « store-and-forward » strategy: the first arrived message takes the ownership of the path and only one message can wait for the path freeing.

The Omega network is only blocking when the processors are asking for a communications size higher than the network throughput enables.

A message is sent according to the following way between the emitting processor P and the receiving processor Q:

- let it be $q_1q_2...q_p$ the binary decomposition of the processor number Q of the network;

- the processor P sends the message on its high output if $q_1$ is worth 0 or its low output if $q_1$ is worth 1 ;

- the p-1 following processors that are successively receiving the message resend it according to the same procedure by consecutively examining every bits $q_2...q_p$ of the remaining of the decomposition of the processor number to reach ;

- the last reached processor is the receiver of the message.

This procedure is independent from the location of the processor P in the network.

Whatever is the target system, some algorithms of the KDTREE software have been adapted in order to be efficiently implemented on this computing architecture.

The adaptation mainly bears on reducing the use of linked lists at the node level in a tree because they increase the size of communications that limits the use of system configuration with a high number of processors. It is for instance the case when searching adjacencies where adjacency lists are temporally stored at the level of terminal nodes.

### 10.5.3. Conclusion about these experimentations

Although it has been tried to find a common approach for parallelizing the KDTREE software on these two target systems, the Connection Machine and the T.NODE computer, two distinct approaches have been used in practice:

- one on asynchronous machine, where the basement algorithmic has been not very much modified and the data structures kept ;

- the other one on synchronous machine, where the algorithmic must have been adapted and the data structuring model reviewed.

The synchronism of the Connection Machine does not allow to keep the implicit compression of $2^k$-trees and needs to replace them by pyramidal structures where all the branches are fully developed until the resolution asked for processing.

It must be seen the Connection Machine as a cellular machine which is more appropriated to process trees as sequential codes than linked lists.

| CONNECTION MACHINE | T.NODE |
|---|---|
| **Synchronous computer** | **Asynchronous computer** |
| **Distributed memory and serial communication systems** | |
| **Hypercube network** | **Logarithmic network (*)** |
| **Electronically integrated routing integrating a collision processing** | **Routing simulated by software with collision avoidance (*)** |
| **Communications based on memory access** | **Communications based on memory access (*)** |
| **Integrated memory access from the host computer to the parallel processor memory** | **Simulated memory access from the host computer to the parallel processor memory (*)** |
| **SIMD programming** | **SPMD programming** |
| **Pyramidal data representation based on tree codes** | **Tree-like data representation based on linked list** |
| **Node mapping using a linear addressing code (one node per processor)** | **Dichotomous node mapping (a branch per processor)** |
| **Breadth -first tree traversal and host recursion** | **Breadth-first then depth-first tree traversal and local recursion** |
| | **(*) : works done during the software porting** |

**Figure 18 : Comparison of implementation approaches used on two distributed memory systems**

# Annexes

# A. Example of a command file

## A.1. Description of the command file

The image is stored on disk with the name shut256.bar is read, and then displayed in the window named Image_Extraction. It is a binary view of the American spatial shuttle (image 256 rows x 256 columns).

A translation is applied on this image upwards on the right. The result is displayed in a new window (Translation).

Then a series of mathematical morphology operations starts (homotopic transformations):

- an erosion is applied on the original image (Erosion), then the result is displayed over the original image colored in blue (Superposition) ;

- a dilation is applied on the original image (Dilatation), then displayed in the background of the original colored image and its eroded version (Superposition_Erosion_DiIatation).

The blue remaining part is the boundary of the original image, deleted by the erosion, extended to its exo-boundary by the dilation.

Then median transforms that are also topological transforms are successively performed:

- a median filtering is applied on the original image (Median_Filtering). The modified points are displayed above the original image (Superposition_Filtering) ;

- from the original image the median set is computed.

This one is then shown (Median_Set), then dispalyed over the original image (Superposition_Median_Set).

Then it is starting a regional analysis sequence : the adjacency analysis is performed on the original image and the connected components are labeled (Extraction_of_components). Each of the five first components appears differently colored (one window per component) in the position and direction with which they have been observed (Component_i).

It is applied on these components an attribute calculus and it is looked for representations independent from the observation reference frame:

- the moments of components are computed, then the Eigen tree of each comonent is built ;

- each component is once more displayed in its Eigen reference frame (Normalized_component_i).

At the end, it is simulated the building of a data base for pattern recognition:

- the Eigen components are gathered into a same tree;

- two windows are deleted and the result is displayed. (Superposition_of_Components) ;

- the data base and its convex hull are displayed together. (Superposition_Convex_Hull).

After seconds of waiting, the three windows are withdrawn.

What has been performed in a plane can also been applied to a space of higher dimension: it is then doing statistical data analysis.

## A.2. Images displayed by the command file

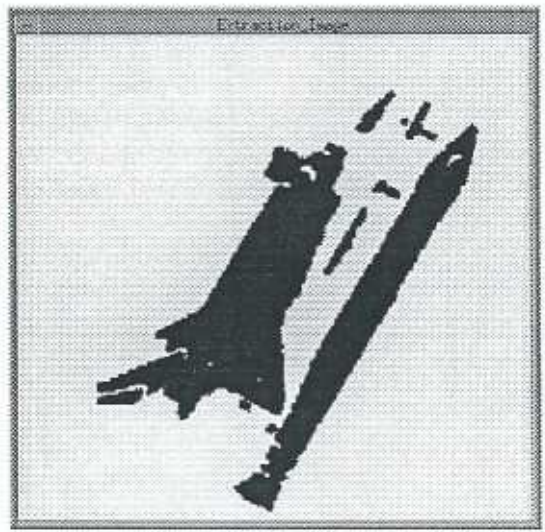

**Figure 19 : Original image**

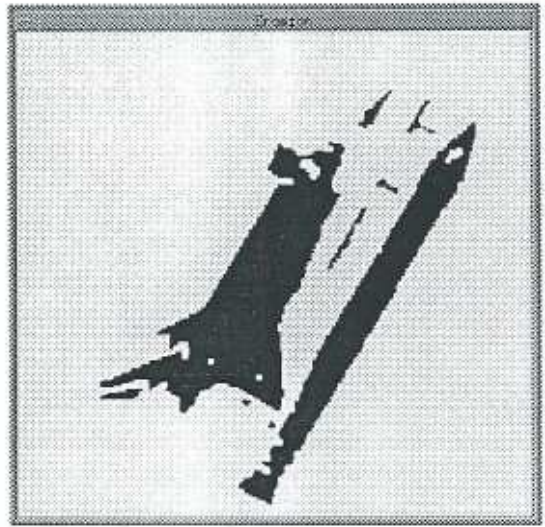

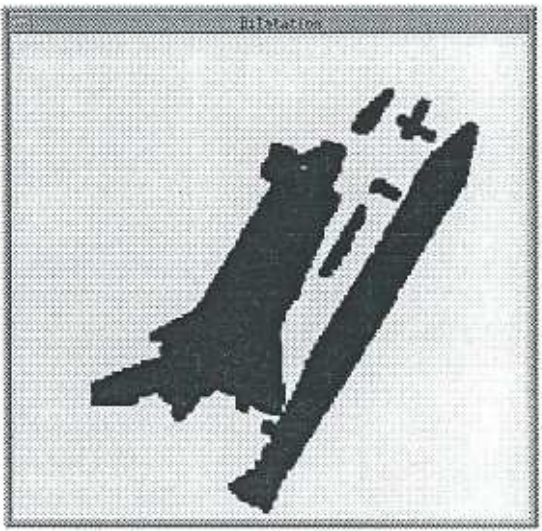
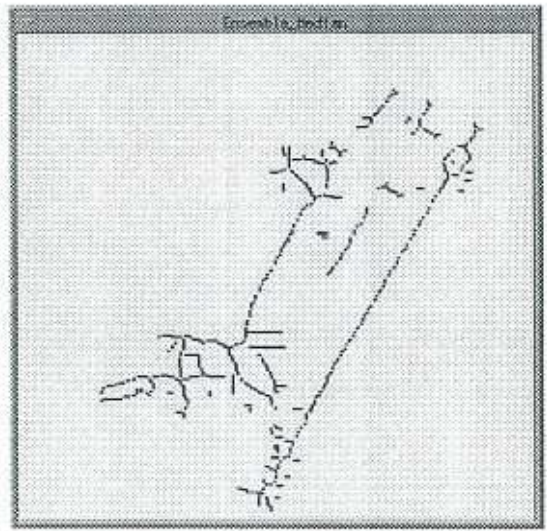

**Figure 20 : Filtering**

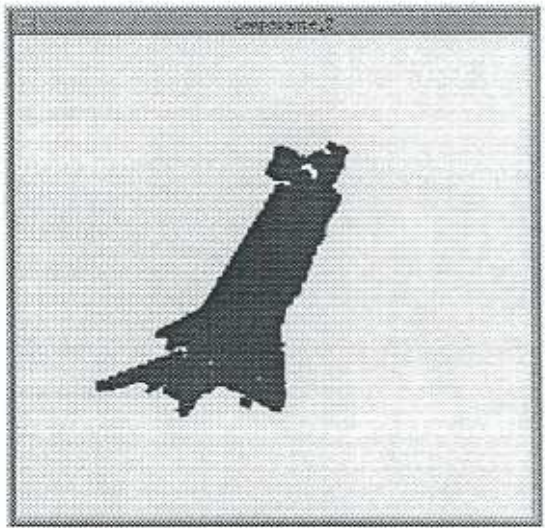 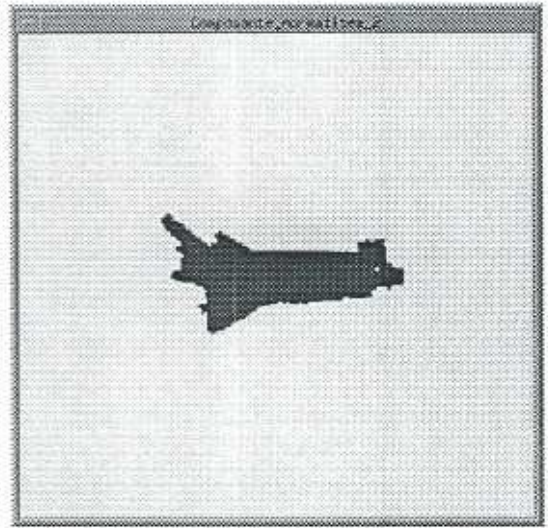

Figure 21 : Component images

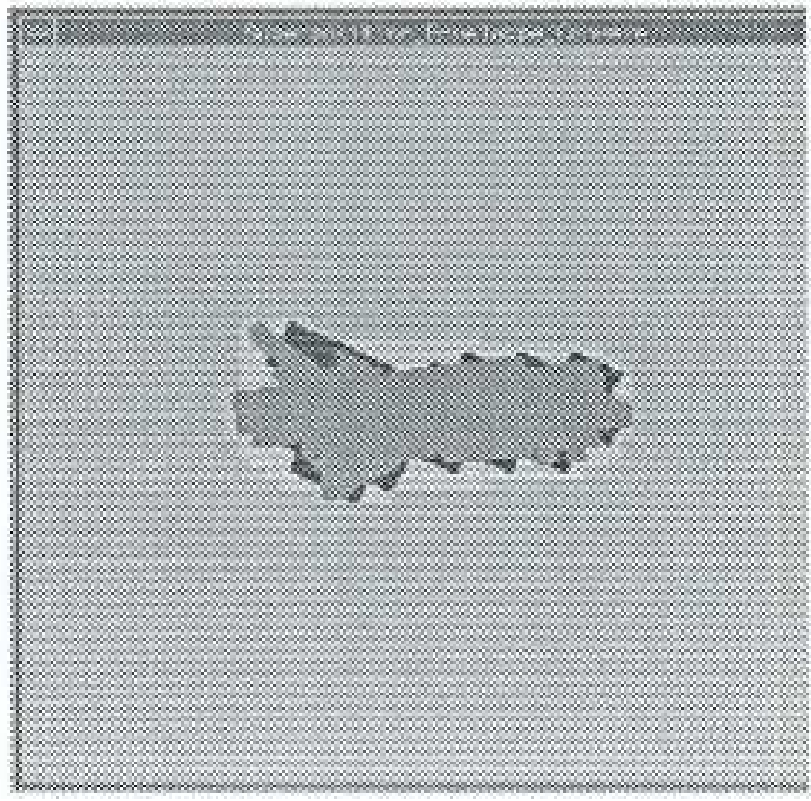

Figure 22 : Component superposition with their convex hull

## A.3. Listing of the command file

```
/* KDTREE: demonstration command file */
/* Demonstration of planar image processing */
/* initialization and image read */
    KDINVR(PRECIS, MQINTG, 0, 8);
    KDMDVR(PRECIS, MQINTG, 0, 8);
    KDSWGR(375);
    KDWBGR;
    KDINGR;
    KDMVGR(5, 20);
    KDERGR;
    KDMSGR(Image_Extraction);
    KDRDBT(shu256) = shuttle;
    KDQTGR(shuttle, PRECIS);
/* image translation */
    KDVRCI(2, 0.25, 0.25) = vectrs;
    KDTRAN(shuttle, vectrs, 2, PRECIS, PRECIS) = shutrs;
    KDINGR;
    KDMVGR(395, 20);
    KDMSGR(Translation);
    KDQTGR(shutrs, PRECIS);
/* erosion */
    KDASS(shuttle, 2, PRECIS) = shu60e;
    KD0ERO(shu60e, 2, PRECIS);
    KDNWGR(1);
    KDERGR;
    KDMSGR(Erosion);
```

```
    KDQTGR(shu60e, PRECIS);

    KDASS(shuttle, 2, PRECIS) = shu60;

    KDCOLT(shu60, 3, 2, PRECIS) = shu60c;

    KDINGR;

    KDMVGR(785, 20);

    KDERGR;

    KDMSGR(Superposition);

    KDQTGR(shu60c, PRECIS);

    KDQTGR(shu60e, PRECIS);

 /* dilatation */

    KDASS(shuttle, 2, PRECIS) = shu6Ort;

    KD0DIL(shu60d, 2, PRECIS);

    KDINGR;

    KDMVGR(5, 485);

    KDERGR;

    KDMSGR(Dilatation);

    KDQTGR(shu60d, PRECIS);

    KDNWGR(2);

    KDERGR;

    KDMSGR(Superposition_Erosion_Dilatation);

    KDQTGR(shu60d, PRECIS);

    KDQTGR(shu60c, PRECIS);

    KDQTGR(sgu60e, PRECIS);

/* filtering */

    KDASS(shuttle, 2, PRECIS) = shu60f;

    KD0MDF(shu60f, 2, PRECIS);

    KDNWGR(3);

    KDERGR;
```

```
    KDMSGR(Median_Set);

    KDNWGR(1);

    KDERGR;

    KDMSGR(Median_Filtering);

    KDQTGR(shu60f, PRECIS);

    KDEXCL(shu60c, shu60f, 2, PRECIS) = shu60xof;

    KDNWGR(2);

    KDERGR;

    KDMSGR(Superposition_Filtering);

    KDQTGR(shu60c, PRECIS);

    KDQTGR(shu60xof, PRECIS);

/* median set */

    KDASS(shuttle, 2, PRECIS) = nav60em;

    KDEMED(shu60em, 1, 2, PRECIS);

    KDNWGR(3);

    KDQTGR(shu60em, PRECIS);

    KDNWGR(2);

    KDERGR;

    KDMSGR(Superposition_Median_Set);

    KDQTGR(shu60c, PRECIS);

    KDQTGR(shu60em, PRECIS);

/* regional analysis */

    KDASS(shuttle, 2, PRECIS) = shu60;

    KD1ANR(shu60, 2, PRECIS);

    KDCCLB(shu60) = list;

    KDBSGT(list, shu60);

    KDINGR;

    KDMVGR(395, 485);
```

```
KDMSGR(Extraction_of_components);

KDERGR;

KDNWGR(0);

KDMSCR(Extraction_of_components) ;

KDERGR;

KDNWGR(1);

KDMSGR(Extraction_of_components);

KDERGR;

KDNWGR(2);

KDMSGR(Extraction_of_components);

KDERGR;

KDNWGR(3);

KDMSGR(Extraction_of_components);

KDERGR;

KDEXSG(list, 2) = comp1;

KDCOLT(comp1, 3, 2, PRECIS) = comp1;

KDNWGR(0);

KDMSGR(Component_1);

KDQTGR(comp1, PRECIS);

KDEXSG(list, 2) = comp2;

KDCOLT(comp2, 4, 2, PRECIS) = comp2;

KDNWGR(1);

KDMSGR(Component_2);

KDQTGR(comp2, PRECIS);

KDEXSG(list, 2) = comp3;

KDCOLT(comp3, 5, 2, PRECIS) = comp3;

KDNWGR(2);

KDMSGR(Component_3);
```

```
    KDQTGR(comp3, PRECIS);

    KDEXSG(list, 2) = comp4;

    KDCOLT(comp4, 6, 2, PRECIS) = comp4;

    KDNWGR(3);

    KDMSGR(Component_4);

    KDQTGR(comp4, PRECIS);

    KDEXSG(list, 2) = comp5;

    KDCOLT(comp5, 7, 2, PRECIS) = comp5;

    KDNWGR(4);

    KDMSGR(Component_5);

    KDQTGR(comp5, PRECIS);

/* moment calculus and Eigen trees */

    KDMOMG(comp1, 2, PRECIS) = list;

    KDCTRM(list, 2) = list;

    KDNRMG (list, 2) = list;

    KDLSST(list);

    KDMOMG(comp1, 2, PRECIS) = list;

    KDCTRM(list, 2) = list;

    KDNRMR(list, moment, matrix, 2);

    KDAPTR(comp1, moment, matrix, 2, PRECIS, PRECIS) = eigtree1;

    KDNWGR(0);

    KDERGR;

    KDMSGR(Normalized_component_1);

    KDQTGR(eigtree1, PRECIS);

    KDMOMG(comp2, 2, PRECIS) = list;

    KDCTRM(list, 2) = list;

    KDNRMR(list, moment, matrix, 2);

    KDAPRR(comp2, moment, matrix, 2, PRECIS, PRECIS) = eigtree2;
```

KDNWGR(1);

KDERGR;

KDMSGR(Normalized_component_2);

KDQTGR(eigtree2, PRECIS);

KDMOMG(comp3, 2, PRECIS) = list;

KDCTRM(list, 2) = list;

KDNRMR(list, moment, matrix, 2);

KDAPRR(comp3, moment, matrix, 2, PRECIS, PRECIS) = eigtree3;

KDNWGR(2);

KDERGR;

KDMSGR(Normalized_component_3);

KDQTGR(eigtree3, PRECIS);

KDMOMG(comp4, 2, PRECIS) = list;

KDCTRM(list, 2) = list;

KDNRMR(list, moment, matrix, 2);

KDAPRR(comp4, moment, matrix, 2, PRECIS, PRECIS) = eigtree4;

KDNWGR(3);

KDERGR;

KDMSGR(Normalized_component_4);

KDQTGR(eigtree4, PRECIS);

KDMOMG(comp5, 2, PRECIS) = list;

KDCTRM(list, 2) = list;

KDNRMR(list, moment, matrix, 2);

KDAPRR(comp5, moment, matrix, 2, PRECIS, PRECIS) = eigtree5;

KDNWGR(4);

KDERGR;

KDMSGR(Normalized_component_5);

KDQTGR(eigtree5, PRECIS);

```
/* simulation of the building of a data base */
    KDCPYR(0.0, BLANC, 2) = database;
    KDUNIO(database, eigtree1, 2, PRECIS) = database;
    KDUNIO(database, eigtree2, 2, PRECIS) = database;
    KDUNIO(database, eigtree3, 2, PRECIS) = database;
    KDUNIO(database, eigtree4, 2, PRECIS) = database;
    KDUNIO(database, eigtree5, 2, PRECIS) = database;
    KDNWGR(2);
    KDERGR;
    KDMSGR(Superposition_Convex_Hull);
    KDNWGR(3);
    KDDSGR;
    KDNWGR(4);
    KDDSGR;
    KDNWGR(1);
    KDERGR;
    KDMSGR(Convex_Hull);
    KDNWGR(0);
    KDERGR;
    KDMSGR(Superposition_of_Components);
    KDQTGR(database, PRECIS);
/* convex hull computation */
    KDMDVR(PRECIS, MQINTG, 0, 6);
    KDSUPY(database, 2, PRECIS) = hull;
    KDCVXH(hull, 2, PRECIS);
    KDMDVR(PRECIS, MQINTG, 0, 8);
    KDNWGR(1);
    KDQTGR(hull, PRECIS);
```

```
KDNWGR(2);

KDQTGR(hull, PRECIS);

KDQTGR(database, PRECIS);

KDPAUS(30);

 KDNWGR(2);

KDDSGR;

KDNWGR(1);

KDDSGR;

KDNWGR(0);

KDDSGR;

KDEND;
```

## B. List of KDTREE software commands

## *C. Complementary algorithms in hierarchical modeling*

# 1. Tree generation in inductive limit

root :              root of the tree to be enriched

vector:             head of the real vector to be added

{minspc, maxspc} :  space lower and higher faces

dimens :            dimension of the modeling space

precis :            computation precision

{minold, maxold} :  lower and higher faces of the previous space

{minnew, maxnew} :  lower and higher faces of the new space

dold:               diagonal of the previous space

dnew:               diagonal of the new space

level :             level reached in the tree

depth :             computation depth

{minlft, maxlft} :  lower and higher faces of left son

{minrgt, maxrgt} :  lower and higher faces of right son

lftson :            left son of the root

rgtson :            right son of the root

## 1.1. Creation of a tree in inductive limit

PROCEDURE kdctil (vector, root, minspc, maxspc, dimens)

BEGIN

    root <- *tree*(<u>black</u>)

    minspc <- *copy*(vector)

    maxspc <- *copy* (vector)

END

## 1.2. Addition of a no normalized vector to a tree

PROCEDURE kdavil (root, vector, minspc, maxspc, dimens, precis)

BEGIN

    IF (NOT kdinvs (vector, minspc, maxspc, dimens)) THEN DO

        /* the vector is outside from the current space */

        /* computation of the new space limits */

        CALL kdcnls (vector, minspc, maxspc, minnew ,maxnew, dimens)

        /* extension of the present tree up to the new space limits */

        root <- kdetls (root, minspc, maxspc, minnew ,maxnew, dimens, precis)

        minspc <- minnew

        maxspc <- maxnew

    END

    /* vector normalization */

    FOR nudim = 1 TO dimens DO

        vector (nudim) <- (vector (nudim)- minspc (nudim))/( maxspc (nudim)- minspc (nudim))

    END

    /* addition of the vector to the tree */

    CALL kdarvt (root, vector, dimens, precis)

END

## 1.3. Computation of the new limits of the space

PROCEDURE kdcnls (vector, minold, maxold, minnew, maxnew, dimens)
BEGIN

    /* initialization of the new space limits */

    FOR nudim = 1 TO dimens DO

        minnew(nudim) <- MIN(vector (nudim), minold(nudim))

        maxnew(nudim) <- MAX(vector (nudim), maxold(nudim))

    END

    /* computation of the diagonal of the new space */

    dold <- 0

    dnew <- kdspdg (minnew, maxnew, dimens)

    /* iterative search for space limits */

    WHILE (dnew<>dold) DO

        /* update of the space coordinates */

        FOR nudim = 1 TO dimens DO

            minnew(nudim) <- dnew*(FLOOR(minnew(nudim)/dnew))

            maxnew(nudim) <- dnew*(CEIL(maxnew(nudim)/dnew))

        END

        /* update of the diagonal space */

        dold <- dnew

        dnew <- kdspdg (minnew, maxnew, dimens)

    END

END

## 1.4. Extension of the tree up to the new limits of the space

FUNCTION kdetls (root, minold, maxold, minnew, maxnew, dimens, precis)

BEGIN

    /* initialization of the computing parameters */

    depth <- dimens*precis

    level <- 0

    /* recursive computing unit */

    FUNCTION kdetls (root [,minold, maxold], minnew ,maxnew[, dimens, precis], level [, depth])

    BEGIN

        /* position evaluation of the old space compared to the new one*/

        inclus <- true

        IF ((NOT kdinvs (minold, minnew, maxnew, dimens))

        OR (NOT kdinvs (maxold, minnew, maxnew, dimens))) THEN included <- false

        equal <- kdeqbs (minold, maxold, minnew, maxnew, dimens)

        IF ((NOT equal) AND (included)) THEN DO

            /* dividing the new space */

            nudim <- MOD(level, dimens)+1

            minlft <- *copy*(minnew)

            maxlft<- *copy* (maxnew)

            minrgt <- *copy* (minnew)

            maxrgt <- *copy* (maxnew)

            maxlft(nudim) <- (minnew(nudim)+maxnew(nudim))/2

            minrgt(nudim) <- (minnew(nudim)+maxnew(nudim))/2

            lftson <- kdetls (root, minlft, maxlft, level +1)

            rgtson <- kdetls (root, minrgt, maxrgt, level +1)

            kdetls <- *nodes union* (lftson, rgtson)

    END

```
            ELSE DO
                IF (equal) THEN DO
                    /* the old space has been localized in the new space:
                    hanging the tree linked to the current node */
                    kdetls <- root
                END
                ELSE kdetls <- tree(white)
            END
            delete vectors(minold, maxold)
    END
    /* pruning the new tree at the computing precision */
    root <- kdass (kdetls, dimens, precis)
    delete tree (kdetls)
    kdetls <- root
END
```

## 1.5. Inclusion test of the vector in the space

```
FUNCTION kdinvs (vector, minspc, maxspc, dimens)
BEGIN
    kdinvs <- true
    FOR nudim = 1 TO dimens DO
        IF ((vector (nudim) < minspc (nudim)) OR (vector (nudim) > maxspc (nudim)))
        THEN kdinvs <- false
    END
END
```

## 1.6. Equality test between a block and the space

```
FUNCTION kdeqbs (minblo, maxblo ,minspc, maxspc, dimens)
BEGIN
    kdeqbs <- true
    FOR nudim = 1 TO dimens DO
        IF ((minblo(nudim)<> minspc (nudim))OR (maxblo(nudim)<> maxspc (nudim)))
        THEN kdeqbs <- false
    END
END
```

## 1.7. Computation of the space diagonal

```
FUNCTION kdspdg (minspc, maxspc, dimens)
  BEGIN
    kdspdg <- 0
    FOR nudim = 1 TO dimens DO
        IF (|maxspc (nudim) - minspc (nudim)| > kdspdg)
        THEN kdspdg <- | maxspc (nudim)- minspc (nudim)|
      END
    kdspdg <- 2**(CEIL(LOG2(kdspdg)))
END
```

## 1.8. Union of two trees in inductive limit

PROCEDURE kdunil (root1, min1, max1, root2, min2, max2, root, minnew, maxnew, dimens, precis)

BEGIN

    /* copy of operand trees */

    croot1 <- *copy* (root1)

    croot2 <- *copy* (root2)

    /* equality test of the two spaces */

    IF (kdeqbs (min1, max1, min2, max2, dimens) THEN_DO

        minnew <- min1

        maxnew <- max1

    END
    ELSE DO

        IF (kdinvs (min1, min2, max2, dimens) AND kdinvs (max1, min2, max2, dimens))

        THEN DO

            /* first space included in the second one */

            minnew <- min2

            maxnew <- max2

            root1 <- kdetls(croot1, min1, max1, minnew, maxnew, dimens, precis)

        END
        ELSE DO

            IF (kdinvs (min2, min1, max1, dimens) AND kdinvs (max2, min1, max1, dimens)) THEN DO

                /* second space included in the first one */

                minnew <- min1

                rnaxnew <- max1

                root2 <- kdetls (croot2, min2, max2, minnew, maxnew, dimens, precis)

            END

            ELSE DO

                /* no inclusion */

                /* compute new space limits */

                CALL kdcnew (min1, max1, min2, max2, minnew, maxnew, dimens)

root1 <- kdetls (croot1 , min1, max1, minnew, maxnew, dimens, precis)

                    root2 <- kdetls (croot2, min2, max2, minnew, maxnew, dimens, precis)

            END

        END

END

/* union of the two resulting trees */

root <- kdunio(croot1, croot2, dimens, precis)

*delete tree* (croot1)

*delete tree* (croot2)

END

## 1.9. Intersection two trees in inductive limit

PROCEDURE kdinil (root1, min1, max1, root2, min2, max2, root, minnew, maxnew, dimens, precis)

    This procedure is identical to kdunil except that in place of kdunio is called kdintr.

## 1.10. Exclusion two trees in inductive limit

PROCEDURE kdexil (root1, min1, max1, root2, min2, max2, root, minnew, maxnew, dimens, precis)

    This procedure is identical to kdunil except that in place of kdunio is called kdexcl.

## 1.11. Difference two trees in inductive limit

PROCEDURE kddfil (root1, min1, max1, root2, min2, max2, root, minnew, maxnew, dimens, precis)

    This procedure is identical to kdunil except that in place of kdunio is called kddiff.

## 1.12. Computation of the new space limits

PROCEDURE kdcnew (min1, max1, min2, max2, minnew, maxnew, dimens)

BEGIN

    /* initialization of the new space limits* /

    FOR nudim = 1 TO dimens DO

        rninnew (nudirn) <- MIN (min1(nudirn), min2(nudirn))

        maxnew (nudim) <- MAX (max1(nudim), max2(nudim))

    END

    /* computation of the new space diagonal */

    dold <- 0

    dnew <- kdspdg (minnew, maxnew, dimens)

    /* iterative search for space limits */

    WHILE (dnew<>dold) DO

        FOR nudim = 1 TO dimens DO

            minnew (nudim) <- dnew * (FLOOR (rninnew(nudim)/dnew))

            maxnew (nudim) <- dnew * (CEIL (maxnew(nudim)/dnew))

        FIN

        /* diagonal update */

        dold <- dnew

        dnew <- kdspdg (minnew, maxnew, dimens)

    END

END

## 2. Set integration and differentiation

root :              root of the tree to be processed

root1, root2 :      roots of the sub-trees to be compared

dimens :            dimension of the space

dimgen :            dimension of generation

precis :            computing precision

level :             level reached in the tree

depth :             computing depth

roogen :            node leading to the regeneration

axisnb :            axis number

## 2.1. Computation of the hypograph of a tree along one of the dimensions of the modeling space

```
PROCEDURE kdhypg (root, dimgen, dimens, precis)
BEGIN
    /* initialization of the generation */
    level <- 0
    /* recursive computing unit */
    PROCEDURE kdhypg (root [, dimgen], level)
    BEGIN
        /* looking for node couples for initializing the generation* /
        IF (NOT terminal(root)) THEN DO
            IF ((level+1)<>dimgen) THEN DO
                /* descending and searching for initial couples */
                CALL kdhypg (left son (root), level+1)
                CALL kdhypg (right son(root), level+1)
            END
            ELSE DO
                /* generation of the graph along the requested dimension */
                CALL kdhypd (left son(root ), right son(root), dimgen , dimens, precis )
            END
        END
        merge (root)
    END
END
```

## 2.2. Tree traversal with regeneration of minimal nodes along the generation direction

```
PROCEDURE kdhypd(root1, root2, dimgen, dimens, precis)
BEGIN
    /* computing parameter initialization */
    depth <- precis*dimens
    level <- dimgen
    /* recursive computing unit */
    PROCEDURE kdhypd (root1, root2[ ,dimgen, dimens], level)
    BEGIN
        IF ((level<>depth) AND ((NOT white (root1)) OR (NOT white (root2))))
        THEN DO
            IF (terminal (root1)) THEN divide (root1)
            IF (terminal (root2)) THEN divide (root2)
            IF ((MOD (level, dimens)+1)<>dimgen) THEN DO
                /* direction orthogonal to the generation direction */
                CALL kdhypd (left son (root1), left son (root2), level+1)
                CALL kdhypd (right son (root1), right son (root2), level+1)
            END
            ELSE DO
                /* direction parallel to the generation direction */
                /* sub-trees descent and regeneration of minimal nodes */
                roogen <- right son (root2)
                IF (NOT white (roogen)) THEN DO
                    IF (NOT black (left son (root2)))
                    THEN CALL kdhypd(left son(root2), roogen, level+1)
                    IF (NOT black (right son (root1)))
```

```
                              THEN CALL kdhypd(right son(root1), roogen, level+l)
                              IF (NOT black (left son (root1)))
                              THEN CALL kdhypd(left son(root1), roogen, level+l)
                    END
                    roogen <- left son(root2)
                    IF (NOT white (roogen)) THEN DO
                              IF (NOT black (right son (root1)))
                              THEN CALL kdhypd(right son(root1), roogen, level+1)
                              IF (NOT black (left son (root1)))
                              THEN CALL kdhypd(right son(root1), roogen, level+1)
                    END
                    roogen <- right son(root1)
                    IF (NOT white (roogen)) THEN DO
                              IF (NOT black (left son(root1)))
                              THEN CALL kdhypd(left son(root1), roogen, level+1)
                    END
               END
          END
          ELSE DO
               /* regeneration of the hidden node */
               IF (NOT black (root2)) THEN blacken (root1)
          END
          /* sub-trees merge on traversal return */
          merge (root1)
          merge (root2)
     END
END
```

## 2.3. Computation of the epigraph of a tree along one of the directions of the modeling space

PROCEDURE kdepig(root, dimgen, dimens, precis)

     This procedure is identical to kdhypg, except that in place of kdhypd is called the procedure kdepid.

## 2.4. *Tree traversal with regeneration of maximal nodes along the generation direction*

PROCEDURE kdepid(root1, root2, dimgen, dimens precis)

BEGIN

    /* traversal initialization */

    depth <- precis*dimens

    level <- dimgen

    /* recursive computing unit  */

    PROCEDURE kdepid (root1, root2[ ,dimgen, dimens], level)

    BEGIN

        IF ((level<>depth) AND ((NOT *white* (root1)) OR (NOT *white* (root2))))

        THEN DO

            IF (*terminal* (root1)) THEN *divide* (root1)

            IF (*terminal* (root2)) THEN *divide* (root2)

            IF ((MOD (level, dimens)+1<>dimgen) THEN DO

                /* direction orthogonal to the generation direction */

                CALL kdepid (*left son* (root1), *left son* (root2), level+1)

                CALL kdepid (*right son* (root1), *right son* (root2), level+1)

            END

        ELSE DO

            /* direction parallel to the generation direction */

            /* sub-trees descent and regeneration of maximal nodes */

            roogen <- *left son* (root1)

            IF (NOT *white* (roogen)) THEN DO

                IF (NOT *black* (*right son* (root1)))

                THEN CALL kdepid(roogen, *right son*(root1), level+1)

                IF (NOT *black*(*left son*(root2)))

                THEN CALL kdepid(roogen, *left son*(root2), level+1)

                IF (NOT *black*(*right son*(root2)))

```
                        THEN CALL kdepid(roogen, right son(root2), level+1)
                END
            roogen <- right son(root1)
            IF (NOT white (roogen)) THEN DO
                    IF (NOT black(left son(root2)))
                    THEN CALL kdepid(roogen, left son(root2), level+1)
                    IF (NOT black(right son(root2)))
                    THEN CALL kdepid(roogen, right son(root2), level+1)
            END
            roogen <- left son(root2)
            IF (NOT white (roogen)) THEN DO
                    IF (NOT black(right son(root2)))
                    THEN CALL kdepid(roogen, right son(root2), level+1)
            END
        END
    END
    ELSE DO
        /* regeneration of the hidden node */
        IF (NOT black (root1)) THEN blacken (root2)
    END
    /* sub-trees merge on traversal return */
    merge (root1)
    merge (root2)
  END
END
```

## 2.5. Filling the boundary of an object

FUNCTION kdfill (root, dimens, precis)

BEGIN

/* computation of the intersection of hypographs and epigraphs along all the directions of the space */

kdremp <- *tree*(black)

FOR axisnb=1 TO dimens DO

    root1 <- kdremp

    root2 <- kdass (root, dimens, precis)

    CALL kdhypg (root2, axisnb, dimens, precis)

    kdremp <- kdinter (root1, root2, dimens, precis)

    *delete tree*(root1)

    *delete tree* (root2)

    root1 <- kdremp

    root2 <- kdass (root, dimens, precis)

    CALL kdepig(root2, axisnb, dimens, precis)

    kdremp <- kdinter (root1, root2, dimens, precis)

    *delete tree* (root1)

    *delete tree* (root2)

END

END

# 3. Convex hull of a set

| | |
|---|---|
| root : | root of the tree to be processed |
| root1, root2 : | roots of convex sub-hulls |
| dimens : | space dimension |
| precis : | computing precision |
| level : | level reached in the tree |
| depth : | computing depth |
| minblo, maxblo : | lower and higher faces of a block |
| minlft, maxlft: | left son faces |
| minrgt, maxrgt : | right son faces |
| veclft, vecrgt : | block centers associated to the root sons |
| diametr : | diameter of a block |
| cvxseghd: | convex segment list |
| xmin, xmax | segment origin and end |
| dimsep : | dimension of separation of lower and higher hulls |
| hultyp : | low or high type of a hull |
| symvec : | symmetry vector |
| symtyp : | symmetry type associated to an axis |
| vectr1, vectr2 : | centers of blocks |
| lgbloc : | side length of a block |
| insgbl : | inclusion indicator of a segment in a block |
| inters : | intersection indicator between a segment and a block |

## 3.1. Computation of the convex hull of a set

PROCEDURE kdcvxh (root, dimens, precis)

BEGIN

    /* intialization of computing parameters */

    depth <- dimens*precis,  level <- 0

    minblo <- *vector*(0,0,...,0)

    maxblo <- *vector*(1,1,...,1)

    /* recursive computing unit */

    PROCEDURE kdcvxh (root, minblo, maxblo[, dimens],level[, depth])

    BEGIN

        IF ((NOT *terminal*(root) AND (level<>depth)) THEN DO

            CALL divide a block by halves (minblo, maxblo, minlft, maxlft, minrgt, maxrgt)

            CALL kdcvxh (*left son*(root), *link*(minlft), *link*(maxlft), level+1)

            CALL kdcvxh (*right son*(root), *link*(minrgt), *link*(maxrgt), level+1)

        END

        /* black terminal nodes are convex */

        /* ascending building of the convex hull */

        IF (NOT *terminal*(root)) THEN DO

            IF ((NOT *white*(*left son*(root))) AND (NOT *white*(*right son*(root))))

            THEN DO

                /* finding the lower and higher hulls orthogonally to the dividing direction */

                CALL hull search initialization (*left son*(root), <u>inf</u>, level+1 ,depth)

                CALL hull search initialization (*right son*(root), <u>sup</u>, level+1, depth)

                /* computation of the covering of the two hulls by convex segments */

                CALL center and diameter computation of the root sons (minblo, maxblo, veclft, vecrgt, diametr, dimens)

                cvxseghd <- computation of the hull covering (*left son*(root), *right son*(root), veclft, vecrgt, diametr, level+1, depth)

                *delete vectors* (veclft, vecrgt, diametr)

                /* digitalization of the list of convex segments */

     WHILE (NOT *empty queue* (cvxseghd)) DO

      xmin <- *extraction from a vector* (cvxseghd)

      xmax <- *extraction from a vector* (cvxseghd)

      CALL segment digitalization (root, minblo, maxblo, xmin, xmax, dimens, level, depth)

     END

    END

   END

   *delete vectors* (minblo, maxblo)

  END

END

Multidimensional Hierarchical Modeling : Tome 2        Page 143

## 3.2. Search initialization of lower and higher hulls according to a given direction

PROCEDURE kdieis (root, hultyp, dimsep, dimens, level, depth)
BEGIN

    /* initialization of computing parameters */

    symvec <- *vector* (<u>sym</u>, <u>neutral</u>, ..., <u>neutral</u>)

    /* recursive computing unit */

    PROCEDURE kdieis (root, [hultyp, dimsep, dimens,] level[,depth])

    BEGIN

        IF ((NOT *terminal* (root)) AND (level<>depth)) THEN DO

            IF ((MOD (level, dimens) + 1) = dimsep)

            THEN symsep <- <u>true</u>

            ELSE symsep <- <u>false</u>

            CALL kdenis (*left son* (root), *right son* (root), hultyp, symsep, *link* (symvec), level+1, depth)

            /* depth first tree traversal*/

            CALL kdieis (*left son* (root), level+1)

            CALL kdieis (*right son* (root), level+1)

            /* marking no terminal nodes on traversal return*/

            IF ((*value* (*left son* (root))=hultyp) OR (*value*(*right son*(root))=hultyp)) THEN *value*(root) <- hultyp

        END

    END

    *delete* (symvec)

END

## 3.3. Looking for points belonging to the lower or higher hull of a set, orthogonally to a space direction

PROCEDURE kdenis (root1, root2, hultyp, symsep, symvec, level, depth)

BEGIN

    /* recursive computing unit */

    PROCEDURE kdenis (root1, root2[, hultyp, symep,]  symvec, level[, depth])

    BEGIN

      IF ((*link* (root1)<>*link*(root2)) AND (level<>depth) THEN DO

        /* tree down traversal looking for $d_1$-adjacencies */

        IF (*black* (root1)) THEN *divide* (root1)

        IF (*black* (root2)) THEN *divide* (root2)

        symtyp <- *value* (symvec)

        IF (symtyp = <u>neutral</u>) THEN DO

          /* descent orthogonally to the symmetry axis */

          CALL kdenis (*left son (*root1), *left son* (root2), *link* (symvec) , level+1)

          CALL kdenis (*right son* (root1), *right son* (root2), *link* (symvec), level+1)

        ELSE

          /* descent parallel to the symmetry axis */

          CALL kdenis (*right son* (root1), *left son* (root2), *link* (vecsvm), level+1)

      END

      /* marking non terminal nodes on traversal return */

      IF (NOT *terminal* (root1)) THEN

        IF ((*value* (*left son* (root1)=hultyp) OR (*value* (*right son*(root1)=hultyp)) THEN *value*(root1) <- hultyp

      IF (NOT *terminal* (root2)) THEN

        IF ((*value* (*left son* (root2))=hultyp) OR (*value*(*right son*(root2))=hultyp)) THEN *value*(root2) <- hultyp

    END

```
            ELSE DO
                /* precision reached: identification of nodes belonging to the hull */
                IF (link (root1)<>link(root2)) THEN DO
                    /* one of the points is at the boundary*/
                    IF ((NOT white (root1)) AND (white (root2))) THEN
                        IF ((symsep AND (hultyp=sup) OR (NOT symsep))
                        THEN value (rooti) <- hultyp
                    IF ((white (root1)) AND (NOT white (root2))) THEN
                        IF ((symsep AND (hultyp=inf) OR (NOT symsep))
                        THEN value (root2) <- hultyp
                END
            END
        END
END
```

## 3.4. Computation of a covering of lower and higher hulls by convex segments

FUNCTION kdrcvx (root1, root2, vectr1, vectr2, lgbloc, level, depth)

BEGIN

    /* recursive computing unit */

    FUNCTION kdrcvx (root1, root2, vectr1, vectr2, lgbloc, level[ ,depth])

    BEGIN

      IF (((NOT *terminal* (root1)) OR (NOT *terminal* (root2))) AND (level<>depth))

      THEN DO

        /* parallel descent of the two sub-trees towards the lower and higher hulls */

        index <- 0

        IF (*value* (*left son* (root1)) = inf) THEN index <- +1

        IF (value (*right son* (root1)) = inf) THEN index <- +2

        IF (*value* (*left son* (root2)) = sup) THEN index <- +4

        IF (*value* (*right son* (root2)) = sup) THEN index <- +8

        ACCPORDING TO index DO :

          5 : BEGIN

            *value* (vectr1) <- -*value*(lgbloc)

            *value* (vectr2) <- -*value*(lgbloc)

            *value* (lgbloc) <- *value*(lgbloc)/2

            kdrcvx <- kdrcvx (*left son*(root1), *left son*(root2), *link*(vectr1), *link*(vectr2), *link*(lgbloc), level+1)

            *value* (lgbloc) <- *value*(lgbloc)*2

            *value* (vectr2) <- +*value*(lgbloc)

            *value* (vectr1) <- +*value*(lgbloc)

          END

          6, 9,10 : /* same principle */

          7 : BEGIN

            *value* (vectr1) <- -*value*(lgbloc)

     *value* (vectr2) <- -*value*(lgbloc)

     *value* (lgbloc) <- -*value*(lgbloc)/2

     kdrcvx1 <- kdrcvx (*left son*(root1), *left son*(root2), *link*(vectr1), *link*(vectr2), *link*(lgbloc), level+1)

     value (vectr1) <- +value(lgbloc)*4

     kdrcvx2 <- kdrcvx (*left son*(root1), *left son*(root2), *link*(vectr1), *link*(vectr2), *link*(lgbloc), level+1)

     *value* (lgbloc) <- *value*(lgbloc)*2

     *value* (vectr2) <- +*value*(lgbloc)

     *value* (vectr1) <- -*value*(lgbloc)

     kdrcvx <- *concantenate* (kdrcvx1, kdrcvx2)

    END

    11, 13, 14, 15:/* same principle */

   END

  END

  ELSE DO

    /* precision reached: creation of a convex segment */

    xinf <- *copy* (vectr1)

    xsup <- *copy* (vectr2)

    kdrcvx <- *concantenate* (xinf, xsup)

  END

 END

END

## 3.5. Digitalization of a convex segment

PROCEDURE kddcvx (root, minblo, maxblo, xmin, xmax, dimens, level, depth)
BEGIN

    /* recursive computing unit */

    PROCEDURE kddcvx (root, minblo, maxblo, xmin, xmax[, dimens], level[, depth])
    BEGIN

        /* evaluation of the segment position compared to a block */

        CALL kdposb(insgbl, inters, xmin, xmax, minblo, maxblo, dimens)

        IF (inters AND (level <> depth)) THEN DO

            IF (insgbl) THEN DO

                /* segment included in the block */

                IF (NOT *black* (root)) THEN DO

                    /* tree traversal looking for white nodes */

                    IF (*terminal* (root)) THEN *divide* (root)

                    CALL kddivb (minblo, maxblo, minlft, maxlft, minrgt, maxrgt)

                    CALL kddevx (*left son*(root), minlft, maxlft, *link*(xmin), *link*(xmax), level+1)

                    CALL kddcvx (*right son*(root), minrgt, maxrgt, *link*(xmin), *link*(xmax), level+1)

                END

            ELSE

                /* once more node examination after segment division */

                CALL kddivs (xmin, xmax, xmin1, xmin2, xmax2, dimens)

                CALL kddcvx (root, minblo, maxblo, *link*(xmin1), *link*(xmax1), level)

                CALL kddcvx (root ,minblo, maxblo, *link*(xmin2), *link*(xmax2), level)

            END

        END

```
            ELSE
                    /* precision reached: the block is blackened */
                    IF (inters) THEN DO
                            IF (NOT terminal (root)) THEN  delete tree (root)
                            root <- black
                    END
            END
            /* marking erase of nodes belonging to the hulls */
            value (root) <- nil
            IF (NOT terminal (root)) THEN merge (root)
    END
END
```

## 3.6. Position evaluation of a segment compared to a block

PROCEDURE kdposb (insgbl, inters, xmin, xmax, minblo, maxblo, dimens)
BEGIN

    /* inclusion tests of the two segment ends */

    inxmin <- <u>true</u>,          inxmax <- <u>true</u>

    hdxmin <- xmin,          hdxmax <- xmax

    hdmnbl <- minblo,         hdmxbl <- maxblo

    FOR nudim = 1 TO dimens DO

        IF ((*value*(hdxmin) < *value*(hdmnbl)) OR (*value*(hdxmin) > *value*(hdmxbl))) THEN inxmin <- <u>false</u>

        IF ((*value*(hdxmax) < *value*(hdmnbl)) OR (*value*(hdxmax) > *value*(hdmxbl))) THEN inxmax <- <u>false</u>

        hdxmin <- *link*(hdxmin), hdxmax <- *link(*hdxmax)

        hdminbl <- *link*(hdmnbl), hdmxbl <- *link*(hdmxbl)

    END

    /* evaluation of the segment inclusion in the block */

    IF ((inxmin) AND (inxmax)) THEN insgbl <- <u>true</u> ELSE insgbl <- <u>false</u>

    /* evaluation of the segment intersection with the block */

    IF ((inxmin) OR (inxmax)) THEN inters <- <u>true</u> ELSE inters <- <u>false</u>

END

## 3.7. Computation of the centers and the diameters of the root sons

PROCEDURE kdcdfr (minblo, maxblo, veclft, vecrgt, diametr, dimens)

BEGIN

    hdmin <- minblo, hdmax <- maxblo

    diametr <- *create a vector*

    veclft <- *create a vector*

    vecrgt <- *create a vector*

    FOR nudim = 1 TO dimens DO

        *insert at queue end* (diametr, ( *value*(hdmax) - *value*(hdmin)))

        *insert at queue end* (velft, (*value*(hdmin) + *value*(hdmax))/2)

        *insert at queue end* (vecrgt, (*value*(hdmin) + *value*(hdmax))/2)

    END

    larg <- *extract from queue head* (diametr)

    larg <- larg/2

    *insert at queue end* (diametr, larg)

    x <- *extract from queue head* (veclft)

    *insert at queue end* (veclft, (x - larg))

    x <- *extract from queue head* (vecrgt)

    *insert at queue end* (vecrgt, (x + larg))

END

## *3.8. Division of a block by halves*

PROCEDURE kddivb (minblo, maxblo, minlft, maxlft, minrgt, maxrgt)

BEGIN

    minlft <- *copy* (minblo)

    maxlft <- *copy* (maxblo)

    minrgt <- *copy* (minblo)

    maxrgt <- *copy* (maxblo)

    /* creation of median planes according to the current direction*/

    *value*(maxlft) <- (*value*(minblo) + *value*(maxblo))/2

    *value*(minrgt) <- (*value*(minblo) + *value*(maxblo))/2

END

## *3.9. Division of a segment by halves*

PROCEDURE kddivs (xmin, xmax, xmin1, xmax1, xmin2, xmax2, dimens)

BEGIN

    xmin1 <- *create a vector*,    xmax1 <- *create a vector*

    xmin2 <- *create a vector*,    xmax2 <- *create a vector*

    hdxmin <- xmin, hdxmax <- xmax

    FOR nudim = 1 TO dimens DO

        *insert at queue end* (xmin1, value(hdxmin))

        *insert at queue end* (xmax1, (*value*(hdxmin) + *value*(hdxmax)/2)

        *insert at queue end* (xmin2, (*value*(hdxmin) + *value*(hdxmax)/2)

        *insert at queue end* (xmax2, *value*(hdxmax))

        hdxmin <- *link* (hdxmin), hdxmax <- *link* (hdxmax)

    END

END

# 4. Adjacency analysis

root :                  root of the tree to be processed

root1, root2 :          symmetrical nodes in the tree

dimens :                space dimension

precis :                computing precision

depth :                 computing depth

level :                 level reached in the tree

symvec :                symmetry vector

symvec2 :               copy of a symmetry vector

typesym :               type of symmetry associated to the axis

## 4.1. Search for $d_1$-adjacencies in space objects

```
PROCEDURE kd1anr(root, dimens, precis)
BEGIN
    /* search initialization */
    depth <- dimens*precis
    level <- 0
    /* recursive computing unit */
    PROCEDURE kd1anr (root, level[ ,depth])
    BEGIN
        IF ((NOT terminal (root)) AND (level<>depth)) THEN DO
            /* search for adjacencies stemmed from the reached non terminal block */
            symvec <- vector (sym, neutral, ..., neutral)
            CALL kd1asn (left son (root), right son (root), link (symvec), level+1, depth)
            /* depth-first tree traversal*/
            CALL kd1anr (left son (root), level+1)
            CALL kd1anr (right son (root), level+1)
            delete vector (symvec)
        END
        tree valuation (root)
    END
END
```

## 4.2. Search for $d_1$-adjacencies according to a given symmetry vector

PROCEDURE kd1asn(root1, root2, symvec, level , depth)

BEGIN

    /* recursive computing unit */

    PROCEDURE kd1asn(root1, root2, symvec, level[, depth])

    BEGIN

        IF (((NOT *terminal* (root1)) OR (NOT *terminal* (root2))) AND (level<>depth))

        THEN DO

            typesym <- *value*(symvec)

            IF (typesym=<u>neutral</u>) THEN DO

                /* descent orthogonally to the symmetry axis */

                CALL kd1asn (*left son* (root1), *left son* (root2), *link* (symvec), level+1)

                CALL kd1asn (*right son* (root1), *right son* (root2), *link* (symvec), level+1)

            END

            ELSE DO

                /* descent parallel to the symmetry axis */

                CALL kd1asn (*right son* (root1), *left son* (root2), *link* (symvec), level+1)

            END

        END

        ELSE DO

            /* precision reached: generation of the adjacencies for the nodes belonging to space objects */

            IF ((NOT white (root1)) AND (NOT white (root2))) THEN DO

                IF (*nil* (*value* (root1))) THEN *value* (root1) <- *creation of an adjacency list*

                *register the adjacency* (*value*(root1), root2)

                IF (*nil* (*value* (root2))) THEN *value* (root2) <- *creation of an adjacency list*

                *register the adjacency* (*value*(root2), root1)

            END

        END

END

END

## 4.3. Search for $d_\infty$-adjacencies in space objects

```
PROCEDURE kd0anr(root, dimens, precis)
BEGIN
    /* search initialization */
    depth <- dimens*precis
    level <- 0
    /* recursive computing unit */
    PROCEDURE kd0anr (root, level[ ,depth])
    BEGIN
        IF ((NOT terminal (root)) AND (level<>depth)) THEN DO
            /* search for adjacencies stemmed from the reached non terminal block */
            symvec <- vector (sym, neutral, ..., neutral)
            CALL kd0asn (left son (root), right son (root), link (symvec), level+1, depth)
            /* depth-first tree traversal */
            CALL kd0anr (left son (root), level+1)
            CALL kd0anr (right son (root), level+1)
            delete vector (symvec)
        END
        tree valuation (root)
    END
END
```

## 4.4. Search for $d_\infty$-adjacencies according to a given symmetry vector

PROCEDURE kd0asn(root1, root2, symvec, level, depth)

BEGIN

    /* recursive computing unit */

    PROCEDURE kd0asn(root1, root2, symvec, level[, depth])

    BEGIN

        IF (((NOT *terminal* (root1)) OR (NOT *terminal* (root2))) AND (level<>depth))

        THEN DO

            symvec2 <- *copy* (symvec)

            typesym <- *value* (symvec)

            IF (typesym=<u>neutral</u>) THEN DO

                /* descent orthogonally to the symmetry axis */

                CALL kd0asn (*left son* (root1), *left son* (root2), *link* (symvec2), level+1)

                CALL kd0asn (*right son* (root1), *right son* (root2), *link* (symvec2), level+1)

                /* creation of crossed symmetries */

                *value* (symvec2) <- <u>sym</u>

                CALL kd0asn (*left son* (root1), *right son* (root2), *link* (symvec2), level+1)

                *value* (symvec2) <- <u>antisym</u>

                CALL kd0asn (*right son* (root1), *left son* (root2), *link* (symvec2), level+1)

            END

            /* descent parallel to the symmetry axis */

            IF (typesym = <u>sym</u>)

            THEN CALL kd0asn (*right son* (root1), *left son* (root2), *link* (symvec2), level+1)

            IF (typesym = <u>antisym</u>)

            THEN CALL kd0asn (*left son* (root1), *right son* (root2), *link* (symvec2), level+1)

            *delete vector*(symvec2)

    END

          ELSE DO

               /* precision reached: generation of the adjacencies for the nodes belonging to space objects */

               IF ((NOT *white* (root1)) AND (NOT *white* (root2))) THEN DO

                   IF (*nil* (*value* (root1))) THEN *value* (root1) <- *creation of an adjacency list*

                   *register the adjacency* (*value*(root1), root2)

                   IF (*nil* (*value* (root2))) THEN *value* (root2) <- *creation of an adjacency list*

                   *register the adjacency* (*value*(root2), root1)

               END

          END

     END

END

# 5. Homotopic transforms

| | |
|---|---|
| root : | root of the tree to be processed |
| root1, root2 : | symmetrical nodes in the tree |
| dimens : | space dimension |
| precis : | computing precision |
| depth : | computing depth |
| level : | level reached in the tree |
| symvec : | symmetry vector |
| symvec2 : | copy of the symmetry vector |
| typesym : | type of symmetry associated to the axis |

## *5.1. Whitening the faces of the unitary hypercube*

```
PROCEDURE kdsclo (root, dimens, precis)
BEGIN
    depth <- dimens*precis
    level <- 0
    symvec <- vector (sym, neutral, ..., neutral)
    FOR nudim=1 TO dimens DO
        CALL kdwhfc (root, left, symvec, level, depth)
        CALL kdwhfc (root, right, symvec, level, depth)
        rotate vector (symvec)
    END
    delete vector (symvec)
END
```

## 5.2. Whitening the points belonging to one of the faces of the unitary hypercube

```
PROCEDURE kdwhfc (root, side, symvec, level, depth)
BEGIN
    / * recursive computing unit* /
    PROCEDURE kdwhfc (root[ , side],  symvec, level[ , depth])
    BEGIN
        IF ((NOT white (root)) AND (level<>depth)) THEN DO
            /* descent for looking for faces to be whitened */
            IF (terminal (root)) THEN divide (root)
            typesym <- value(symvec)
            IF (typesym=neutral) THEN DO
                /* descent orthogonally to the face normal */
                CALL kdwhfc (left son (root), link (symvec), level+1)
                CALL kdwhfc (right son (root), link (symvec), level+1)
            END
            ELSE DO
                /* descent parallel to the face normal */
                IF (side=left)
                THEN CALL kdwhfc (left son (root), link (symvec), level+1)
                ELSE CALL kdwhfc (right son (root), link (symvec), level+1)
            END
        END
        ELSE IF (level=depth) THEN DO
            IF (NOT terminal (root)) THEN delete tree (root)
            whiten (root)
        END
    END
END
```

## 5.3. Compute the $d_1$-boundary of a set

```
FUNCTION kd1bnd (root, dimens, precis)

BEGIN

    /* construction initialization */

    depth <- dimens*precis, level <- 0

    /* recursive computing unit */

    FUNCTION kd1bnd (root, level[, depth])

    BEGIN

        IF ((NOT terminal (root)) AND (level<>depth)) THEN DO

            /* looking for adjacencies in the space background* /

            symvec <- vector (sym, neutral, ..., neutral)

            CALL kd1frn (left son (root), right son (root), link (symvec), level+1, depth)

            /* depth-first tree traversal */

            lftson <- kd1bnd (left son (root), level+1)

            rgtson <- kd1bnd (right son (root), level+1)

            kd1bnd <- nodes union (lftson, rgtson)

            delete vector (symvec)

        END

        ELSE DO

            /* boundary computation */

            IF (value (root)=black) THEN kd1bnd <- tree (black) ELSE kd1bnd <- tree (white)

            value (root) <- nil

        END

        /* merge of sons' nodes on return */

        IF (NOT terminal (kd1bnd)) THEN merge (kd1bnd)

        IF (NOT terminal (root)) THEN merge (root)

    END

END
```

## 5.4. Search and marking of nodes $d_1$-boundary of the set

PROCEDURE kd1frn (root1, root2, symvec, level, depth)

BEGIN

/* recursive computing unit */

PROCEDURE kd1frn (root1, root2, symvec, level[ , depth])

BEGIN

IF (((NOT *terminal* (root1)) OR (NOT *terminal* (root2)))

AND (level<>depth)) THEN DO

/* descent and looking for boundary points */

IF (*black* (root1)) THEN *divide* (root1)

IF (*black* (root2)) THEN *divide* (root2)

symvec2 <- *copy* (symvec)

typesym <- *value* (symvec)

IF (typesym=<u>neutral</u>) THEN DO

/* descent orthogonally to the symmetry axis */

CALL kd1frn (*left son* (root1), *left son* (root2), *link* (symvec2), level+1)

CALL kd1frn (*right son* (root1), *right son* (root2), *link* (symvec2), level+1)

END

ELSE DO

/* descent parallel to the symmetry axis */

CALL kd1frn (*right son* (root1), *left son* (root2), *link* (symvec2), level+1)

END

*delete* (symvec2)

END

```
            ELSE DO
                /* precision reached: marking boundary nodes */
                IF (link (root1) <> link (root2)) THEN DO
                    IF ((NOT white (root1)) AND white (root2)) THEN value (root1) <- black
                    IF (white (root1) AND (NOT white (root2))) THEN value (root2) <- black
                END
            END
        END
END
```

## 5.5. Compute the $d_\infty$-boundary of a set

FUNCTION kd0bnd(root, dimens ,precis)

This function is identical to kd1bnd, except that in place of the procedure kd1frn, it is called the procedure kd0frn.

## 5.6. Search and marking of nodes $d_\infty$ - boundary of the set

PROCEDURE kd0frn(root1, root2, symvec, level, depth)

This procedure is identical to kd1frn, except that the decent for searching for boundary points is an adjacency search according to $d_\infty$ (cf. kd0asn).

## 5.7. Erosion according to $d_1$ of space objects

```
PROCEDURE kd1ero(root, dimens, precis)
BEGIN
    /* computation parameter initialization */
    depth <- dimens*precis
    level <- 0
    /* recursive computing unit */
    PROCEDURE kd1ero(root, level[, depth])
    BEGIN
        IF ((NOT terminal (root)) AND (level<>depth)) THEN DO
            /* looking for adjacencies with the space background */
            symvec <- vector (sym, neutral, ..., neutral)
            CALL kd1ern (left son (root), right son (root), link (symvec), level+1)
            /* depth-first tree traversal */
            CALL kd1ero (left son (root), level+1)
            CALL kd1ero (right son (root), level+1)
            delete vector (symvec)
        END
    END
    /* change the object background color */
    CALL kdccmq (root, level, depth)
END
```

## 5.8. Search and marking of $d_1$ - boundary nodes to be deleted

PROCEDURE kd1ern(root1, root2, symvec, level, depth)

BEGIN

    /* recursive computing unit */

    PROCEDURE kd1ern(root1, root2, symvec, level[, depth])

    BEGIN

        IF (((NOT *terminal* (root1)) OR (NOT *terminal* (root2)))

        AND (level<>depth)) THEN DO

            /* descent and looking for boundary points */

            IF (*black* (root1)) THEN *divide* (root1)

            IF (*black* (root2)) THEN *divide* (root2)

            symvec2 <- *copy* (symvec)

            typesym <- *value* (symvec)

            IF (typesym=<u>neutral</u>) THEN DO

                /* descent orthogonally to the symmetry axis */

                CALL kd1ern (*left son* (root1), *left son* ( root2), *link* (symvec2), level+1)

                CALL kd1ern (*right son* (root1), *right son* (root2), *link* (symvec2), level+1)

            END

            ELSE DO

                /* descent parallel to the symmetry axis */

                CALL kd1ern (*right son* (root1), *left son* (root2), *link* (symvec2), level+1)

            END

            *delete vector* (symvec2)

    END

```
            ELSE DO
                /*  precision reached: marking boundary nodes */
                IF (link (root1)<>link (root2)) THEN DO
                    IF ((NOT white (root1)) AND white (root2)) THEN value (root1) <- white
                    IF (white (root1) AND (NOT white (root2))) THEN value (root2) <- white
                END
            END
        END
    END
```

## 5.9.  Color change of marked nodes in a tree

```
PROCEDURE kdccmq(root, level, depth)
BEGIN
    /* recursive computing unit */
    PROCEDURE kdccmq(root, level[, depth])
    BEGIN
        IF ((NOT terminal (root)) AND (level<>depth)) THEN DO
            /* depth-first tree traversal */
            CALL kdccmq (left son (root), level+1)
            CALL kdccmq (right son (root), level+1)
        END
        /* color change on the recursive return */
        IF (value (root)<>nil) THEN DO
            IF (NOT terminal (root)) THEN delete tree (root)
            link (root) <- value (root)
            value (root) <- nil
        END
        /* merge of sons' nodes on return */
        IF (NOT terminal (root)) THEN merge (root)
    END
END
```

## 5.10. Erosion according to $d_\infty$ of space objects

PROCEDURE kd0ero(root, dimens, precis)

This procedure is identical to kd1ero, except that in place of the procedure kd1ern, is called the procedure kd0ern.

## 5.11. Search and marking $d_\infty$-boundary nodes to be deleted

PROCEDURE kd0ern(root1, root2, symvec, level, depth)

This procedure is identical to kd1ern, except that the descent for looking for boundary points is an adjacency search according to $d_\infty$ (cf. kd0asn).

## 5.12. Dilation according to $d_1$ of space objects

```
PROCEDURE kd1dil(root, dimens, precis)
BEGIN
    /* computer parameter initialization */
    depth <- dimens*precis
    level <- 0
    /* recursive computing unit */
    PROCEDURE kd1dil(root, level[, depth])
    BEGIN
        IF ((NOT terminal (root)) AND (level<>depth)) THEN DO
            /* looking for adjacencies with the space background */
            CALL kd1dib (left son (root), right son (root), link (symvec), level+1)
            /* depth-first tree traversal */
            CALL kd1dil (left son (root), level+1)
            CALL kd1dil (right son (root), level+1)
            delete vector (symvec)
        END
    END
    /* change the object exo-background color */
    CALL kdccmq (root, level, depth)
END
```

## 5.13. Search and marking of $d_1$-exo-boundary nodes to be created

PROCEDURE kd1dib(rootl, root2, symvec, level, depth)

BEGIN

    /* recursive computing unit */

    PROCEDURE kd1dib (root1, root2, symvec, level[, depth])

    BEGIN

        IF (((NOT *terminal* (root1)) OR (NOT *terminal* (root2))) AND (level<>depth)) THEN DO

            /* descent and looking for boundary points */

            IF (*white* (root1)) THEN *divide* (root1)

            IF (*white* (root2)) THEN *divide* (root2)

            symvec2 <- *copy* (symvec)

            typesym <- *value* (symvec)

            IF (typesym=<u>neutral</u>) <u>THEN</u> <u>DO</u>

                /* descent orthogonally to the symmetry axis */

                CALL kd1dib (*left son* (root1), *left son* (root2), *link (*symvec2), level+1)

                CALL kd1dib (*right son* (root1), *right son* (root2), *link* (symvec2), level+1)

            END

            ELSE DO

                /* descent parallel to the symmetry axis */

                CALL kd1dib (*right son* (root1), *left son* (root2), *link* (symvec2), level+1)

            END

            *delete vector* (symvec2)

    END

```
        ELSE DO
            /* precision reached: marking exo-boundary nodes */
            IF (link (root1)<>link (root2)) THEN DO
                IF ((NOT white (root1)) AND white (root2)) THEN value(root2) <- black
                IF (white (root1) AND (NOT white (root2))) THEN value (root1) <- black
            END
        END
    END
END
```

## 5.14. Dilation according to $d_\infty$ of space objects

PROCEDURE kd0dil(root, dimens, precis)

This procedure is identical to kd1dil, except that in place of the procedure kd1dib, is called the procedure kd0dib.

## 5.15. Search and marking $d_\infty$-exo-boundary nodes to be created

PROCEDURE kd0dib (root1, root2, symvec, level, depth)

This procedure is identical to kd1dib, except that the descent for looking for boundary points is an adjacency search according to $d_\infty$ (cf.kd0asn).

## 5.16. Opening according to $d_1$ of space objects

PROCEDURE kd1ouv(root, dimens, precis)

BEGIN

    CALL kd1ero(root, dimens, precis)

    CALL kd1dil(root, dimens, precis)

END

## 5.17. Closing according to $d_1$ of space objects

PROCEDURE kd1fer(root, dimens, precis)

BEGIN

    CALL kd1dil(root, dimens, precis)

    CALL kd1ero(root, dimens, precis)

END

## 5.18. Opening according $d_\infty$ of space objects

PROCEDURE kd0ouv(root, dimens, precis)

## 5.19. Closing according $d_\infty$ of space objects

PROCEDURE kd0fer(root, dimens, precis)

    These procedures are identical to kd1ouv and kd1fer, except that they are calling kd0ero and kd0dil, in place of kd1ero and kd1dil.

# 6. Median transforms

| | |
|---|---|
| root : | root of the tree to be processed |
| root1, root2 : | symmetrical nodes in the tree |
| dimens : | space dimension |
| precis : | computing precision |
| depth : | computing depth |
| level : | level reached in the tree |
| symvec : | symmetry vector |
| symvec2 : | copy of the symmetry vector |
| typesym : | type of symmetry associated to the axis |
| dimmed : | median manifold dimension |
| cnxdeg : | connectivity degree of a median set |
| axisnb : | space axis number |

## 6.1. Median filtering according to $d_1$ of space objects

PROCEDURE kd1mdf(root, dimens, precis)

BEGIN

/* splitting of the boundary and the exo-boundary */

CALL kd1fmr(root, dimens, precis)

/* marking the node to be filtered */

CALL kd1mnf(root, dimens, precis)

END

## 6.2. Splitting of the $d_1$-boundary and the $d_1$-exo-boundary of a set

PROCEDURE kd1mfr(root, dimens, precis)

BEGIN

/* computing parameter initialization */

level <- 0, depth <- dimens*precis

symvec <- *vector*(sym, neutral, ..., neutral)

/* recursive computing unit */

PROCEDURE kd1mfr(root, level[ ,depth])

BEGIN

IF ((NOT *terminal*(root)) AND (level <> depth))THEN DO

/* looking for and splitting d1-boundary and d1-exo-boundary nodes */

CALL kd1mcf (*left son* (root), *right son* (root), symvec, level+1)

/* depth-first tree traversal */

CALL kd1mfr (*left son* (root), level+1)

CALL kd1mfr (*right son* (root), level+1)

END

END

*delete vector* (symvec)

END

## 6.3. Search and splitting $d_1$-boundary et $d_1$-exo-boundary nodes of the set

```
PROCEDURE kd1mcf(root1, root2, symvec, level, depth)
BEGIN
    /*  recursive computing unit */
    PROCEDURE kd1mcf (root1, root2, symvec, level[ , depth])
    BEGIN
        IF ((link (root1) <> link (root2)) AND (level <> depth)) THEN DO
            IF (terminal (root1)) THEN divide (root1)
            IF (terminal (root2)) THEN divide (root2)
            symvec2 <- copy (symvec)
            typesym <- value (symvec)
            IF (typesym = neutral) THEN DO
                /* descent orthogonally to the symmetry axis */
                CALL kd1mcf (left son (root1), left son(root2), symvec2, level+1)
                CALL kd1mcf (right son (root1), right son (root2), symvec2, level+1)
            END
            IF (typesym = sym) THEN DO
                /* descent parallel to the symmetry axis */
                CALL kd1mcf (right son (root1), left son (root2), symvec2, level+1)
            END
            delete vector (symvec2)
        END
    END
END
```

## 6.4. Marking nodes to be modified by median filtering according to $d_1$

PROCEDURE kd1mnf(root, dimens, precis)

BEGIN

    /* computing parameter initialization */

    level <- 0

    depth <- dimens * precis

    symvec <- *vector* (<u>sym</u>, <u>neutral</u>, ..., <u>neutral</u>)

    /* recursive computing unit */

    PROCEDURE kd1mnf(root, level[, depth])

    BEGIN

        IF ((NOT *terminal*(root)) AND (level <> depth)) THEN DO

            /* non terminal block reached */

            CALL kd1fmg (*left son* (root), *right son* (root), symvec, level + 1)

            /* depth-first tree traversal */

            CALL kd1mnf (*left son* (root), level + 1)

            CALL kd1mnf (*right son* (root), level + 1)

        END

    END

    /* modify the color of the object boundary */

    CALL kdccfm(root, level, depth)

    *delete vector* (symvec)

END

## 6.5. Search and counting $d_1$-adjacencies among the nodes to be filtered

```
PROCEDURE kd1fmg(root1, root2, symvec, level, depth)
BEGIN
    /* recursive computing unit */
    PROCEDURE kd1fmg (root1, root2, symvec, level[, depth])
    BEGIN
        IF (((NOT terminal (root1)) OR (NOT terminal (root2))) AND (level <> depth))
        THEN DO
            symvec2 <- copy (symvec)
            typesym <- value (symvec)
            IF (typesym = neutral) THEN DO
                /* descent orthogonally to the symmetry axis */
                CALL, kd1fmg (left son (root1), left son (root2), symvec, level+1)
                CALL kd1fmg (right son (root1), right son (root2) , symvec, level+1)
            END
            IF (typesym = sym) THEN DO
                /* descent parallel to the symmetry axis */
                CALL kd1fmg (right son (root1), left son (root2), symvec, level+1)
            END
            delete vector (symvec2)
        END
        ELSE DO
            /* precision reached: counting colored adjacencies */
            IF (white (root2)) THEN value (root1) <- value (root1) - 1
            ELSE value (root1) <- value (root1) + 1
            IF (white (root1)) THEN value (root2) <- value (root2) - 1
            ELSE value (root2) <- value (root2) + 1
        END
    END
END
```

## 6.6. Modifying the color of the boundary and the exo-boundary, according to the majority color of their neighbors

```
PROCEDURE kdccfm(root, level, depth)
BEGIN
    /* recursive computing unit */
    PROCEDURE kdccfm(root, level[, depth])
    BEGIN
        IF ((NOT terminal (root)) AND (level <> depth)) THEN DO
            /* depth-first tree traversal */
            CALL kdccfm(left son(root), level+1)
            CALL kdccfm(right son(root), level+1)
        END
        /* color modification on recursive return */
        IF (value (root)<> nil) THEN DO
            IF (NOT terminal (root)) THEN delete tree (root)
            IF (value (root)>O) THEN link (root) <- black
            IF (value (root)<O) THEN link (root) <- white
            value (root) <- nil
        END
        /* merge of sons' nodes on return */
         IF (NOT terminal (root)) THEN merge (root)
    END
END
```

## 6.7. Median filtering according to $d_\infty$ of space objects

PROCEDURE kd0mdf (root, dimens, precis)

This procedure is identical to kd1mdf except that in place of kd1fmr and kd1mnf are called the procedures kd0fmr and kd0mnf.

## 6.8. Splitting the $d_\infty$-boundary and the $d_\infty$-exo-boundary of a set

PROCEDURE kd0rnfr(root, dimens, precis)

This procedure is identical to kd1mfr except that in place of kd1mcf is called the procedure kd0mcf.

## 6.9. Search and splitting $d_\infty$-boundary and $d_\infty$-exo- boundary nodes of the set

PROCEDURE kd0mcf(root1, root2, symvec, level, depth)

This procedure is identical to k1lmcf except that the descent for searching for boundary and exo-boundary points is a search for adjacencies according to $d_\infty$.

## 6.10. Marking nodes to be modified by median filtering according to $d_\infty$

PROCEDURE kd0mnf(root, dimens, précis)

This procedure is identical to kd1mnf except that in place of kd1fmg is called the procedure kd0fmg.

## 6.11. Search and counting $d_\infty$-adjacencies among the nodes to be filtered

PROCEDURE kd0fmg(root1, root2, symvec, level, depth)

This procedure is identical to kd1fmg except that the search for adjacencies is done according to $d_\infty$.

## 6.12. Computing the median set of an object

PROCEDURE kdemed(root, dimmed, dimens, precis)

BEGIN

    DO

        /* object thinning */

        nbpoint <- kdamct(root, dimmed, dimens, precis)

    UNTIL (nbpoint=0)

END

## 6.13. Thinning of a space object

FUNCTION kdamct(root, dimmed, dimens, precis)

BEGIN

    cnxdeg <- dimens - dimmed

    CALL kdamfr(root, dimens, precis)

    nblftpt <- kdsupl(root, cnxdeg, dimens, precis)

    CALL kdamfr(root, dimens, precis)

    nbrgtpt <- kdsupr(root, cnxdeg, dimens, precis)

    kdamct <- nblftpt + nbrgtpt

END

## 6.14. Evaluation of the connectivity degree of boundary points with the background

```
PROCEDURE kdamfr(root, dimens, precis)
BEGIN
    /* computing parameter initialization */
    depth <- dimens*precis
    level <- 0
    /* recursive computing unit */
    PROCEDURE kdamfr(root, level[ ,depth])
    BEGIN
        IF ((NOT terminal (root)) AND (level <> depth)) THEN DO
            /* search for adjacencies with the space background */
            symvec <- vector(sym, neutral, ..., neutral)
            CALL kdiamc (left son (root), right son (root), link (symvec), (level MOD (dimens)+1,  level+1, depth)
            /* depth-first tree traversal */
            CALL kdamfr (left son (root), level+1)
            CALL kdamfr (right son(root), level+1)
            delete vector (symvec)
        END
    END
END
```

## 6.15. Evaluation of the connectivity degree of boundary points according to a given direction

PROCEDURE kdiamc(root1, root2, symvec, axisnb, level, depth)

BEGIN

    /* recursive computing unit */

    PROCEDURE kdiamc(root1, root2, vecym[, axisnb],level[, depth])

    BEGIN

        IF ((*link* (root1) <> *link* (root2)) AND (level <> depth)) THEN DO

            /* descent for looking for boundary points*/

            IF (*black* (root1)) THEN *divide* (root1)

            IF (*black* (root2)) THEN *divide* (root2)

            symvec2 <- *copy* (symvec)

            typesym <- *value*(symvec)

            IF (typesym = <u>neutral</u>) THEN DO

                /* descent orthogonally to the symmetry axis */

                CALL kdiamc (*left son* (root1), *left son* (root2), *link* (symvec2), level+1)

                CALL kdiamc (*right son* (root1), *right son* (root2), *link* (symvec2), level+1)

            END

            ELSE DO

                /* descent orthogonally to the symmetry axis */

                CALL kdiamc (*right son*(root1), *left son*(root2), *link*(symvec2), level+1)

            END

            *delete vector*(symvec2)

    END

ELSE DO

/* precision reached: compute of the connectivity degree of boundary points */

IF ((NOT *white* (root1)) AND (*white* (root2)))

THEN *value* (root1) <- *value* (root1) + (4**axisnb)*<u>right</u>

IF ((*white* (root1)) AND (NOT *white* (root2)))

THEN *value* (root2) <- *value* (root2) + (4** axisnb)*<u>left</u>

END

END

END

## 6.16. Suppression of left boundary points weakly connected with the background

```
FUNCTION kdsupl(root, cnxdeg, dimens, precis)
BEGIN
    /* computing parameter initialization */
    depth <- dimens*precis
    level <- 0
    /* recursive computing unit */
    FUNCTION kdsupl (root[, cnxdeg, dimens], level[, ,depth])
    BEGIN
        IF ((NOT terminal(root)) AND (level <> depth)) THEN DO
            /* depth-first tree traversal */
            suplft <- kdsupl (left son(root), level+1)
            suprgt <- kdsupl (right son(root), level+1)
            kdsupl <- suplft + suprgt
        END
        ELSE DO
            /* precision reached: connectivity evaluation of boundary points */
            IF (NOT white(root)) THEN DO
                connex <- 0
                front <- false
                FOR axisnb=1 TO dimens DO
                    IF (value (root) AND left) THEN front <- true
                    IF (value(root) AND (left + right)) THEN connex <- +1
                    value(root) <- value(root)/4
                END
```

```
                    /* removal of a weakly connected point */
                    IF ((front) AND (connex < cnxdeg)) THEN DO
                            IF (NOT terminal (root)) THEN delete tree (root)
                            whiten (root)
                            kdsupl <- 1
                    END
                    ELSE kdsupl <- 0
              END
              value (root) <- nil
         END
         /* merge of sons' nodes on return */
         IF (NOT terminal (root)) THEN merge (root)
      END
END
```

## 6.17. Suppression of right boundary points weakly connected with the background

FUNCTION kdsupr(root, cnxdeg, dimens, precis)

   This function is identical to kdsupl, except that the boundary indicator is only validated for right connected points with the help of the instruction:

   IF (*value* (root) AND <u>right</u>) THEN front <- <u>true</u>

## 6.18. Computation of the intrinsic dimension of a set

FUNCTION kdidim(root, dimens, precis)

BEGIN

    CALL kdevdi(root, dimens, precis)

    kdidim <- kdcdim(root, dimens, precis)

END

## 6.19. Evaluation of the intrinsic dimension of the set points

PROCEDURE kdevdi(root, dimens, precis)

BEGIN

    /* computing parameter initialization */

    depth <- dimens*precis

    level <- 0

    /* recursive computing unit */

    PROCEDURE kdevdi (root, level[, depth])

    BEGIN

        IF ((NOT *terminal* (root)) AND (level <> depth)) THEN DO

            /* search for adjacencies between points of the set */

            symvec <- *vector* (sym, neutral, ..., neutral)

            CALL kddcps (*left son*(root), *right son*(root), *link* (symvec), (level MOD dimens)+1, level+1, depth)

            /* depth-first tree traversal */

            CALL kdevdi (*left son* (root), level+1)

            CALL kdevdi (*right son* (root), level+1)

            *delete vector* (symvec)

        END

    END

END

## 6.20. Evaluation of the connectivity degree of the set points

PROCEDURE kddcps(root1, root2, symvec, axisnb, level, depth)

BEGIN

    /* recursive computing unit */

    PROCEDURE kddcps(root1, root2, symvec[, axisnb], level[, depth])

    BEGIN

        IF (((NOT *terminal* (root1)) OR (NOT *terminal* (root2))) AND (level <> depth)) THEN DO

            /* descent and looking for set points*/

            symvec2 <- *copy* (symvec)

            typesym <- *value* (symvec)

            IF (typesym = <u>neutral</u>) THEN DO

                /* descent orthogonally to the symmetry axis */

                CALL kddcps (*left son* (root1), *left son* (root2), *link* (symvec2), level+1)

                CALL kddcps (*right son* (root1), *right son* (root2), *link* (symvec2), level+1)

            END

            ELSE DO

                /* descent parallel to the symmetry axis */

                CALL kddcps (*right son* (root1), *left son* (root2), *link* (symvec2), level+1)

            END

            *delete vector* (symvec2)

    END

```
        ELSE DO
            /* precision reached: compute the connectivity degree of the set nodes */
            IF ((NOT white (root1)) AND (NOT white (root2))) THEN DO
                value (root1)  <-  value(root1) + (4**axisnb)*right
                value (root2)  <-  value(root2) + (4**axisnb)*left
            END
        END
    END
END
```

## 6.21. Computation of the intrinsic dimension of the set points

```
FUNCTION kdcdim(root, dimens, precis)
BEGIN
    /* computing parameter initialization */
    depth <- dimens*precis
    level <- 0
    /* recursive computing unit */
    FUNCTION kdcdim(root[, dimens], level[, depth])
    BEGIN
        IF ((NOT terminal (root)) AND (level <> depth)) THEN DO
            /* depth-first tree traversal */
            kdlft <- kdcdim (left son (root), level+1)
            kdrgt <- kdcdim (right son (root), level+1)
            kdcdim <- MAX (kdlft, kdrgt)
        END
        ELSE DO
            /* precision reached: evaluation of the intrinsic dimension of the set point */
            kdcdim <- 0
            IF (NOT white (root)) THEN DO
                FOR axisnb=1 TO dimens DO
                    IF (value (root) AND (left + right)) THEN  kdcdim <- +1
                    value (root) <- value (root)/4
                END
            END
            value (root) <- nil
        END
    END
END
```

# 7. Representation conversion between $2^K$-trees and pyramids

| | |
|---|---|
| root : | root of the tree to be enriched |
| fctdim : | functional dimension |
| dimens : | space dimension |
| precis : | computing precision |
| level : | level reached in the tree |
| depth : | computing depth |
| fctmin, fctmax : | functional minimum and maximum values |
| lftson, rgtson : | left and right sons of the root |
| minroo, maxrac : | lower and higher faces of the root |
| minson, maxson : | faces associated to a son of the root |
| vecpyr : | vector sampled in a pyramid |

## 7.1. Conversion of a 2$^K$-tree into a pyramid

FUNCTION kd2kpy(root, fctdim, dimens, precis)

BEGIN

    /* computing parameter initialization */

    fctmin <- 0., fctmax <- 1.

    depth <- dimens*precis

    level <- 0

    /* recursive computing unit */

    FUNCTION kd2kpy(root[, fctdim], fctmax[, dimens], level[, depth])

    BEGIN

        IF ((NOT *terminal* (root)) AND (level <> depth)) THEN DO

            /* depth-first tree traversal */

            IF ((MOD (level, dimens) + 1 = fctdim) THEN DO

                /* functional (image) dividing */

                lftson <- kd2kpy (*left son*(root), fctmin, (fctmin + fctmax)/2., level+1)

                rgtson <- kd2kpy (*right son*(root), (fctmin + fctmax)/2., fctmax, level+1)

            END

          ELSE DO

              /* support (inverse image) dividing */

              lftson <- kd2kpy (*left son*(root), fctmin, fctmax, level+1)

              rgtson <- kd2kpy (*right son*(root), fctmin, fctmax, level+1)

          END

```
            /* pyramid building on traversal return */
            IF ((MOD (level, dimens) + 1) = fctdim) THEN DO
                /* functional axis, union of sub-pyramids */
                kd2kpy <- kdunio (lftson, rgtson, dimens, precis)
                delete tree(lftson)
                delete tree(rgtson)
            END
            ELSE nodes union (lftson, rgtson)
            /* pyramid back climbing */
            merge (kd2kpy)
        END
        ELSE DO
            /* precision reached */
            IF (NOT white (root))
            THEN kd2kpy <- pyramid ((fctmin + fctmax)/2., black)
            ELSE kd2kpy <- pyramid (nil, white)
        END
    END
END
```

## 7.2. Conversion of a pyramid into a $2^K$-tree

FUNCTION kdpy2k(root, dimens, precis)

BEGIN

    /* computing parameter initialization */

    tree2k <- *tree* (white)

    minroo <- *vector*(0., 0., ..., 0.),    maxroo <- *vector*(1., 1., ..., 1.)

    depth <- dimens*precis,    level <- 0

    /* recursive computing unit */

    PROCEDURE kdpy2k(root[, tree2k], minroo, maxroo, level[, depth])

    BEGIN

        IF (level <> depth) THEN DO

            /* depth-first tree traversal: pyramid sampling */

            minson <- *copy*(minroo),    maxson <- *copy*(maxroo)

            *value* (minson) <- *value*(minroo)

            *value* (maxson) <-(*value*(minrooc) + *value*(maxroo))/2.

            CALL kdpy2k (*left son (*root), *link* (minson), *link* (maxson), level+1)

            *value* (minson) <- (*value*(minroo) + *value*(maxroo))/2.

            *value* (maxson) <- *value*(maxroo)

            CALL kdpy2k (*right son* (root), *link* (minson), *link* (maxson), level+1)

            *delete* (maxson),        *delete*(minson)

        END

        ELSE DO

            /* precision reached: enrichment of the $2^k$-tree */

            vecpyr <- *copy* (maxroo)

            *insert head* (*value*(root), vecpyr)

            kdarvt (tree2k, vecpyr, dimens, precis)

        END

    END

END

## 7.3. Extraction of the support from a pyramid

```
FUNCTION kdsupy(root, dimens, precis)
BEGIN
    /* computing parameter initialization */
    depth <- dimens*precis
    level <- 0
    /* recursive computing unit */
    FUNCTION kdsupy(root, level)
    BEGIN
        IF ((NOT terminal (root)) AND (level <> depth)) THEN DO
            /* depth-first tree traversal */
            lftson <- kdsupy (left son (root), level+1)
            rgtson <- kdsupy (right son (root), level+1)
            kdsupy <- sub-trees union(lftson, rgtson)
        END
        ELSE kdsupy <- tree (color (root))
    END
END
```

## 7.4. Coloring of a $2^K$-tree

FUNCTION kdcolt(root, fctval, dimens, precis)

BEGIN

    /* computing parameter initialization */

    depth <- dimens * precis

    level <- 0

    /* recursive computing unit */

    FUNCTION kdcolt (root, level)

    BEGIN

        IF ((NOT *terminal* (root) AND (level <> depth) THEN DO

            /* depth-first tree traversal */

            lftson <- kdcolt (*left son*(root), level+1)

            rgtson <- kdcolt (*right son*(root), level+1)

            kdcolt <- *sub-trees union*(lftson, rgtson)

        END

        ELSE kdcolt <- *pyramid* (fctval, *color* (root))

    END

END

# 8. Transforms applied on the functional values of a pyramid

root :         root of the pyramid to be processed

dimens :       space dimension

precis :       computing precision

center :       functional average or centering factor

disper :       functional dispersion or scale factor

depth :        computing depth

level :        level reached in the tree

lftfct :       functional value associated to the left son

rgtfct :       functional value associated to the right son

lftavg :       left son average

rgtavg :       right son average

lftdsp :       left son dispersion

rgtdsp :       right son dispersion

## *8.1. Minimum of a pyramid*

FUNCTION kdmipy(root, dimens, precis)

BEGIN

    /* computing parameter initialization */

    depth <- dimens * precis

    level <- 0

    /* recursive computing unit */

    FUNCTION kdmipy(root, level)

    BEGIN

        IF ((NOT *terminal* (root)) AND (level <> depth)) THEN DO

            /* depth-first tree traversal */

            lftfct <- kdmipy(*left son*(root), level+1)

            rgtfct <- kdmipy(*right son*(root), level+1)

            /* compute node minimum */

            IF (lftfct < rgtfct)

            THEN kdmipy <- lftfct

            ELSE kdmipy <- rgtfct

        END

        ELSE DO

            /* evaluation of the reached node value */

            kdmipy <- *functional*(root)

        END

    END

END

## 8.2. Maximum of a pyramid

FUNCTION kdmapy(root, dimens, precis)

Function identical to kdmipy, except that the test:

IF (lftfct < rgtfct) THEN kdmipy <- lftfct ELSE kdmipy <- rgtfct

becomes:

IF (lftfct > rgtfct) THEN kdmapy <- lftfct ELSE kdmapy <- rgtfct

## 8.3. Computation of the center and the dispersion of a pyramid

PROCEDURE kdctdp(root, center, disper, dimens, precis)

BEGIN

    /* computing parameter initialization */

    depth <- dimens * precis

    level <- 0

    /* recursive computing unit */

    PROCEDURE kdctdp(root, center, disper, level)

    BEGIN

        IF ((NOT *terminal* (root)) AND (level <> depth))THEN DO

            /* depth-first tree traversal */

            CALL kdctdp (*left son* (root), lftavg, lftdsp, level+1)

            CALL kdctdp (*right son* (root), rgtavg, rgtdsp, level+1)

            /* compute center and dispersion of the non terminal node */

            center <- (lftavg + rgtavg)/2.

            disper <- (lftdsp + rgtdsp)/2.+(lftavg-rgtavg)**2/4.

        END

        ELSE DO

            /* compute center and dispersion of the reached node */

            center <- *functional*(root)

            disper <- 0.

        END

    END

    IF (disper > 0.) THEN disper <- sqrt (disper) ELSE disper <- 0.

END

## *8.4. Scaling a pyramid*

PROCEDURE kdscal(root, center, disper, dimens, precis)

BEGIN

    /* computing parameter initialization */

    depth <- dimens * precis

    level <- 0

    /* recursive computing unit */

    PROCEDURE kdscal(root, level)

    BEGIN

        IF ((NOT *terminal* (root)) AND (level <> depth)) THEN DO

            /* depth-first tree traversal */

            CALL kdscal (*left son* (root), level+1)

            CALL kdscal (*right son* (root), level+1)

            /* functional value update */

            IF (*functional* (*left son* (root)) > *functional* (*right son* (root)))

            THEN *functional* (root) <- *functional* (*left son* (root))

            ELSE *functional* (root) <- *functional* (*right son* (root))

        END

        ELSE DO

            /* functional value scaling */

            *functional* (root) <- (*functional* (root) - center)/disper

        END

    END

END

# 9. Median transformations of a pyramid

| | |
|---|---|
| root : | root of the tree to be processed |
| dimens : | space dimension |
| precis : | computing precision |
| nbpoint : | number of points |
| root1, root2 : | symmetrical nodes in the tree |
| symvec : | symmetry vector |
| level : | level reached in the tree |
| depth : | computing depth |
| colist : | list of colors associated to the neighbors of a node |
| nbneig : | number of neighbors colored with the current value |
| nbmaj : | number of occurrences of the majority color |

## 9.1. $d_1$-extension of a discrete set

```
PROCEDURE kd1prl(root, dimens, precis)
BEGIN
    DO
        /* set expansion using median filtering */
        nbpoint <- kd1exp(root, dimens, precis)
    UNTIL (nbpoint = 0)
END
```

## 9.2. $d_1$-expansion of a discrete set by median filtering

```
FUNCTION kd1exp(root, dimens, precis)
BEGIN
    CALL kd1cfr(root, dimens, precis)
    kd1exp <- kdmajp(root, dimens, precis)
END
```

## 9.3. Search for colors of $d_1$-neighbors of the boundary

PROCEDURE kd1cfr (root, dimens, precis)

This procedure is identical to kd1anr, except that in place of the procedure kd1asn is called the procedure kd1cbl.

## 9.4. Search for colors of $d_1$-neighbors of the boundary according to a given symmetry vector

PROCEDURE kd1cbl (root1, root2, symvec, level, depth)

This procedure is identical to kd1asn, except that when the precision is reached, in place of registering in an adjacency list the nodes adjacent to the reached nodes, it is stored the colors of the nodes neighboring the boundary nodes:

/* precision reached: store the colors of the nodes neighboring the boundary nodes */
IF ((*white* ((root1)) OR (*white* (root2)) THEN DO

    IF ((*white* (root1)) AND (NOT *white* (root2))) THEN DO

        IF (*nil* (*value* (root1))) THEN *value* (root1) <- *create a color list*

        $store\ color\ in\ list(value(\text{root1}), color(\text{root2}))$

    END

    IF ((NOT *white* (root1)) AND (*white* (root2))) THEN DO

        IF (*nil* (*value* (root2))) THEN *value* (root2) <- *create a color list*

        $store\ color\ in\ list\ (value(\text{root2}), color\ (\text{root1}))$

    END

END

## 9.5. Assignment the majority color of their neighbors to the points

```
FUNCTION kdmajp(root, dimens, precis)

BEGIN

    /* computing parameter initialization */

    depth <- dimens*precis,        level <- 0

    /* recursive computing unit */

    FUNCTION kdmajp(root, level[, depth])

    BEGIN

        IF ((NOT terminal (root)) AND (level <> depth)) THEN DO

            /* depth-first tree traversal */

            lftmaj <- kdmajp (left son(root), level+1)

            rgtmaj <- kdmajp (right son(root), level+1)

            kdmajp <- lftmaj + rgtmaj

        END

        ELSE DO

            /* precision reached: evaluation of the neighboring majority color and
            assignment the color to the node */

            IF (value (root) <> nil) THEN DO

                color (root) <- kdvote(value(root))

                delete (value(root))

                value (root) <- nil

                link (root) <- black

                kdmajp <- 1

            END

            ELSE kdmajp <- 0

        END

        /* merge of sons' nodes on return path */

         IF (NOT terminal (root)) THEN merge (root)

    END
```

END

## 9.6. *Determination of the value providing the maximum number of occurrences in a list*

```
FUNCTION kdvote(colist)
BEGIN
    /* computer parameter initialization */
    index <- link(colist)
    nbmaj <- 0
    /* list analysis */
    WHILE (index <> nil) DO
        /* counting the occurrences of the current value */
        ixcour <- link(index)
        nbneig <- 1
        WHILE (ixcour <> nil) DO
            IF (value (ixcour) = value (index)) THEN nbneig <- +1
            ixcour <- link(ixcour)
        END
        /* determination of the color providing the maximum number of occurrences */
        IF (nbneig > nbmaj) THEN DO
            kdvote <- value(index)
            nbmaj <- nbneig
        END
        /* next color to be processed */
        index <- link(index)
    END
END
```

## 9.7.  $d_\infty$-extension of a discrete set

PROCEDURE kd0prl(root, dimens, precis)

## 9.8.  $d_\infty$-expansion of a discrete set by median filtering

FUNCTION kd0exp(root, dimens, precis)

## 9.9.  Search for colors of $d_\infty$-neighbors of the boundary

PROCEDURE kd0cfr(root, dimens, precis)

## 9.10. Search for colors of $d_\infty$-neighbors of the boundary according to a given symmetry vector

PROCEDURE kd0cbl(root1, root2, symvec, level, depth)

After having adapted the calls to the sub-programs and the handling of $d_\infty$-adjacencies in kd0cbl, these procedures are identical to kd1prl, kd1exp, kd1cfr, kd1cbl.

## 9.11. Median filtering according to $d_1$ of a pyramid

PROCEDURE kd1fmp(root, dimens, precis)

## 9.12. Search for colors $d_1$-neighbors of the inside

PROCEDURE kd1cin(root, dimens, precis)

## 9.13. Search for colors $d_1$- neighbors of the inside according to a given symmetry vector

PROCEDURE kd1cnr(root1, root2, symvec, level, depth)

After having adapted the calls to the sub-programs and the registration of the colors of interior and not boundary nodes, these procedures are identical to à kd1exp, kd1cfr and kd1cbl.

With the handling of $d_\infty$-adjacencies in kd0cnr, the equivalent modules kd0fmp, kd0cin and kd0cnr will provide a median filtering according to $d_\infty$ of a pyramid.